# The Role of Immigrants, Emigrants, and Locals in the Historical Formation of European Knowledge Agglomerations[*]

*Philipp Koch*[1,2,†], *Viktor Stojkoski*[1,3,‡], *César A. Hidalgo*[1,4,5,§]

[1] Center for Collective Learning, ANITI, TSE-R, IAST, IRIT, Université de Toulouse, 31000 Toulouse, France.
[2] EcoAustria – Institute for Economic Research, 1030 Vienna, Austria.
[3] Faculty of Economics, University Ss. Cyril and Methodius in Skopje, North Macedonia.
[4] Alliance Manchester Business School, University of Manchester, United Kingdom.
[5] School of Engineering and Applied Sciences, Harvard University, United States.

**Abstract**

Did migrants make Paris a Mecca for the arts and Vienna a beacon of classical music? Or was their rise a pure consequence of local actors? Here, we use data on more than 22,000 historical individuals born between the years 1000 and 2000 to estimate the contribution of famous immigrants, emigrants, and locals to the knowledge specializations of European regions. We find that the probability that a region develops or keeps specialization in an activity (based on the birth of famous physicists, painters, etc.) grows with both, the presence of immigrants with knowledge on that activity and immigrants with knowledge in related activities. In contrast, we do not find robust evidence that the presence of locals with related knowledge explains entries and/or exits. We address some endogeneity concerns using fixed-effects models considering any location-period-activity specific factors (e.g. the presence of a new university attracting scientists).

# 1. Introduction

Migrants help carry knowledge across space[1–5], shaping the geography of cultural and economic activities[6–10]. But most studies documenting the role of migrants in the diffusion of knowledge use recent data on patents[6,11–19], research[13,20], or product exports[21], or analyze historical spillovers within activities[22–33], leaving questions about the role of migrants in the historical formation of knowledge agglomerations relatively unexplored.

To explore the role of migrants in the historical formation of knowledge agglomerations we use biographic data on more than 22,000 famous individuals—artists, physicists, explorers, philosophers, etc.—living in Europe between the years 1000 and 2000. We use this data to investigate how immigrants, emigrants, and locals explain the probability that famous individuals specialized in an activity—that was not yet present in a region—are born during the next century. That is, we study how the knowledge of migrants and locals contributes to explain, for example, Paris becoming the birthplace of painters and Vienna of composers.

We can explore these questions by creating measures of knowledge spillovers within and between locations and activities. Consider spillovers across locations within the same activity. The knowledge that migrants carry across borders may impact a location's ability to give birth to famous figures in the activity that the migrants specialize in. That is, immigrant mathematicians may increase the probability that a city or region begets famous mathematicians. Similarly, emigrating mathematicians may decrease that probability. To capture such spillovers, we identify whether a region experiences a larger than expected inflow or outflow of famous individuals specialized in an activity.

Now consider spillovers across both locations and activities. Migrants and locals specialized in an activity (e.g. mathematics) can impact a region's ability to give birth to famous figures in a related activity (e.g. physics). To capture such spillovers, we use measures of relatedness[34–38], which exploit information on the colocation of activities to estimate how "cognitively close" a location is to an activity.

During the past decades, measures of relatedness have been validated as robust predictors of the probability that countries, regions and cities enter or exit an activity, such as product exports[34,39,40], technologies[41–47], industries[48–51], and research areas[52–54]. Recent contributions to this literature have focused on unpacking relatedness by considering multiple channels[55–62]. For instance, does industry-specific or occupation-specific knowledge contribute to the growth and survival of firms?[58] Or do value chains or knowledge agglomerations explain the collocation of firms?[56] To the best of our knowledge no study has yet unpacked relatedness in the context of historical migration. Here we use a dataset spanning 1,000 years of history in Europe to explore how the knowledge of immigrants, emigrants, and locals explains the probability that a famous cultural figure specialized in an activity is born in a specific region. This contributes to both, understanding the role of migrants in the geography of knowledge and unpacking relatedness metrics in the context of migration.

Our findings show that migrants play a crucial role in knowledge agglomerations. Specifically, we find that the probability that a European region enters a new activity grows on average by between 1.7 and 4.6 percentage points if that region received an excess number of immigrants specialized in that activity during the last century. Moreover, we find this correlation is enhanced by immigrants specialized in related activities. Similarly, we find the probability that a European region loses one of its existing specializations decreases on average by 5.0 to 10.2 percentage points if that region received an excess number of immigrants specialized in that activity. This correlation is also enhanced by immigrants specialized in related activities. In contrast, we do not find a statistically significant and robust role of the related knowledge of locals (people born in that region) in entries or exits.

To tackle some important endogeneity concerns (migration is often a motivated choice), we employ a highly restrictive fixed effects structure controlling for all possible unobserved factors that are specific to a broad occupational category in a region during a century. These are factors that might affect both, migration patterns and the birth of famous individuals, such as a new university attracting scientists and leading to the birth of more famous scientists in the future, or a prosperous city attracting and begetting more artists. In addition, we control for unobserved factors that are specific to a more granular occupational category in a century which might affect both migration and births (e.g. the emergence of a new technology (e.g. photography) begetting a new occupational category (photographers)). This captures, for instance, that musicians and singers are likely to have different migration and birth patterns across time than other artists such as painters or actors. Lastly, we tackle some concerns of reverse causality by focusing on excess migration and estimating the expected number of migrants in a location. Although we control for multiple possible observed and unobserved factors to limit endogeneity concerns, we want to stress that we are not able to make strictly causal claims.

Together, these findings advance our understanding of the role of immigrants, emigrants, and locals in the historical formation of knowledge agglomerations. They contribute to both, the literature on the role of migrants in knowledge diffusion[1–9,11–18,20–28] and the literature on relatedness[34–62]. Moreover, by developing measures of the related knowledge of migrants, we combine both migration and relatedness in a framework that can be used to study how knowledge spillovers across space and across activities combine in more recent settings. Lastly, this study provides a long-term perspective on the evolution of regional specialisations in Europe, a perspective which is underrepresented in the field of economic geography[63].

## 2. Data & Methods

### 2.1. Data

We use the 2020 version of Pantheon[64], a publicly available dataset including information on famous individuals with a Wikipedia page in more than 15 different language editions. We focus on the 22,847 famous individuals born or died in Europe between the years 1000 and 2000. We choose Pantheon because it assigns individuals using a controlled taxonomy of 101 occupations, such as painter, writer, composer, physicist, chemist, mathematician, etc. Pantheon provides a good sectoral disaggregation compared to other datasets which either have few sectors[65] or use uncontrolled taxonomies with duplicate entries, e.g. film director and movie director[66]. This granularity is needed to construct measures of specialization and relatedness. The full taxonomy and descriptive statistics are provided in the Supplementary Materials (SM) Section 1.1.

We use geographic coordinates to assign the place of birth and death of each biography to European administrative regions (NUTS-2 or regions of similar size for countries outside the EU, e.g. Russian Oblasts, see SM Section 1.2). Figures 1a and 1b show the places of birth and death of all individuals in our dataset within the applied administrative borders. Due to a lack of data on the full trajectory of individuals, we follow the literature investigating migration patterns of famous individuals[65–68] and use places of birth and death as rough proxies for migration. Manual inspection of a random sample of 200 biographies revealed places of death to be a valid proxy of an important living place for around 90 percent of biographies and corresponded to a place of major impact for 75 percent of biographies (SM Section 1.3).

Finally, we assign each individual to a century *t* based solely on his or her year of birth. That is, a famous person who is born in the 18th century in Brussels and died in Paris (in the 18th or 19th century) is considered a local in Brussels and an immigrant in Paris in the 18th century. We choose this approach since we do not have information on the time of migration. We take this into account in the regression models by lagging the independent variables (see also SM Section 1.1).

## 2.2. Descriptive Statistics: Migration & Spatial Concentration Patterns

We find that most of the migration of famous Europeans over the past 1,000 years took place within countries and towards large cities (e.g. from smaller cities in France to Paris). Figures 1c and 1d visualize the migration network. Migration is common among famous individuals. In fact, going back to the 11th century, the share of migrants in our dataset never drops below 65 percent. In the 19th century, almost 80 percent of famous individuals in our dataset died in a different region than the one in which they were born (see Fig. 1e).

These migration patterns are not random but follow a process of preferential attachment, clustering individuals in major cities[65,68–72] and leading to a higher spatial concentration for places of death than birth. For instance, 416 famous individuals were born in Paris in the 19th century, but 934 died there (SM Section 2.1).

We use information entropy $H$ to quantify the spatial concentration of births and deaths across regions. Information entropy (base 2) estimates the number of yes-or-no questions that we would need to answer—on average—to find the place of birth or death of an individual (see SM Section 2.1). If deaths are more concentrated than births, we will need less questions to guess a place of death than one of birth. We can use entropy $H$ to estimate the effective number of places of birth or death as $E=2^H$, which is the number of regions effectively experiencing the birth or death of a famous individual.

Figure 1f shows the effective number of places of birth and death $E$ for each century. Prior to the 15th century, the spatial concentration of famous births and deaths was similar. But starting in the 15th century, places of death have become more spatially concentrated and places of birth more widespread. In fact, by the 19th century famous individuals were effectively born in more than 200 (out of 405) regions across Europe, while they effectively died in only 100 regions (Fig. 1f).

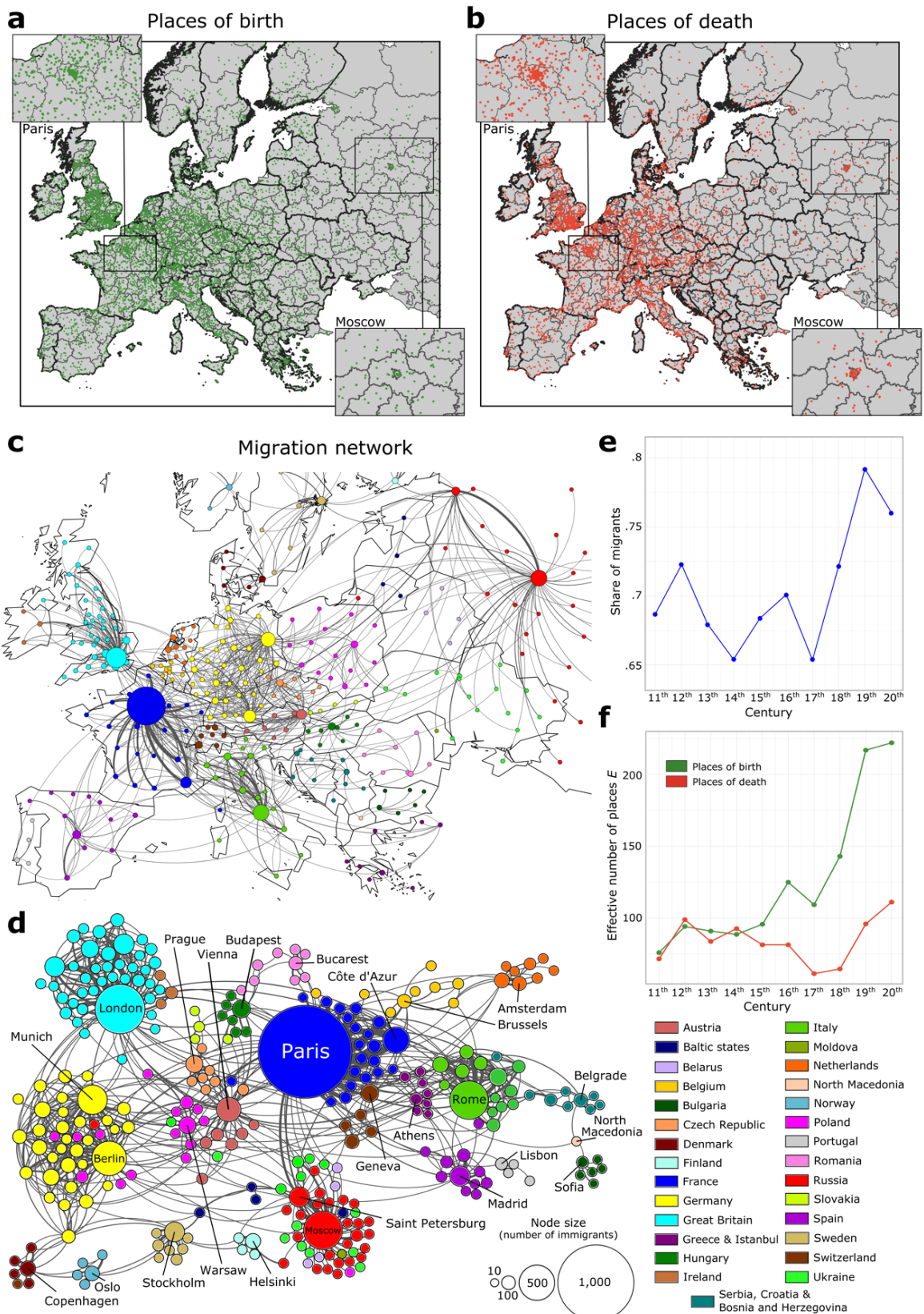

*Figure 1. **Places of birth, places of death and migration patterns of famous individuals in Europe over the past 1,000 years.** (a-b) Maps of **a** places of birth and **b** places of death included in the analysis (NUTS-2 regions for EU, comparable regions for other countries, e.g. oblasts in Russia, see SM Section 1.2). (c-d) Migration network of famous individuals within Europe over the past 1,000 years, using **c** geography or **d** a force-directed algorithm for visualization. The latter reveals that famous individuals tend to move within countries towards large regions. (e) Share of migrants in the dataset per century. (f) Effective number of places of birth and death E derived from Shannon entropy (see SM Section 2.1). Starting in the 15th century, the places of death of famous individuals are more spatially concentrated than their places of birth.*

## 2.3. Methods

### 2.3.1 Relatedness of Immigrants, Emigrants, and Locals

To explore how the knowledge of immigrants, emigrants, and locals shapes the geography of knowledge, we estimate the probability that a region gives birth to a famous individual specialized in an activity as a function of estimates of knowledge spillovers within and between regions and activities.

To capture the knowledge spillovers of migration within the same activity, we calculate the ratio between the observed number of famous immigrants ($N_{ik,t}^{immi}$) or emigrants ($N_{ik,t}^{emi}$) with a certain activity and their expected number (respectively $\widehat{N}_{ik,t}^{immi}$ and $\widehat{N}_{ik,t}^{emi}$), where $i$ denotes the region, $k$ the occupation, and $t$ the century.

Taking the ratio between the observed and expected number of migrants allows us to create measures of excess immigration or excess emigration, and thus, to control for the natural attractiveness of a location and the characteristics of an activity. This is important to address reverse causality concerns, since the effects of migrants could be simply a reflection of local factors making a place attractive for migrants with a certain specialization.

It is worth mentioning that migration decisions can be influenced by multiple local factors. Creatives, for instance, are more likely to move towards places that are already populated by other creatives[26] or potential patrons[73,74]. Geographical and cultural distance[75,76], such as a common language or the presence of fellow countrymen can also play a role[77]. Lastly, migration can also be exogenously forced due to conflict[28] or climate[78]. By focusing on excess migrants, instead of total migrants, in a restrictive fixed-effects model we help mitigate the risks of reverse causality.

Mathematically, this involves taking the ratio between the observed and expected number of immigrants or emigrants:

$$R_{ik,t}^{immi} = \frac{N_{ik,t}^{immi}}{\widehat{N}_{ik,t}^{immi}}$$

$$R_{ik,t}^{emi} = \frac{N_{ik,t}^{emi}}{\widehat{N}_{ik,t}^{emi}}$$

, (1)

where the values are for individuals in region $i$ and activity $k$ born in century $t$.

Here we use two models for the expected number of migrants ($\widehat{N}_{ik,t}$). The first one considers the number of individuals in a location and the number of individuals specialized in an activity.

That is a "bins and balls" model for the expected number of immigrants or emigrants, making Eq. 1 the Revealed Comparative Advantage[79] or Location Quotient, a common measure of specialization:

$$\widehat{N}_{ik,t} = \frac{\sum_k N_{ik,t} \sum_i N_{ik,t}}{\sum_{i,k} N_{ik,t}}. \tag{2}$$

The second model expands on this by taking the attractiveness of a location in a specific activity into account[49]. We model $\widehat{N}_{ik,t}$ using a negative binomial regression where we control for the observed number in the previous century ($N_{ik,t-1}$), the previous specialization of the location in the activity based on famous individuals born there ($S_{ik,t-1}^{births}$),

$$S_{ik,t-1}^{births} = \frac{N_{ik,t-1}^{births}}{\left(\frac{\sum_k N_{ik,t-1}^{births} \sum_i N_{ik,t-1}^{births}}{\sum_{i,k} N_{ik,t-1}^{births}}\right)}. \tag{3}$$

where $N_{ik,t}^{births}$ denotes the number of famous individuals born in location $i$ specialized in activity $k$ in century $t$, and fixed effects for each location-time ($\theta_{it}$) and activity-time ($\vartheta_{kt}$) to account for unobserved factors. That is, we estimate

$$\widehat{N}_{ik,t} = f\big(\alpha_0 + \alpha_1 N_{ik,t-1} + \alpha_2 S_{ik,t-1}^{births} + \theta_{it} + \vartheta_{kt}\big), \tag{4}$$

where $f$ denotes the negative binomial probability density (see SM Section 3.4.1 for results).

If the observed number of immigrants or emigrants in an activity exceeds the expected number, we say that region received excess immigrants, or produced excess emigrants, on that activity and time period.

Next, we create two specialization matrices for famous immigrants ($M_{ik,t}^{immi}$; died "here," but born elsewhere) and emigrants ($M_{ik,t}^{emi}$; born "here," but died elsewhere):

$$M_{ik,t}^{immi} = \begin{cases} 1 & if\ R_{ik,t}^{immi} \geq 1 \\ 0 & otherwise \end{cases}$$

$$M_{ik,t}^{emi} = \begin{cases} 1 & if\ R_{ik,t}^{emi} \geq 1 \\ 0 & otherwise. \end{cases} \tag{5}$$

Figures 2a and b show these two matrices using data for individuals born in the 19th century. The matrices are characterized by a nested structure that we recover by sorting locations by diversity (respectively $\sum_k M_{ik,t}^{immi}$ and $\sum_k M_{ik,t}^{emi}$), and activities by ubiquity (respectively $\sum_i M_{ik,t}^{immi}$ and $\sum_i M_{ik,t}^{emi}$). This structure is typical for matrices summarizing the geography of activities[80,81] (SM Section 2.2), but also, for networks describing species interactions in ecology[82–84].

To capture spillovers across activities we use measures of relatedness[34–38]. Relatedness exploits information on the colocation of activities to estimate their affinity with a location. We create three separate measures of relatedness for immigrants, emigrants, and locals.

These measures build on the specialization matrices described in Eq. 5. This time, however, we need to create specialization matrices for locals, which we define as famous individuals who were born in a region, no matter if they died there or elsewhere. We use this definition because of the large share of migrants among famous individuals (see Fig. 1d), which would reduce our number of observations drastically if we defined locals as individuals who were born and died in the same place. Controlling for the related knowledge of emigrants, however, relatedness based on all births is a valid proxy for the related knowledge of individuals who were born and died in the same region (SM Section 2.3).

That is, as before, we calculate the ratio between observed and expected births of famous individuals:

$$R_{ik,t}^{births} = \frac{N_{ik,t}^{births}}{\widehat{N}_{ik,t}^{births}}. \tag{6}$$

Again, we can apply both the naïve model described in Eq. 2 or estimate the expected number of births given local factors (Eq. 3, see SM Section 3.4.1) before creating binary specialization matrices for locals:

$$M_{ik,t}^{births} = \begin{cases} 1 & if\ R_{ik,t}^{births} \geq 1 \\ 0 & otherwise \end{cases} \tag{7}$$

This matrix also exhibits a nested structure (Fig. 2c).

Next, we define the proximity or similarity between two activities as the minimum of the conditional probability that a location is specialized in both of them[34]:

$$\varphi_{kk',t}^{immi} = \frac{\sum_i M_{ik,t}^{immi} M_{ik',t}^{immi}}{max\left(\sum_i M_{ik,t}^{immi}, \sum_i M_{ik',t}^{immi}\right)},$$

$$\varphi_{kk',t}^{emi} = \frac{\sum_i M_{ik,t}^{emi} M_{ik',t}^{emi}}{max\left(\sum_i M_{ik,t}^{emi}, \sum_i M_{ik',t}^{emi}\right)}, \tag{8}$$

$$\varphi_{kk\prime,t}^{births} = \frac{\sum_i M_{ik,t}^{births} M_{ik\prime,t}^{births}}{max\left(\sum_i M_{ik,t}^{births}, \sum_i M_{ik\prime,t}^{births}\right)},$$

and use these proximities to calculate the relatedness between locations and activities as:

$$\omega_{ik,t}^{immi} = \frac{\sum_{k\prime} M_{ik\prime,t}^{immi} \varphi_{kk\prime,t}^{immi}}{\sum_{k\prime} \varphi_{kk\prime,t}^{immi}},$$

$$\omega_{ik,t}^{emi} = \frac{\sum_{k\prime} M_{ik\prime,t}^{emi} \varphi_{kk\prime,t}^{emi}}{\sum_{k\prime} \varphi_{kk\prime,t}^{emi}}, \qquad (9)$$

$$\omega_{ik,t}^{births} = \frac{\sum_{k\prime} M_{ik\prime,t}^{births} \varphi_{kk\prime,t}^{births}}{\sum_{k\prime} \varphi_{kk\prime,t}^{births}}.$$

These measures quantify how far, for example, immigrants to Paris are from being specialized in archeology, emigrants from Madrid are from being specialized in singing, or locals in Berlin are from being specialized in philosophy.

We note that the relatedness densities calculate with the naïve and binomial model are highly correlated ($R^2$>0.9). So, going forward, we present results using the naïve model and provide additional results using the negative binomial model in the SM (Section 3.4.1).

Since the multiple factors contributing to the colocation of activities can be different when looking at immigration, emigration and births, we create separate measures of proximity ($\varphi_{kk\prime,t}^{immi}$, $\varphi_{kk\prime,t}^{emi}$, $\varphi_{kk\prime,t}^{births}$, Eq. 8). But as a robustness check, we also consider a joint measure of proximity ($\varphi_{kk\prime,t}^{joint}$) using colocation at birth and death (see SM Section 2.5). Nevertheless, we find the separate measures of proximity provide valuable nuance (see SM Figure S6). Consider explorers and military personnel. Explorers and military personnel share many required capabilities such as navigating, planning, commanding etc., that may be explained by local factors such as military academies for education, distance to the sea, recency of a war, or naval technology. Also, exploration teams often involve soldiers and military personnel, which could then become famous as explorers. Hence, explorers and military personnel are likely to share a geographic origin. Yet, since exploration and military campaigns tend to involve different locations, these two activities are less likely to collocate at death. Now consider composers and noblemen. For these two activities, the proximity based on immigration patterns is higher than the proximity based on births. It makes sense that these activities are to some extent related when looking at places of birth: Noblemen are known to be patrons for the arts. Hence, noblemen born in a location will likely create institutions that promote the cultivation of the talent of composers born in this location. But it is also plausible that these activities are even more related when looking at immigration patterns. Given that we observe a disproportional migration flow of noblemen towards a certain location, we can view this location as highly related to composers, since the institutional factors attracting noblemen likely play a role in attracting and cultivating the talent of composers as well.

These examples highlight why we believe that generating separate measures of proximity for immigrants, emigrants and births provides a nuanced perspective that helps unpack relatedness (see SM Section 2.5. for more details).

We illustrate the structure of these proximity networks for immigrants born in the 19th century ($\varphi_{kk',t}^{immi}$, Fig. 2d). A high proximity between two activities indicates similarity or complementarity among them. Like measures of propensity, measures of proximity capture the combined presence of multiple factors that may be contributing to the colocation of two activities. For example, we find a high proximity between biologists and physicians, mathematicians and physicists, and musicians and actors (see Fig. 2d). While the latter may be considered an example of colocation due to high complementarity (musicians and actors may perform together), associations between mathematicians and physicists, or biologists and physicians, may indicate similarity in knowledge or skills.

### 2.3.2 Entries and Exits

We use our measures of relatedness to study the entry and exit of activities in European regions. We do this by estimating logistic models explaining the probability that a region starts to give birth to a disproportionately large number of famous individual specialized in an activity (entries) or stops doing so (exits). That is, a region enters the activity "philosophy" if more philosophers are born there in a certain century than expected, while this has not been the case in the prior century. Similarly, we explain the probability that a region loses an existing specialization. A region exits the activity "physics" if fewer physicists are born there than expected, while this has not been the case in the prior century. The variables $Entry_{ik,t}$ and $Exit_{ik,t}$ emerge directly from the specialization matrix defined in Eq. 7.

Specifically, we define:

$$Entry_{ik,t} = \begin{cases} 1 & if\ M_{ik,t-1}^{births} = 0\ and\ M_{ik,t}^{births} = 1 \\ 0 & otherwise \end{cases},$$

$$Exit_{ik,t} = \begin{cases} 1 & if\ M_{ik,t-1}^{births} = 1\ and\ M_{ik,t}^{births} = 0 \\ 0 & otherwise \end{cases}.$$

(10)

That is, a region *i* enters (exits) an occupation *k* in century *t* if the observed births of famous individuals with that occupation during the considered century is larger (lower) than expected, while this was not the case in the prior century.

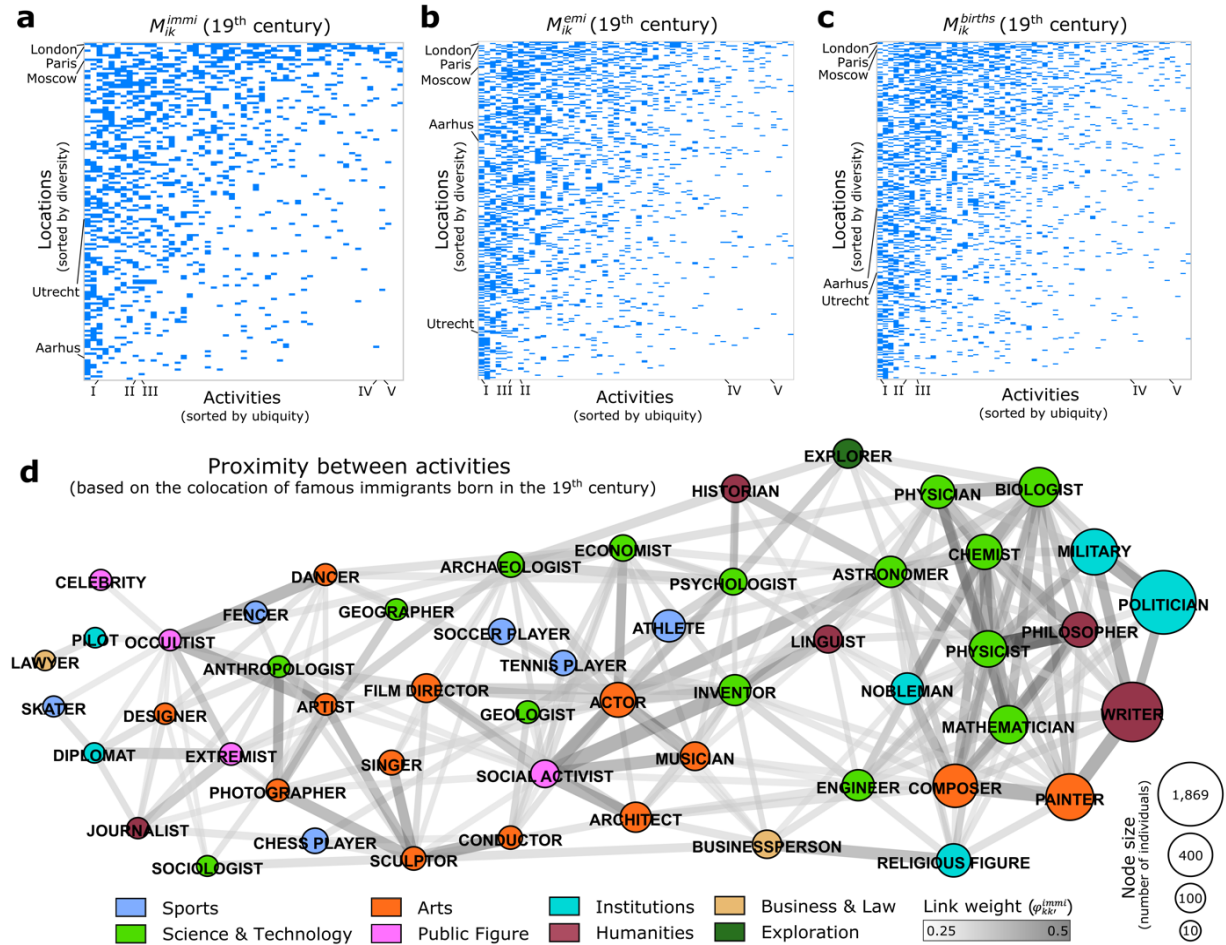

*Figure 2. Nested specialization matrices and the similarity between activities in the 19th century. (a-c)* Specialization matrices based on **a** immigrants, **b** emigrants, and **c** locals in the 19th century (see Eq. 5 and 7). Examples of locations and activities (I: writer, II: mathematician, III: physicist, IV: journalist, V: pilot) are highlighted. *(d)* Proximities between activities based on the colocation of famous immigrants born in the 19th century using the naïve model described in Eq. 2 to determine the expected number of immigrants. Node size is proportional to the number of famous individuals specialized in the respective activity and born in the 19th century.

Defining entries and exits looking at places of birth is a rather conservative approach. For a region to enter an activity, it needs to become a place where the required knowledge to cultivate a certain talent can be absorbed through formal or informal institutions and social ties. Indeed, early exposure to local knowledge in an individual's life is highly relevant in shaping his or her career, both for inventors nowadays[85] and artists centuries ago[86]. A different approach of describing the geography of knowledge would be, for instance, to focus on all individuals living at a certain place. But this would require having data on all places of living.

We explain entries and exits using measures of the presence of immigrants and emigrants in that activity ($M_{ik,t}^{immi}$, $M_{ik,t}^{emi}$) and of the related activities that we can attribute to immigrants, emigrants and locals ($\omega_{ik,t}^{immi}$, $\omega_{ik,t}^{emi}$, $\omega_{ik,t}^{births}$). For instance, a significantly positive correlation between $M_{ik,t}^{immi}$ and entries would point towards migrants bringing into the region the knowledge needed to carry out activity *k*. That is, a high influx of mathematicians would increase the probability that the region begets its own famous mathematicians. This would be consistent with research showing that migrants help carry the knowledge needed to enter an activity[6,11–18,20–28].

Similarly, a significant correlation between $\omega_{ik,t}^{immi}$ and entries would support the idea that the related knowledge brought by migrants also impacts the probability that a region develops a new activity. That is, the knowledge of famous immigrants specialized in mathematics diffuses to related fields, such as physics or chemistry, and increases the probability that a region begets its own physicists or chemists.

Lastly, a significant correlation between $\omega_{ik,t}^{births}$ and entries, after controlling for $\omega_{ik,t}^{emi}$, would indicate that the related knowledge of locals contributes to entering a new activity. That is, a region with many locals already specialized in mathematics has a higher probability of branching into physics or chemistry.

The entry of a region into a new activity could be the result of multiple factors other than migration. For instance, the creation of a new university could attract scientists, and the expansion of a port could create conditions attractive to merchants. We address such endogeneity concerns by using highly restrictive fixed effects models accounting for unobserved factors that could affect both migration and the probability that a region enters an activity. Specifically, we control for these unobserved factors by using fixed effects specific to a broad occupational category, region, and century ($\gamma_{mit}$, i.e. a three-way interaction). Index *m* denotes one of eight broad occupational categories such as "arts", "science & technology", "humanities", or "sports" (see column 1 of Table S1 in the SM).

In addition, we control for unobserved factors affecting both migration and future births that are specific to a more granular occupational category and time ($\delta_{lt}$). Index *l* denotes one of 26 occupation categories, which distinguish, for instance, between social sciences, natural sciences, and engineering within the broad category "science & technology" or music, design and film & theatre within the broad category "arts" (see column 2 in Table S1 in the SM). The latter fixed effects capture, for instance, that the invention of motion picture technology at the end of the 19th century likely affected migration and birth patterns among film directors and actors differently than among other artists, such as painters or sculptors.

We also control for several other observed factors that might correlate with the probability of entry or exit and that are not captured in the fixed effects. This includes an activity's ubiquity (i.e. the number of locations specialized in it) and how close a region already is to having or losing a specialization ($R_{ik,t-1}^{births}$, see Eq. 6). Lastly, we account for knowledge diffusion across space due to other reasons than migration by creating measures of the spatial proximity to other regions with specializations in that specific activity or in related activities (see SM Section 2.4). We provide descriptive statistics and discuss the explanatory variables in more detail in SM Section 3.1.

In sum, we define $Y_{ik,t} = \{Entry_{ik,t}, Exit_{ik,t}\}$ and estimate:

$$P(Y_{ik,t}) = g(\beta_1 M_{ik,t-1}^{immi} + \beta_2 M_{ik,t-1}^{emi}$$
$$+ \beta_3 \omega_{ik,t-1}^{immi} + \beta_4 \omega_{ik,t-1}^{emi} + \beta_5 \omega_{ik,t-1}^{births} \quad , \quad (11)$$
$$+ \alpha' X_{ik,t-1} + \gamma_{mit} + \delta_{lt} + \varepsilon_{ik,t})$$

where $g$ denotes the logistic probability density, $X_{ik,t-1}$ denotes a vector of observed control variables and $\gamma_{mit}, \delta_{lt}$ the fixed effects.

We calculate average marginal effects based on this logistic regression by computing the marginal effect for each data point and taking the average.

## 3. Results

Table 1 and Figure 3 show the relationship between the activities of immigrants, emigrants, and locals and the number of observed entries and exits. For entries (Table 1, columns 1-5), the probability correlates positively with an excess inflow of migrants specialized in an activity during the previous century ($M_{ik,t-1}^{immi} = 1$). Specifically, an excess of immigrants increases the probability of entry on average by 4.6 percentage points (Fig. 3a). Figure 3b plots the probability of entry as a function of $M_{ik,t-1}^{immi}$.

We also find that the probability of entry grows with the related knowledge of immigrants. A standard-deviation-increase of $\omega_{ik,t-1}^{immi}$ increases the probability of entry on average by 5.8 percentage points (Fig. 3a). Figure 3c visualizes the results by plotting the average probability of entry as a function of the relatedness density of immigrants ($\omega_{ik,t-1}^{immi}$). In accordance with the literature[35,87], the average probability of entry grows super-linearly, from 1.1 percent if no related knowledge of famous immigrants is present in a region ($\omega_{ik,t-1}^{immi} = 0$) to 16.4 percent if all related activities are present ($\omega_{ik,t-1}^{immi} = 100$). Moreover, we find a positive correlation ($p < 0.1$) between $\omega_{ik,t-1}^{births}$ and entries, but unlike the estimate of the related knowledge of immigrants, this correlation is not robust (SM Section 3.4).

When we look at exits (Table 1, columns 6-10), we find similar relationships but with the opposite sign. An excess inflow of famous individuals specialized in an activity during the previous century ($M_{ik,t-1}^{immi} = 1$) reduces the probability of exit significantly by 10.2 percentage points on average (Fig. 3d). Also, the related knowledge of immigrants ($\omega_{ik,t-1}^{immi}$) helps prevent losing specialization in an activity. Figure 3e and f visualize these results by plotting the probability of exit as a function of $M_{ik,t-1}^{immi}$ and $\omega_{ik,t-1}^{immi}$, respectively.

*Table 1. Main results of logistic regression models explaining entries and exits of activities.*

|  | Dependent Variable: $Entry_{ik,t}$ | | | | | Dependent Variable: $Exit_{ik,t}$ | | | | |
|---|---|---|---|---|---|---|---|---|---|---|
|  | (1) | (2) | (3) | (4) | (5) | (6) | (7) | (8) | (9) | (10) |
| $M_{ik,t-1}^{immi}$ | 0.334*** | 0.303*** | 0.336*** | 0.331*** | 0.300*** | -0.603*** | -0.584*** | -0.591*** | -0.587*** | -0.571*** |
|  | (0.080) | (0.075) | (0.086) | (0.080) | (0.076) | (0.127) | (0.134) | (0.120) | (0.126) | (0.126) |
| $M_{ik,t-1}^{emi}$ | 0.115 | 0.045 | 0.106 | 0.121 | 0.018 | 0.310 | 0.330 | 0.233 | 0.306 | 0.291 |
|  | (0.261) | (0.278) | (0.261) | (0.255) | (0.270) | (0.240) | (0.232) | (0.216) | (0.222) | (0.203) |
| $\omega_{ik,t-1}^{immi}$ |  | 0.027*** |  |  | 0.028*** |  | -0.067*** |  |  | -0.064*** |
|  |  | (0.006) |  |  | (0.007) |  | (0.016) |  |  | (0.011) |
| $\omega_{ik,t-1}^{emi}$ |  |  | -0.006 |  | -0.024 |  |  | -0.048 |  | -0.025 |
|  |  |  | (0.012) |  | (0.019) |  |  | (0.038) |  | (0.063) |
| $\omega_{ik,t-1}^{births}$ |  |  |  | 0.011 | 0.027* |  |  |  | -0.059*** | -0.034 |
|  |  |  |  | (0.008) | (0.015) |  |  |  | (0.018) | (0.041) |
| Further controls | ✓ | ✓ | ✓ | ✓ | ✓ | ✓ | ✓ | ✓ | ✓ | ✓ |
| *Fixed effects:* | | | | | | | | | | |
| Broad categ.-region-century | ✓ | ✓ | ✓ | ✓ | ✓ | ✓ | ✓ | ✓ | ✓ | ✓ |
| Category-century | ✓ | ✓ | ✓ | ✓ | ✓ | ✓ | ✓ | ✓ | ✓ | ✓ |
| Observations | 3944 | 3944 | 3944 | 3944 | 3944 | 1051 | 1051 | 1051 | 1051 | 1051 |
| Pseudo-R² | 0.213 | 0.214 | 0.213 | 0.213 | 0.215 | 0.224 | 0.230 | 0.226 | 0.226 | 0.232 |
| BIC | 9537.0 | 9539.4 | 9545.0 | 9544.5 | 9553.1 | 3619.6 | 3618.0 | 3623.4 | 3623.3 | 3628.8 |

The fixed effects in these models are highly restrictive, amounting to more than 700 parameters in columns (1)-(5) and more than 350 parameters in columns (6)-(10). All regions included in the regression model exhibit a minimum number of births and migrants such that measures of specialization and relatedness are defined (see SM Section 2.2). Standard errors are clustered by region and period. The full regression tables with all control variables are provided in SM Sections 3.2 and 3.3. * p < 0.1, ** p < 0.05, *** p < 0.01

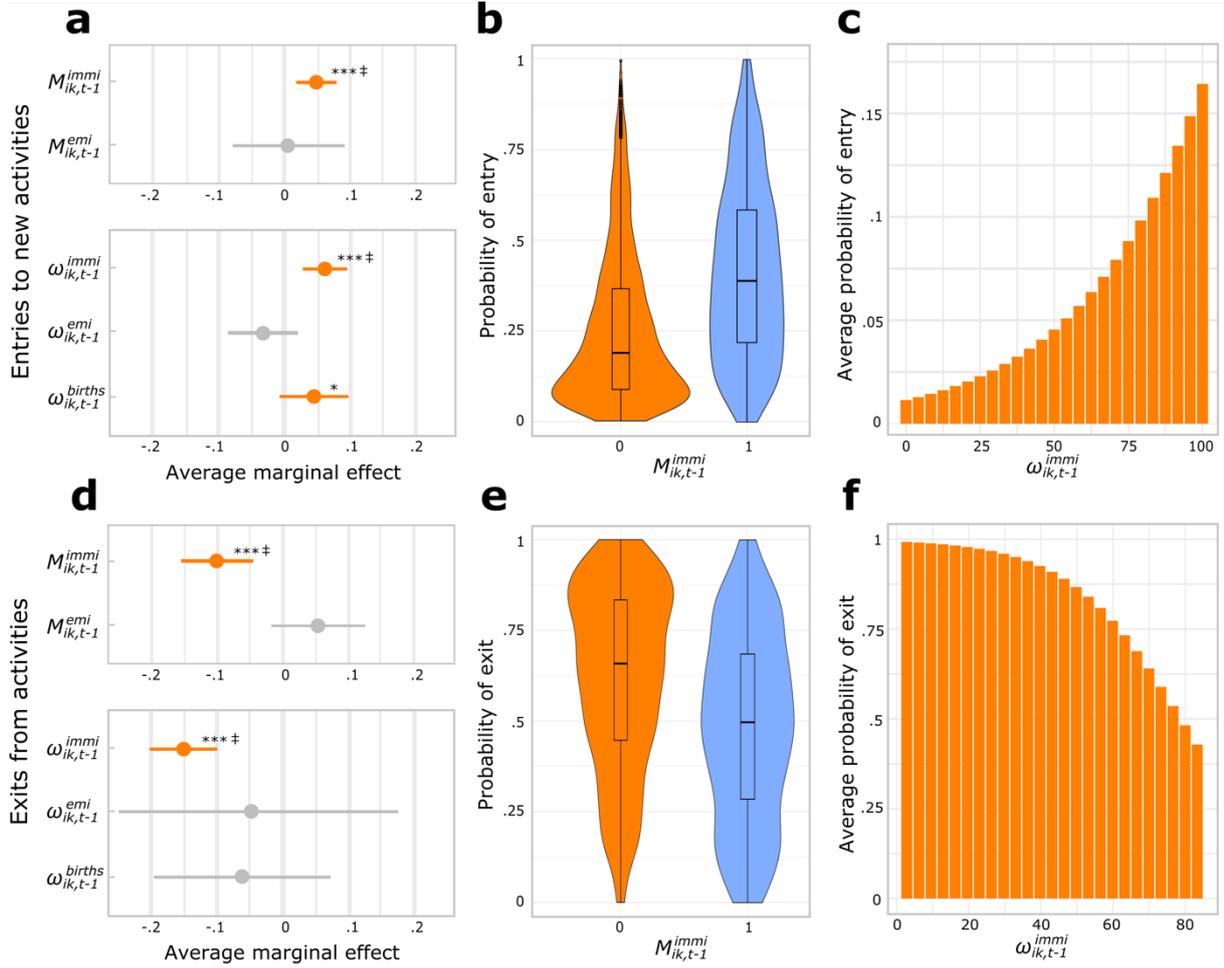

*Figure 3*. ***Visualization of main results.*** *(**a**) Average marginal effects on the probability of entry to new activities based on the logistic regression model in Table 1, column 5. $M_{ik,t-1}^{immi} = 1$ increases the probability of entry on average by 4.6 percentage points, while a standard-deviation-increase of $\omega_{ik,t-1}^{immi}$ correlates with an increase in the average probability of entry of 5.8 percentage points. (**b-c**) Probability of entry to a new activity as a function of **b** $M_{ik,t-1}^{immi}$ and **c** the immigrants' related knowledge, $\omega_{ik,t-1}^{immi}$. (**d**) Average marginal effects on the probability of exit from activities based on the logistic regression model in Table 1, column 10. $M_{ik,t-1}^{immi} = 1$ reduces the probability of exit on average by 10.2 percentage points, while a standard-deviation-increase of $\omega_{ik,t-1}^{immi}$ correlates with a reduction in the average probability of exit of 15.1 percentage points. (**e-f**) Probability of exit from an existing area of specialization as a function of **e** $M_{ik,t-1}^{immi}$ and **f** the immigrants' related knowledge, $\omega_{ik,t-1}^{immi}$. Notes: \* p < 0.1, \*\* p < 0.05, \*\*\* p < 0.01; Average marginal effects are computed by taking the average of the marginal effects across observations; Lines indicate 95% confidence interval; ‡ denotes robustness of the results (SM Section 3.4)*

These results are robust to estimating the expected number of immigrants, emigrants, and locals in Eq. 1 and 6 using the negative binomial regression model described in Eq. 3 (see SM Section 3.4.1). By accounting for local factors, we are able to obtain a more accurate estimate of the expected number of immigrants, emigrants, and locals and, thus, of a disproportionate migration flow. This mitigates some of the endogeneity concerns.

The highly restrictive fixed effects specification, however, reduces the number of observations in the regression model. To assure the robustness of our results, we estimate the logistic regression models with several less restrictive specifications. This also allows us to include observed variables previously captured by the fixed effects, such as urban population[88,89] or a location's diversity of activities. We find that the knowledge of immigrants remains a significant predictor for both entries and exits (SM Section 3.4.2). We acknowledge that, over such long periods, travel times are not constant but decrease with improvements in infrastructure and/or technology. Hence, we allow for century-specific effects of spatial proximity, $\rho_{ik,t-1}^{M}$ and $\rho_{ik,t-1}^{\omega}$, leaving our results unchanged (SM Section 3.4.3). Also, our sample of famous individuals is not balanced over time. Our findings, however, are robust to excluding the 20$^{th}$ century from the analysis as well as looking at the 20$^{th}$ century alone (SM Section 3.4.4). Moreover, our results do not change if we redefine entries and exits as the first or last birth of a famous individual with a specific occupation in a location instead of developing or losing specialization in an activity (SM Section 3.4.5). In addition, we explore the explanatory power of interaction terms between various relatedness densities on entries, following the literature on migrants as agents of structural change[6–8]. We find a significant, but quantitatively negligible negative interaction term between $\omega_{ik}^{immi}$ and $\omega_{ik}^{births}$, indicating that the related knowledge of immigrants and locals are weak substitutes (SM Section 3.4.6). Also, it may be that our findings of knowledge spillovers are different for different activities. We explore potential heterogenous effects by estimating our regression model separately for aggregate occupational categories (SM Section 3.4.7). We find, for instance, a stronger correlation of the presence of immigrants specialized in the same activity ($M_{ik,t-1}^{immi}$) on entries in sciences and public institutions, and an increased correlation of related knowledge of immigrants ($\omega_{ik}^{immi}$) in humanities and sports. Another source of heterogeneity can be city size when size plays a relevant role in generating knowledge spillovers (SM Section 3.4.8). We find that most entries take place in large cities, and thus, the effects of migration and relatedness are mainly urban. Whereas for exits, we see that spillovers across activities are more important in larger cities. But the probability of exiting an activity in small cities grows massively with the emigration of individuals specialized in the same activity, pointing towards a pronounced role of talent loss in shaping regional specialisations of small agglomerations. This finding relates to the recent literature on left-behind places[90,91].

Lastly, although the ratio of observed above expected numbers (Eq. 1 and 6) fulfils the purpose of controlling for size and reverse causality, these models are opaque, not telling us whether our results are driven by changes in the observed or expected number (or both). Hence, we run our main regression model including all terms of the ratio as a robustness check (SM Section 3.4.9). We find that the observed number of immigrants with a specific occupation ($N_{ik,t-1}^{immi}$) correlates positively with future entries and negatively with future exits, confirming our main results with composite indices. One additional immigrant to a region with a specific occupation correlates with an average increase in the probability of entry by 1.68 percentage points and a reduction in the probability of exit by 5.04 percentage points (SM Section 3.4.9).

## 4. Discussion

Labor mobility and migration are core tenets of the United States and the European Union, because policymakers intuit that migrants carry knowledge across space and activities[1–5]. Yet, despite multiple studies documenting the role of migrants in the diffusion of knowledge[6,11–33], there is little historical quantitative evidence of the role of migrants in the historical evolution of knowledge agglomerations.

Here, we used biographic data on more than 22,000 famous individuals—sculptors, composers, politicians, chemists, etc.—living in Europe between the years 1000 and 2000 to explore how the knowledge of immigrants, emigrants, and locals explains the probability that a region enters or exits an activity.

Our findings show that migrants play a crucial role in the historical geography of knowledge. Specifically, we find that the probability that a European region enters a new activity grows with the presence of immigrants with knowledge on that activity. Also, using measures of relatedness[34–38], we find that this correlation is enhanced by spillovers across related activities. Put differently, the probability that a region begets famous mathematicians grows with an excess immigration of mathematicians and with immigrants from related fields, such as physics or chemistry. Similarly, we find that the probability that a European region loses one of its existing areas of specialization decreases with the presence of immigrants specialized in that activity and in related activities. However, we do not find that locals with related knowledge play the same statistically significant and robust role in entries or exits.

These findings advance our understanding of the evolution of European agglomerations over the past millennium and of the role of migrants and locals therein. Specifically, we find robust evidence that European agglomerations did not only evolve path-dependently[92], but also that they benefited from spillovers generated by the migration of famous individuals. This supports the literature on the role of migrants in the diffusion of knowledge[1–8,11–18,20–28] and contributes to the literature on relatedness[34–38] explaining changes in specialization patterns[39–54].

Migrants are known agents of structural change enabling the development of unrelated activities[6–8]. Our findings differ slightly from that by emphasizing migration as a channel of related diversification and path-dependent development, adding to the literature unpacking the principle of relatedness[55–62]. Recently, this intersection between evolutionary economic geography, regional diversification, and migration has been identified as a promising field of research[10]. We contribute methodologically to this literature by disentangling relatedness measures for immigrants, emigrants, and locals. These novel measures make it possible to explore how knowledge spillovers across space and across activities combine (SM Section 2.5). Lastly, this study provides a long-term perspective on the evolution of regional specialisations in Europe, a perspective which has been underrepresented in economic geography[63].

Unfortunately, we do not observe the mechanisms explaining the entry or exit of regions in activities. There are, however, several potential mechanisms responsible for these results, which can be subsumed as horizontal and vertical socialization[93]. For instance, immigrating physicists could teach at a university, leading to a local flourishing of the field of physics and increasing the probability that a famous physicist emerges in the future. Also, immigrating physicists may bring new ideas and approaches with them, which can stimulate creative thinking and cross-pollination of ideas among local scientists in related fields such as chemistry or mathematics. This could lead to the development of new methods as well as new ways of thinking about problems, which could in turn contribute to an increased probability of giving birth to famous chemists or mathematicians in the future. The mechanisms may be different in other activities such as the arts or humanities. The presence of immigrating musicians may create a critical mass of artists, making it profitable to build cultural infrastructure due to economies of scale[26], from which artists in related activities such as singers, composers or dancers benefit as well. Shedding light on these different mechanisms is a promising avenue for future research.

Our study has also other limitations. First, we observe only a small and highly mobile subset of the overall population. That is, 22,000 of the most famous individuals living in Europe over the past 1,000 years. A more comprehensive dataset would allow for a more accurate and granular estimation of a location's related knowledge and the geography of activities. Indeed, we suspect that the limited sample is a likely reason for why we do not observe a statistically significant and robust relevance for locals in shaping the historical geography of knowledge. That being said, the related knowledge of locals plays a significant role in several specifications, for instance if estimating the expected number of famous individuals to define specialisations (SM Section 3.4.1) or for large cities (SM Section 3.4.8). Continuing to investigate the role of locals in the historical geography of knowledge can be an interesting avenue for future research.

Second, we do not observe the full migration trajectory of individuals, but only their place of birth and place of death. Although this approach follows the literature[65–68] and provides a good proxy of migration (SM Section 1.3), more detailed data on where famous individuals lived and when could provide a better analytical basis to explore the evolution of agglomerations[94,95]. Indeed, based on a small number of famous individuals living between 1450 and 1750, it is estimated that they moved on average 3.72 times during their lifetime[96]. Third, we focus only on Europe. So, it may be that the principles behind the historical geography of knowledge uncovered here are different for other parts of the world. Lastly, migration is influenced by multiple factors such as geography and culture[75,76], agglomeration[26], patrons[73,74] or conflict[28,78], evoking reverse causality and endogeneity concerns in our study. We tackled these concerns by using highly restrictive fixed-effects and estimating the expected number of immigrants, emigrants, and locals to define specializations. Despite these efforts, we want to stress that we are not able to make strictly causal claims, a task that can be challenging using historical observational data.

Yet, despite these limitations, our study provides evidence of migration playing a central role in the evolution of European knowledge agglomerations. Also, while being a historical study, our study concerns a topic that is highly relevant in today's economic policy. The effects of migration on local economies have been debated intensively, both in academia[9,97–99] and in policy circles[100,101]. Our findings add to this debate by showing that the immigration of high-skilled individuals correlates with entering and exiting specialisations of regions. Yet, our results can neither be interpreted causally nor tell us whether these findings remain for migration that is incentivized by policy instruments, since we observe migration involving multiple forces, from forced displacement due to war, to organic forms of migration.

**Acknowledgements**

We thank two anonymous referees, Andrea Morrison, Eva Coll, Jesús Crespo Cuaresma, Ron Boschma, Andrea Belmartino, the attendees of the 2022 Economic Geography PhD school in Utrecht, the attendees of the WICK#10 PhD Workshop in Economics of Innovation, Complexity and Knowledge in Turin, the attendees of the 13[th] Geoffrey J.D. Hewings Regional Economics Workshop in Vienna, members of the Complexity Science Hub and EcoAustria in Vienna, and the members of the Center for Collective Learning for valuable feedback.

**Funding**

This project was supported by the Agence Nationale de la Recherche [grant number ANR-19-P3IA-0004], the 101086712-LearnData-HORIZON-WIDERA-2022-TALENTS-01 financed by the European Research Executive Agency (REA), and the European Lighthouse of AI for Sustainability [grant number 101120237-HORIZON-CL4-2022-HUMAN-02].

**Disclosure statement**

The authors declare no competing interests.

**Data and materials availability**

All data are available in the main text or the supplementary materials.

# Supplementary Materials for:
# The Role of Immigrants, Emigrants, and Locals in the Historical Formation of European Knowledge Agglomerations

Philipp Koch[1,2], Viktor Stojkoski[1,3], César A. Hidalgo[1,4,5]

[1] Center for Collective Learning, ANITI, TSE-R, IAST, IRIT, Université de Toulouse, 31000 Toulouse, France.
[2] EcoAustria – Institute for Economic Research, 1030 Vienna, Austria.
[3] Faculty of Economics, University Ss. Cyril and Methodius in Skopje, North Macedonia.
[3] Alliance Manchester Business School, University of Manchester, United Kingdom.
[4] School of Engineering and Applied Sciences, Harvard University, United States.

## Content

# 1. Data

## 1.1. Pantheon

The main data source for our analysis is the 2020 version of the Pantheon dataset (Yu et al., 2016), which is publicly available at pantheon.world. It contains information on more than 88,000 famous individuals with more than 15 language editions in Wikipedia worldwide. We restrict our sample to the years 1000 to 2000, since the number of observations in the overall dataset increases after the year 1000 and becomes less volatile (see Figure S1). Also, as described in the main manuscript, we focus on continental Europe and, thus, only include individuals who are born or have died in Europe. The reasoning behind this restriction is the need to have an as comprehensive picture of the structure of famous individuals in a region as possible. Due to an arguable Western bias in Wikipedia and the selected time horizon, we restrict our sample to Europe. Overall, this reduces our sample to 22,847 individuals.

We follow the occupation taxonomy by Yu et al. (2016), which differentiates in total between 101 occupations of 27 categories and 8 broad categories. Table S1 describes the taxonomy and displays the number of famous individuals born or died in Europe between 1000 and 2000 with the respective occupation. Politicians (5,233), writers (2,817) and painters (1,126) are the most common occupations of famous figures in the past millennium.

Occupations are assigned to individuals based on the occupation that made them famous. For instance, Marie Curie is considered a physicist in our dataset, since she won the Nobel Prize in physics prior to her Nobel Prize in chemistry. Angela Merkel is considered a politician, despite her academic career in chemistry. A more detailed description of this approach is given in Yu et al. (2016).

A consistent occupation classification is a key element for our analysis, since we want to describe the geography of knowledge based on this classification. In fact, more comprehensive data sources for notable people would be available, such as Freebase from Google or a very recently published database (Laouenan et al., 2022), which contains information on 2.29 million notable individuals. Unfortunately, these data sources are not sufficiently consistent with respect to their occupation classification. For example, the database by Laouenan et al. (2022) distinguishes between almost 5,000 occupations, but these are not unique. Nonetheless, these data sources are very promising avenues for future research, potentially enabling the analysis of the historical geography of knowledge based on notable figures beyond Europe.

The number of observations in or dataset increases with time (Figure S1). While we have data on 284 individuals born in the 11$^{th}$ century, this number increases to 7,483 in the 20$^{th}$ century (see Table S2). Due to this imbalance, we perform robustness checks for period subsamples in Section 3.4.3.

To obtain a comprehensive picture of the knowledge of individuals living in a location given the unbalanced sample (Table S2), the period of observation in our study is centuries. Specifically, we assign individuals to centuries based on their year of birth. A famous individual is, for instance, assigned to the 17$^{th}$ century if he or she is born between 1600 and 1699. Splitting the sample into smaller time periods such as decades or half-centuries would prohibit us from estimating measures of specialization or relatedness in all periods due to small numbers of observation.

We do not observe the full migration trajectory of individuals, which is why we use places of birth and death as a proxy for migration (see Section 1.3). Interestingly, migration is very common among notable individuals in the dataset. 75.1 percent of individuals in the dataset die in a different region they are born in. Also, migration among famous individuals has become more prevalent over time. While in the 11$^{th}$ century 31.3 percent of individuals died in a different region than they were born in, this is only the case for 20.8 percent in the 19$^{th}$ century (see Table S2 and Figure 1e in the main text).

Individuals are assigned to centuries based solely on their year of birth. That is, a famous person who is born in the 18$^{th}$ century in Brussels and has later died in Paris is considered a local in Brussels and an immigrant in Paris in the 18$^{th}$ century. Even if the individual dies in the 19$^{th}$ century, the person is assigned an immigrant in the 18$^{th}$ century. We choose this approach, since we do not have information on the time of migration. It may be that the considered individual moved to Paris in the later stages of his or her life, but it may also be the case that the migration took place as a child. We do not believe that this definition is problematic in our analysis, given the lag structure in our regression model and the length of the chosen period of observation, i.e. centuries.

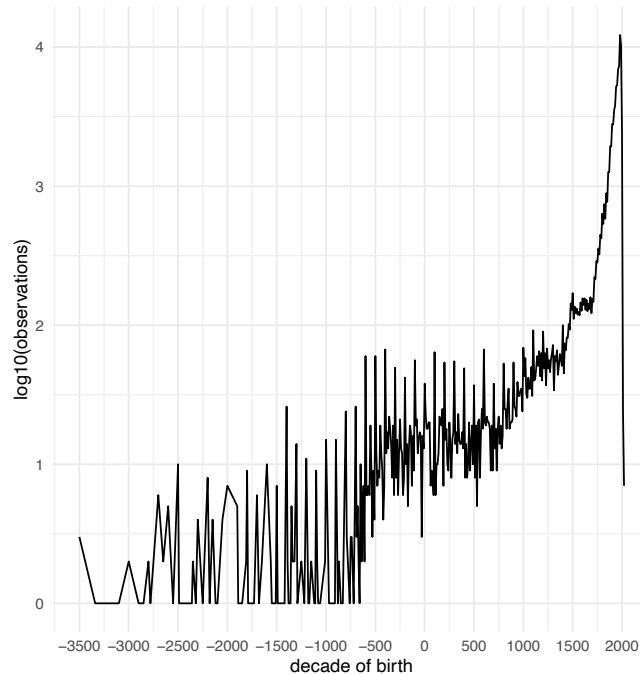

*Figure S1.* Number of observations in the overall Pantheon dataset (Yu et al., 2016) by decade of birth

*Table S1.* Occupation taxonomy following Yu et al. (2016) and number of individuals born and/or deceased in Europe between the years 1000 and 2000

| Broad Category | Category | Occupation | Obs. abs. | Obs. in % |
|---|---|---|---|---|
| Arts | Dance | DANCER | 46 | 0,2% |
| | Design | ARCHITECT | 300 | 1,3% |
| | | DESIGNER | 42 | 0,2% |
| | | COMIC ARTIST | 25 | 0,1% |
| | | FASHION DESIGNER | 14 | 0,1% |
| | Film and Theatre | ACTOR | 969 | 4,2% |
| | | FILM DIRECTOR | 421 | 1,8% |
| | Fine Arts | PAINTER | 1126 | 4,9% |
| | | SCULPTOR | 122 | 0,5% |
| | | PHOTOGRAPHER | 47 | 0,2% |
| | | ARTIST | 33 | 0,1% |
| | Music | COMPOSER | 889 | 3,9% |
| | | MUSICIAN | 390 | 1,7% |
| | | SINGER | 352 | 1,5% |
| | | CONDUCTOR | 64 | 0,3% |
| Business & Law | Business | BUSINESSPERSON | 158 | 0,7% |
| | | PRODUCER | 13 | 0,1% |
| | Law | LAWYER | 24 | 0,1% |
| Exploration | Explorers | EXPLORER | 295 | 1,3% |
| | | ASTRONAUT | 41 | 0,2% |
| Humanities | History | HISTORIAN | 184 | 0,8% |
| | Language | WRITER | 2817 | 12,4% |
| | | LINGUIST | 117 | 0,5% |
| | | JOURNALIST | 48 | 0,2% |
| | Philosophy | PHILOSOPHER | 563 | 2,5% |
| Institutions | Government | POLITICIAN | 5233 | 22,9% |
| | | NOBLEMAN | 551 | 2,4% |
| | | DIPLOMAT | 24 | 0,1% |
| | | PUBLIC WORKER | 8 | 0,0% |
| | Military | MILITARY PERSONNEL | 917 | 4,0% |
| | | PILOT | 33 | 0,1% |
| | Religion | RELIGIOUS FIGURE | 811 | 3,6% |
| Public Figure | Activism | SOCIAL ACTIVIST | 197 | 0,9% |
| | Media Personality | CELEBRITY | 31 | 0,1% |
| | | PRESENTER | 8 | 0,0% |
| | | MODEL | 8 | 0,0% |
| | Outlaws | EXTREMIST | 69 | 0,3% |
| | | OCCULTIST | 31 | 0,1% |
| | | PIRATE | 20 | 0,1% |
| | | MAFIOSO | 12 | 0,1% |
| Science & Technology | Computer Science | COMPUTER SCIENTIST | 21 | 0,1% |
| | Engineering | ENGINEER | 208 | 0,9% |
| | Invention | INVENTOR | 205 | 0,9% |
| | Math | MATHEMATICIAN | 549 | 2,4% |
| | | STATISTICIAN | 6 | 0,0% |
| | Medicine | PHYSICIAN | 331 | 1,5% |
| | Natural Sciences | BIOLOGIST | 600 | 2,6% |
| | | PHYSICIST | 418 | 1,8% |
| | | CHEMIST | 330 | 1,4% |
| | | ASTRONOMER | 285 | 1,2% |
| | | ARCHAEOLOGIST | 79 | 0,3% |
| | | GEOLOGIST | 46 | 0,2% |
| | Social Sciences | ECONOMIST | 128 | 0,6% |
| | | PSYCHOLOGIST | 98 | 0,4% |
| | | GEOGRAPHER | 44 | 0,2% |
| | | ANTHROPOLOGIST | 36 | 0,2% |
| | | SOCIOLOGIST | 26 | 0,1% |
| | | POLITICAL SCIENTIST | 10 | 0,0% |
| Sports | Individual Sports | ATHLETE | 389 | 1,7% |
| | | RACING DRIVER | 260 | 1,1% |
| | | CYCLIST | 136 | 0,6% |
| | | CHESS PLAYER | 123 | 0,5% |
| | | TENNIS PLAYER | 79 | 0,3% |
| | | SKATER | 47 | 0,2% |
| | | WRESTLER | 43 | 0,2% |
| | | FENCER | 42 | 0,2% |
| | | BOXER | 39 | 0,2% |
| | | GYMNAST | 37 | 0,2% |
| | | SKIER | 37 | 0,2% |
| | | SWIMMER | 25 | 0,1% |
| | | MOUNTAINEER | 23 | 0,1% |
| | | TABLE TENNIS PLAYER | 6 | 0,0% |
| | Team Sports | SOCCER PLAYER | 963 | 4,2% |
| | | COACH | 32 | 0,1% |
| | | HOCKEY PLAYER | 22 | 0,1% |
| | | BASKETBALL PLAYER | 18 | 0,1% |
| | | HANDBALL PLAYER | 13 | 0,1% |

*Table S2. Number of famous individuals and share of locals by century*

| Period | No. of observations | Share of individuals that died in the same region they were born in |
|---|---|---|
| 11th century | 284 | 31.3% |
| 12th century | 339 | 27.7% |
| 13th century | 377 | 32.1% |
| 14th century | 422 | 34.6% |
| 15th century | 784 | 31.6% |
| 16th century | 1,109 | 29.9% |
| 17th century | 1,231 | 34.6% |
| 18th century | 2,516 | 27.9% |
| 19th century | 8,301 | 20.8% |
| 20th century | 7,483 | 24.0% |

## 1.2. Administrative regions

For aggregation purposes, we assign individuals to regions based on their geocoded places of birth and death. We use NUTS-2 regions for countries in the European Union and the European Free Trade Association. For other countries in Europe, we use administrative regions of comparable size. Specifically, these are oblasts for Russia, Ukraine and Belarus, federal entities for Bosnia and Herzegovina, and the whole country for Kosovo and Moldova (see Figure S2). Shape files are publicly available online for NUTS regions (see e.g. ec.europa.eu) as well as for administrative regions of other countries (see e.g. gadm.org).

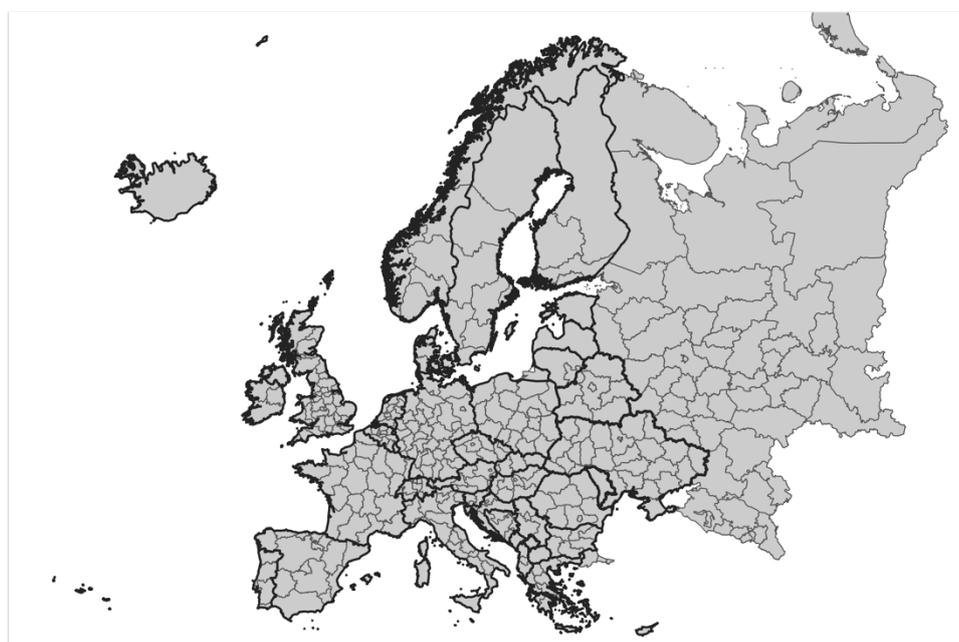

*Figure S2. Administrative regions applied in the analysis. Bold lines mark country borders.*

## 1.3. Using places of birth and death as proxy for migration

We use places of birth and death as a proxy for migration, following the literature using similar data to describe migration movements (Laouenan et al., 2022; Schich et al., 2014; Serafinelli & Tabellini, 2022). Figures 1c and 1d in the main manuscript show the network of migration based on this proxy.

We check whether this proxy is, in fact, meaningful by randomly drawing ~200 individuals from the dataset, paying attention to representativeness across periods. We read the Wikipedia article for each famous individual to determine whether a relation to the place of death exists, which would qualify as migration. We differentiate between (1) having any relation to the place of death (i.e. living there for a considerable amount of time, having noteworthy social connections with multiple visits there, or, in case of politicians and noblemen, reigning over the region) and (2) having a major relation to the place of death. The latter is the case if the place was one of the individual's main places of living, if the famous individual taught at a university there etc.

We find that in 181 out of 202 cases ($\hat{p} = 0.896$, 95% CI: [0.854, 0.938]), the famous individual had a relation to his or her place of death. Hence, only in 10% of observations the place of death is arbitrary. Also, we find that in 151 out of 202 cases ($\hat{p} = 0.748$, 95% CI: [0.688, 0.807]), the famous individual had a major relation to his or her place of death. These results indicate that using place of birth and death as a proxy for migration is a valid approach. The sampled data is available upon request.

It is important to note that we do not claim that the place of death is the only and most relevant place of impact. Famous individuals, who seem to be highly mobile (see Table S2), are likely to stay at multiple cities during their lifetime. These stays, however, are not random. Famous individuals tend to spend time at places, where several individuals with the same specialization are already staying. This is a result of the negative binomial regression used to estimate the expected number of migrants and locals in Table S7.

Due to the observation that migration follows previous specialization patterns, we argue that if the estimates we find for the role of immigrants were affected by using place of birth and death as a proxy for migration and not observing the full migration trajectory, they would tend to be downward biased rather than upward biased.

### 1.4. Population

We augment our analysis with publicly available population data on more than 2,000 European cities going back to 700 AD (Bairoch et al., 1988; Buringh, 2021). We use the coordinates provided in the dataset to assign cities to regions. This enables us to aggregate the population data by region.

## 2. Methods

### 2.1. Information entropy

Due to migration flows towards cities and the tendency of agglomeration, places of death are spatially more concentrated than places of birth. Figure S3 shows the number of births and deaths of famous individuals for the ten most populated regions in the 19[th] century. For example, 416 famous individuals were born in Paris in the 19[th] century, but 934 died there.

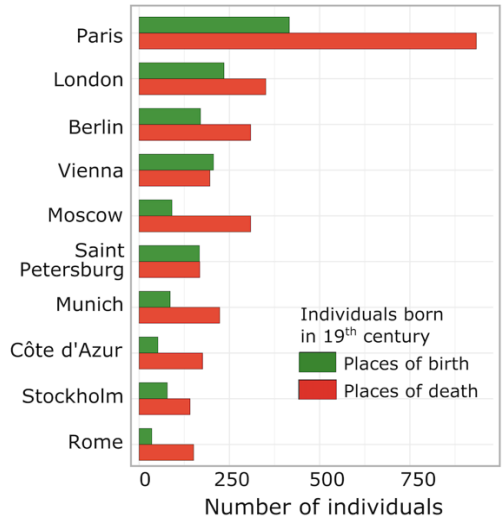

*Figure S3. Number of births and deaths of famous individuals born in the 19[th] century in the ten most common regions of death.*

We quantify the amount of spatial concentration by calculating the effective number of places of birth and places of death using information entropy.

Let $N_{i,t}$ denote the number of famous individuals in region *i* and century *t*, then entropy *H* is given by

$$H_t = -\sum_{i=1}^{I} \frac{N_{i,t}}{\sum_i N_{i,t}} log_2 \left( \frac{N_{i,t}}{\sum_i N_{i,t}} \right) \qquad (S1)$$

Intuitively, $H_t$ is the average number of minimum yes/no questions one has to ask to guess a famous individual's region of birth or death.

The effective number of places is then given by $2^{H_t}$. This measure gives the number of places as if they were equally common. The higher the effective number of places, the lower the spatial concentration.

Figure 1f in the main manuscript plots the effective number of places of birth and places of death per century.

## 2.2. Adjacency matrices

To calculate the relatedness density, we transform our dataset into binary specialization matrices per century. We define that a location is specialized in an occupation if it exhibits a larger number of famous biographies in the respective occupation than expected. As described in the main manuscript, we employ two different approaches to the expected number of biographies. The first one is a naïve "bins and balls" model and identical to the Revealed Comparative Advantage or Location Quotient. The second approach consists of estimating the expected number of immigrants, emigrants and locals in a negative binomial regression model, taking local factors into account.

Then, we create specialization matrices based on immigrants (born somewhere else, but died here), emigration (born here, but died somewhere else) and locals (born here). We define the matrix $M_{ik}^j$ for $j = \{immi, emi, births\}$ as

$$M_{ik}^j = \begin{cases} 1 & if \ \frac{N_{ik,t}^j}{\widehat{N}_{ik,t}^j} \geq 1 \\ 0 & otherwise \end{cases} \quad (S2)$$

Prior to the calculation of the matrix $M_{ik}^j$ we remove regions and occupations with very few observations, since they can distort the specialization matrix. Specifically, we remove regions and occupations with not more than 5 famous individuals in a century, i.e. $\sum_k N_{ik}^j \leq 5$ and $\sum_i N_{ik}^j \leq 5$, respectively. For the 11th to 15th century, we employ a less restrictive cutoff, i.e. $\sum_k N_{ik}^j \leq 3$ and $\sum_i N_{ik}^j \leq 3$, due to fewer observations. Additionally, we remove individuals with the occupation "companion".

Sorting these specialization matrices by diversity and ubiquity reveals their nested structure (see Fig. 2a-c in the main manuscript).

## 2.3. The related knowledge of locals

We define locals as famous individuals who were born in a region, no matter if they died there or elsewhere. We use this definition because of the large share of migrants among famous individuals (see Table S2), which would reduce our number of observations drastically if we defined locals as individuals who were born and died in the same place. Here, we show that the relatedness density based on all famous individuals born is a valid proxy for the related knowledge of individuals that have been born and died in the same region, after controlling for the related knowledge of emigrants.

To show this, we create a measure of relatedness for locals (born and died here) in analogy to the other relatedness measures based on the naïve model of the expected number:

$$M_{ik}^{locals} = \begin{cases} 1 & if \ \frac{N_{ik}^{locals}/\sum_k N_{ik}^{locals}}{\sum_i N_{ik}^{locals}/\sum_{i,k} N_{ik}^{locals}} \geq 1 \\ 0 & otherwise \end{cases}$$

$$\varphi_{kk\prime,t}^{locals} = \frac{\sum_i M_{ik,t}^{locals} M_{ik\prime,t}^{locals}}{max\left(\sum_i M_{ik,t}^{locals}, \sum_i M_{ik\prime,t}^{locals}\right)} \tag{S3}$$

$$\omega_{ik,t}^{locals} = \frac{\sum_{k\prime} M_{ik\prime,t}^{locals} \varphi_{kk\prime,t}^{locals}}{\sum_{k\prime} \varphi_{kk\prime,t}^{locals}}$$

Then, we estimate the following linear regression:

$$\omega_{ik,t}^{locals} = \alpha_1 \omega_{ik,t}^{births} + \alpha_2 \omega_{ik,t}^{emi} + \gamma_i + \delta_t + \varepsilon_{ik} \tag{S4}$$

Figure S4 shows the correlation between the fitted values based on the regression and the real values of $\omega_{ik,t}^{locals}$. The correlation between the real and fitted values is high ($R^2$=0.65), indicating that $\omega_{ik,t}^{births}$ controlling for $\omega_{ik,t}^{emi}$ is a valid proxy for $\omega_{ik,t}^{locals}$. This result is also robust for including all covariates of the logistic regression models described in Section 3.1.

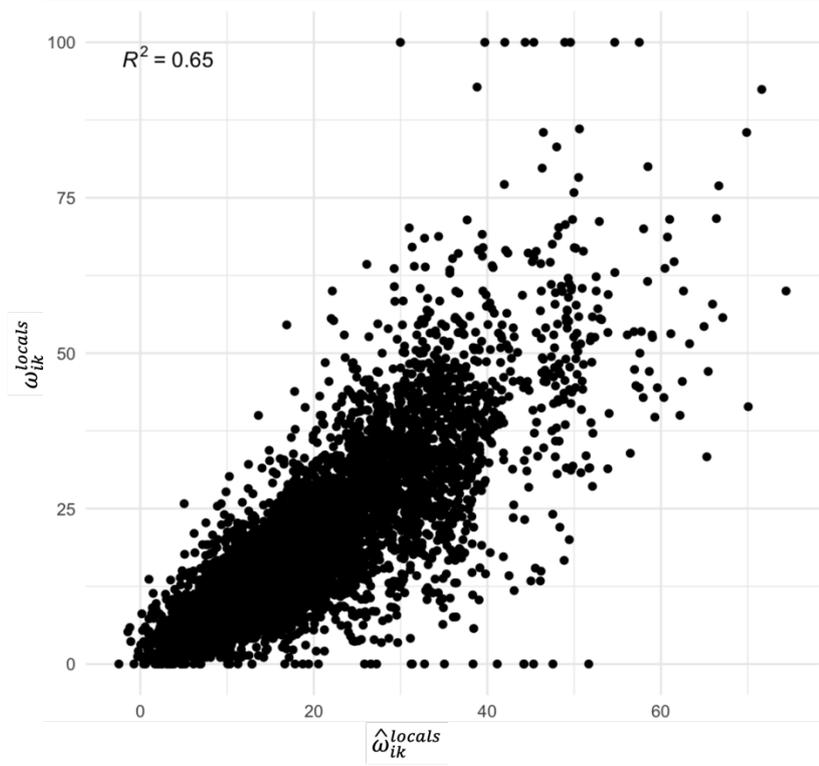

*Figure S4. Correlation between fitted and real values for $\omega_{ik,t}^{locals}$*

## 2.4. Spatial lags

To control for other means of knowledge diffusion across space than migration, we create spatial lags. Specifically, we differentiate between the availability of knowledge in the same activity (do geographically proximate regions have a specialization in that activity?), and related activities (do geographically proximate regions have specializations in related activities?).

To do so, we measure the distance between all the administrative regions depicted in Figure S2 (technically, their centroids) and transform them into a proximity matrix $W_{i'i}$. Let $d_{i'i}$ denote the distance between regions $i'$ and $i$. Then,

$$W_{i'i} = \begin{cases} 1/d_{i'i} & if\ i' \neq i \\ 0 & otherwise \end{cases} \tag{S5}$$

We then define the spatial lag with respect to specializations in the same activity as the region's average proximity to regions with a specialization in that activity:

$$\rho_{ik,t}^M = \frac{\sum_i W_{i'i} M_{ik,t}^{births}}{\sum_i W_{i'i}} \tag{S6}$$

Similarly, we define the spatial lag with respect to relatedness:

$$\rho_{ik,t}^{\omega} = \frac{\sum_i W_{i'i} \omega_{ik,t}^{births}}{\sum_i W_{i'i}} \tag{S7}$$

### 2.5. Elaboration on proximity measures

Measures of proximity capture the combined presence of multiple factors that may be contributing to the colocation of two activities. We create separate measures of proximity for immigrants, emigrants and locals (Eq. 8 in the main manuscript), since the factors driving the colocation of activities may be different for immigrants, emigrants and locals.

In this chapter, we explore the differences between these proximity measures and argue that creating separate measures provides more nuance in quantifying knowledge spillovers than using a joint proximity measure.

We explore the differences in proximity measures based on the co-location of immigrants, emigrants and locals by taking the average across time (e.g. $\bar{\varphi}_{kk'}^{births} = \frac{1}{T}\sum_t \varphi_{kk',t}^{births}$). Figure S5 shows the correlation between $\bar{\varphi}_{kk'}^{immi}$, $\bar{\varphi}_{kk'}^{emi}$ and $\bar{\varphi}_{kk'}^{births}$. All proximity measures are significantly correlated with each other. For instance, $\bar{\varphi}_{kk'}^{emi}$ and $\bar{\varphi}_{kk'}^{births}$ are highly correlated with an R-squared of 0.79 (Figure S5 B).

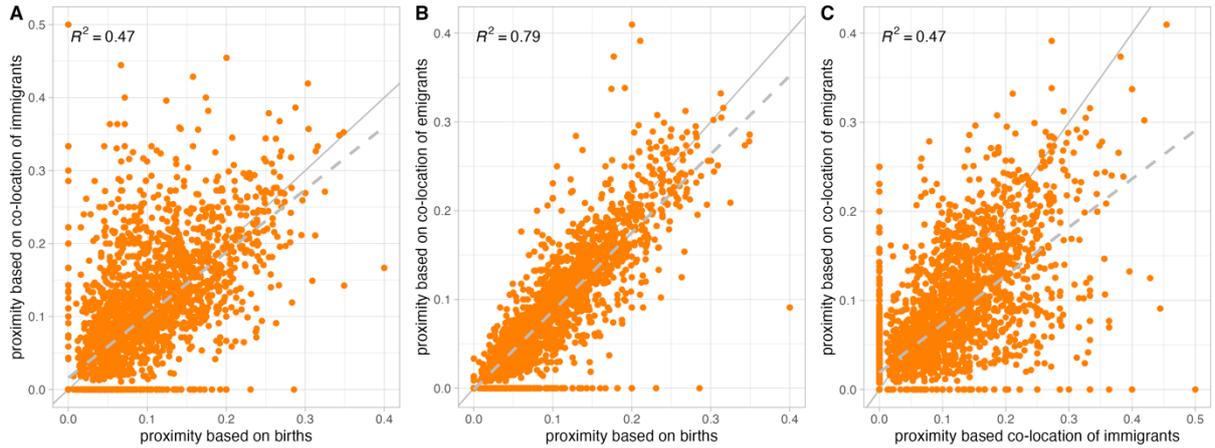

*Figure S5. Correlation between the three different proximity measures*

To explore this in more detail, we take a closer look at the correlation between $\bar{\varphi}_{kk'}^{immi}$ and $\bar{\varphi}_{kk'}^{births}$. For the purposes of visualization, we restrict the sample to combinations of activities that existed in at least five centuries (Figure S6).

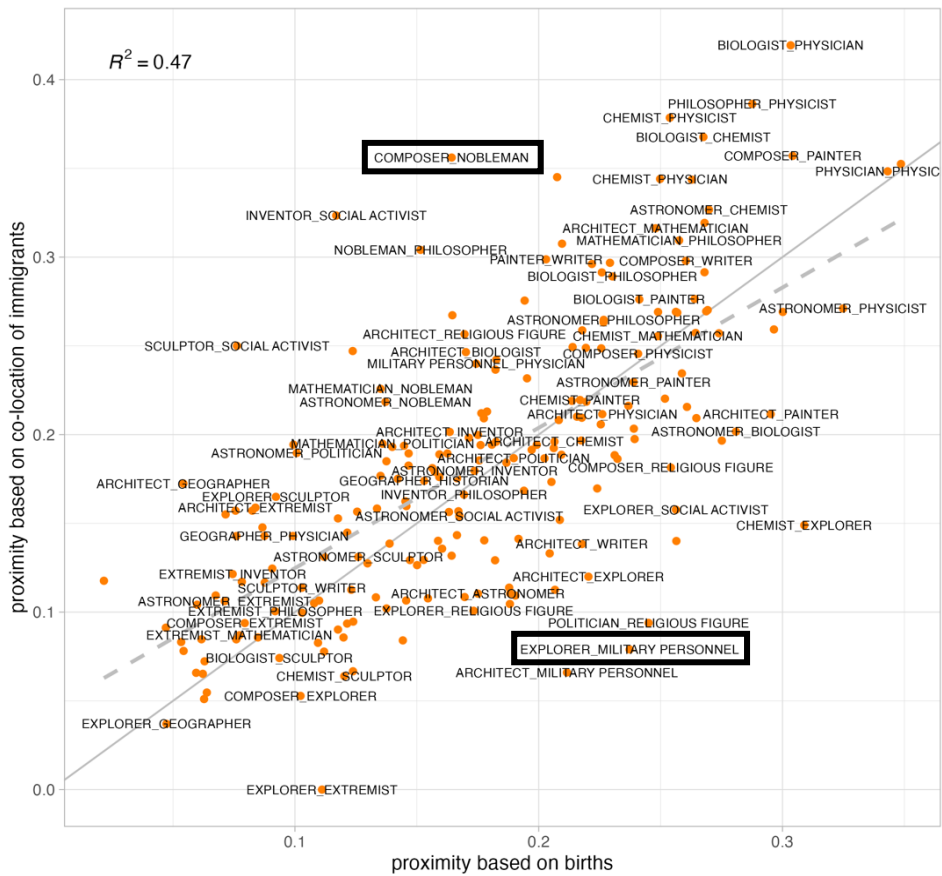

*Figure S6. Correlation between the proximity based on births and the co-location of immigrants*

Two relevant examples (marked in the figure) emerge from this comparison.

Consider explorers and military personnel. These are highly related activities if looking at places of birth, but are distant for places of immigration. Indeed, explorers and military personnel share many required capabilities such as navigating, planning, commanding etc. (the famous Portuguese explorer Duarte Pacheco Pereira is an impersonation of this proximity). That these activities frequently co-occur in births may be explained by the fact that the institutional environment (e.g. educational structures, location at sea, defensive needs) may promote the cultivation of both these talents. If many explorers are born in a location, it, hence, makes sense that this location is related to military personnel. In contrast, the factors contributing to the immigration of explorers and military personnel seem to be less similar.

Now consider composers and noblemen. For these two activities, the proximity based on immigration patterns is higher than for the colocation of births. It makes sense that these activities are to some extent related when looking at places of birth: Noblemen are known to be patrons for the arts. Hence, noblemen born in a location will likely create institutions that promote the cultivation of the talent of composers born in this location. But it is also plausible that these activities are even more related if looking at immigration patterns. Given that we observe a disproportional migration flow of noblemen towards a certain location, we can view this location as highly related to composers, since the institutional factors attracting noblemen likely play a role in attracting and cultivating the talent of composers as well.

These examples show that separate measures of proximity provide a nuanced perspective on the relationships between activities.

Another approach is to create one joint proximity measure that does not differentiate between immigrants, emigrants and locals. To assess the robustness of our results with respect to the chosen proximity measure, we create a joint proximity measure based on the adjacency matrix $R_{ik,t}^{joint}$:

$$R_{ik,t}^{joint} = \frac{N_{ik,t}^{births} + N_{ik,t}^{deaths}}{\widehat{N}_{ik,t}^{births} + \widehat{N}_{ik,t}^{deaths}}$$

with

$$\widehat{N}_{ik,t} = \frac{\sum_k N_{ik,t} \sum_i N_{ik,t}}{\sum_{i,k} N_{ik,t}}$$

After binarizing this adjacency matrix to create $M_{ik,t}^{joint}$ we create the proximity measure $\varphi_{kk',t}^{joint}$:

$$\varphi_{kk',t}^{joint} = \frac{\sum_i M_{ik,t}^{joint} M_{ik',t}^{joint}}{\max\left(\sum_i M_{ik,t}^{joint}, \sum_i M_{ik',t}^{joint}\right)}$$

Lastly, we use this joint proximity measure to create relatedness densities for immigrants, emigrants and locals:

$$\omega_{ik,t}^{immi} = \frac{\sum_{k'} M_{ik',t}^{immi} \varphi_{kk',t}^{joint}}{\sum_{k'} \varphi_{kk',t}^{joint}}$$

$$\omega_{ik,t}^{emi} = \frac{\sum_{k'} M_{ik',t}^{emi} \varphi_{kk',t}^{joint}}{\sum_{k'} \varphi_{kk',t}^{joint}}$$

$$\omega_{ik,t}^{births} = \frac{\sum_{k'} M_{ik',t}^{births} \varphi_{kk',t}^{joint}}{\sum_{k'} \varphi_{kk',t}^{joint}}$$

We use this relatedness measures in our main regression model (Eq. 11 in the main manuscript). The results (provided in the table below) are in-line with our previous findings: $\omega_{ik,t}^{immi}$ correlates positively with future entries and negatively with exits (significant at p<0.1), while the relatedness densities based on emigrants *($\omega_{ik,t}^{emi}$)* or locals *($\omega_{ik,t}^{births}$)* are insignificant. Also, the point estimates for $\omega_{ik,t}^{births}$ are lower than in our main results (Table S5 and Table S6), indicating that our finding of no robust effect for the related knowledge of locals is not driven by using separate proximity measures.

*Table S3. Regression results using a joint measure of proximity.*

| Dependent Variable: | $Entry_{ik,t}$ | $Exit_{ik,t}$ |
|---|---|---|
| | (1) | (2) |
| $M_{ik,t-1}^{immi}$ | 0.352*** | -0.897*** |
| | (0.036) | (0.103) |
| $M_{ik,t-1}^{emi}$ | 0.198 | -0.011 |
| | (0.297) | (0.216) |
| $\omega_{ik,t-1}^{immi}$ | 0.029* | -0.104* |
| | (0.016) | (0.063) |
| $\omega_{ik,t-1}^{emi}$ | 0.027 | -0.071 |
| | (0.030) | (0.086) |
| $\omega_{ik,t-1}^{births}$ | 0.018 | -0.010 |
| | (0.043) | (0.081) |
| $ubiquity_{k,t-1}$ | 0.008** | -0.051*** |
| | (0.003) | (0.012) |
| $\rho_{ik,t-1}^{M}$ | -0.425 | 5.904*** |
| | (0.459) | (1.342) |
| $\rho_{ik,t-1}^{\omega}$ | 0.065*** | 0.249 |
| | (0.018) | (0.173) |
| $R_{ik,t-1}$ | 0.286** | 0.003 |
| | (0.119) | (0.035) |
| FE: Broad categ.-region-period | Y | Y |
| FE: Category-period | Y | Y |
| Observations | 3944 | 1051 |
| Pseudo-$R^2$ | 0.217 | 0.238 |
| BIC | 9607.3 | 3662.8 |

Standard errors are clustered by region and period. * $p < 0.1$, ** $p < 0.05$, *** $p < 0.01$

## 3. Results

### 3.1. Logistic regression models explaining entries and exits and descriptive statistics

As described in the manuscript, we estimate logistic models to explain entries and exits in an activity using measures of the knowledge of immigrants and emigrants in that activity ($M_{ik,t}^{immi}$, $M_{ik,t}^{emi}$) and of the related knowledge that we can attribute to immigrants, emigrants and locals ($\omega_{ik,t}^{immi}$, $\omega_{ik,t}^{emi}$, $\omega_{ik,t}^{births}$).

To reduce endogeneity concerns because of omitted variables, we control for several other observed and unobserved factors that might influence the probability of entry or exit.

We control for an occupation's ubiquity, $\sum_i M_{ik}^{births}$ (i.e. the number of locations that are specialized in the respective occupation), since it may be easier to develop specializations in ubiquitous occupations. Also, we control for knowledge diffusion due to other channels than migration captured in spatial lags, $\rho_{ik,t-1}^{M}$ and $\rho_{ik,t-1}^{\omega}$ (see Section 2.4). Lastly, our definition of entries and exits can be sensitive to borderline cases. The expected number of births may already be very close to the observed number before entering, which increases the probability of entering. Hence, we control for the ratio between the observed and expected number of births in the previous period ($R_{ik,t}^{births} = \frac{N_{ik,t}^{births}}{\widehat{N}_{ik,t}^{births}}$).

We use fixed effects to account for unobserved heterogeneity: A city may set up a university affecting both migration and future births of famous scientists. A city may also become a capital attracting politicians, journalists, or military personnel. A city may become more prosperous or increase its level of education affecting migration and future births in a field. We control for these unobserved factors by using fixed effects specific to a broad category (8 broad categories: arts, science & technology, institutions etc.; see column 1 of Table S1 for the occupation taxonomy) in a region in a century ($\gamma_{mit}$). In addition, we control for unobserved factors affecting both migration and future births that are specific to a more granular occupation category and time ($\delta_{lt}$). Index $l$ denotes one of 26 occupation categories, which distinguish, for instance, between social sciences, natural sciences and engineering within the broad category "science & technology" or music, design and film & theatre within the broad category "arts" (see column 2 in Table S1). These fixed effects capture, for instance, that the invention of motion picture technology at the end of the 19th century likely affected migration and birth patterns among film directors and actors differently than among other occupations within the same broad category of arts, such as composers or musicians.

In less restrictive specifications in Section 3.4, we also add further observed control variables that have been included in the fixed effects in the main specification. That includes the number of occupations a location is specialized in (diversity, $\sum_k M_{ik}^{births}$), since the probability of entry or exit likely grows with the number of occupations already present in the respective location. Furthermore, we control for a region's population (Section 1.3) at the beginning of the century ($pop_{i,t}$), because we suspect a correlation between population size and the probability of entering or exiting an activity.

Defining $Y_{ik,t} = \{Entry_{ik,t}, Exit_{ik,t}\}$, we estimate the following logistic regression model

$$\begin{aligned} P(Y_{ik,t}) = g(&\beta_1 M_{ik,t-1}^{immigrants} + \beta_2 M_{ik,t-1}^{emigrants} \\ &+ \beta_3 \omega_{ik,t-1}^{immigrants} + \beta_4 \omega_{ik,t-1}^{emigrants} \\ &+ \beta_5 \omega_{ik,t-1}^{births} + \alpha_2 ubiquity_{k,t-1} + \alpha_4 \rho_{ik,t-1}^{M} \\ &+ \alpha_5 \rho_{ik,t-1}^{\omega} + \alpha_6 R_{ik,t-1}^{births} + + \gamma_{mit} + \delta_{lt} \\ &+ \varepsilon_{ik,t}) \end{aligned} \quad (S8)$$

where $g$ denotes the logistic probability density.

**Descriptive statistics**

Table S4 provides the summary statistics for the variables used in the regression models (based on the naïve model for the expected number of immigrants, emigrants and locals). Each entry refers to a unique combination of region, occupation and period.

The relatedness densities based on immigrants, emigrants and locals correlate with each other but show a considerable amount of variance (see Figure S7).

*Table S4. Descriptive statistics*

| Variable | N | Mean | Std. Dev | Minimum | 25th pc. | Median | 75th pc. | Maximum |
|---|---|---|---|---|---|---|---|---|
| $\omega_{ik}^{births}$ | 43,273 | 17.83 | 10.84 | 0 | 9.65 | 15.67 | 23.87 | 100 |
| $\omega_{ik}^{immi}$ | 18,597 | 17.54 | 12.31 | 0 | 8.32 | 14.67 | 23.59 | 100 |
| $\omega_{ik}^{emi}$ | 36,985 | 17.88 | 10.59 | 0 | 9.89 | 15.85 | 23.67 | 100 |
| $M_{ik}^{births}$ | 43,814 | 0.16 | 0.36 | 0 | 0 | 0 | 0 | 1 |
| $M_{ik}^{immi}$ | 18,663 | 0.17 | 0.37 | 0 | 0 | 0 | 0 | 1 |
| $M_{ik}^{emi}$ | 37,264 | 0.15 | 0.36 | 0 | 0 | 0 | 0 | 1 |
| $Entry_{ik}$ | 15,818 | 0.17 | 0.37 | 0 | 0 | 0 | 0 | 1 |
| $Exit_{ik}$ | 3,794 | 0.64 | 0.48 | 0 | 0 | 1 | 1 | 1 |
| $N_{ik}^{births}$ | 43,814 | 0.42 | 1.60 | 0 | 0 | 0 | 0 | 64 |
| $N_{ik}^{emi}$ | 43,814 | 0.27 | 0.94 | 0 | 0 | 0 | 0 | 31 |
| $N_{ik}^{immi}$ | 43,814 | 0.27 | 1.73 | 0 | 0 | 0 | 0 | 120 |
| $diversity_i$ | 43,814 | 8.33 | 4.43 | 1 | 5 | 8 | 10 | 30 |
| $ubiquity_k$ | 43,814 | 37.45 | 34.33 | 0 | 12 | 22 | 54 | 149 |
| $\rho_{ik,t}^{\omega}$ | 43,295 | 17.76 | 5.58 | 6.87 | 13.39 | 17.06 | 21.73 | 68.14 |
| $\rho_{ik,t}^{M}$ | 43,814 | 0.16 | 0.13 | 0 | 0.05 | 0.11 | 0.24 | 0.88 |

*Note*: Each observation in the underlying dataset refers to a certain location $i$, occupation $k$ and time $t$.

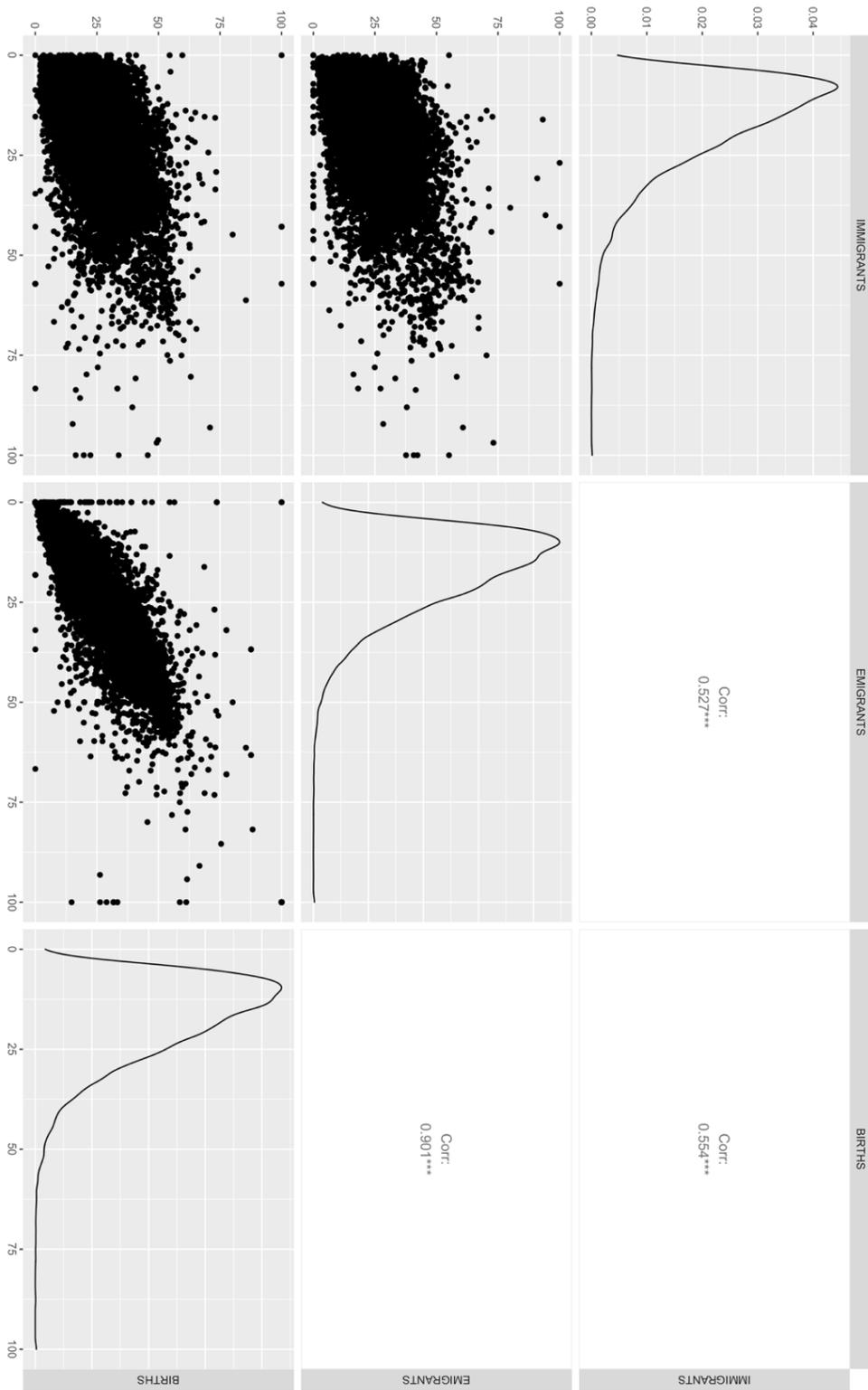

*Figure S7. Correlations between different relatedness density metrics*

## 3.2. Main regression tables explaining entries to new activities

Table S5 shows the main results for the logistic regression model explaining entries to new activities estimating the underlying expected number of immigrants, emigrants and locals with the "bins and balls" model of the Revealed Comparative Advantage. Columns 2-6 of Table S5 correspond to columns 1-5 of Table 1 in the main text.

As mentioned in the main text we find a positive correlation between entries to a specific activity and a disproportionate inflow of famous individuals with knowledge in that activity. Also, the related knowledge of immigrants correlates positively with the probability of future entries.

The control variables behave mostly as expected. We find a positive correlation between the probability of entry and the occupation's ubiquity. Thus, it is easier to enter a ubiquitous (and thus less complex) activity. Also, following the definition of entries, being closer to the threshold of a specialization ($R_{ik,t-1}$) increases the probability of entry.

*Table S5. Main results of logistic regressions explaining the entry to new activities*

|  | \multicolumn{6}{c}{Dependent Variable: $Entry_{ik,t}$} |
|---|---|---|---|---|---|---|
|  | (1) | (2) | (3) | (4) | (5) | (6) |
| $M^{immi}_{ik,t-1}$ |  | 0.334*** | 0.303*** | 0.336*** | 0.331*** | 0.300*** |
|  |  | (0.080) | (0.075) | (0.086) | (0.080) | (0.076) |
| $M^{emi}_{ik,t-1}$ |  | 0.115 | 0.045 | 0.106 | 0.121 | 0.018 |
|  |  | (0.261) | (0.278) | (0.261) | (0.255) | (0.270) |
| $\omega^{immi}_{ik,t-1}$ |  |  | 0.027*** |  |  | 0.028*** |
|  |  |  | (0.006) |  |  | (0.007) |
| $\omega^{emi}_{ik,t-1}$ |  |  |  | -0.006 |  | -0.024 |
|  |  |  |  | (0.012) |  | (0.019) |
| $\omega^{births}_{ik,t-1}$ |  |  |  |  | 0.011 | 0.027* |
|  |  |  |  |  | (0.008) | (0.015) |
| $ubiquity_{k,t-1}$ | 0.011*** | 0.011*** | 0.010*** | 0.011*** | 0.010*** | 0.010*** |
|  | (0.003) | (0.003) | (0.003) | (0.003) | (0.003) | (0.003) |
| $\rho^M_{ik,t-1}$ | -0.261 | -0.289 | -0.311 | -0.286 | -0.294 | -0.312 |
|  | (0.562) | (0.576) | (0.559) | (0.574) | (0.583) | (0.582) |
| $\rho^\omega_{ik,t-1}$ | 0.096** | 0.091** | 0.075 | 0.096* | 0.081* | 0.071 |
|  | (0.046) | (0.046) | (0.049) | (0.050) | (0.044) | (0.047) |
| $R^{births}_{ik,t-1}$ | 0.332*** | 0.236* | 0.295** | 0.228* | 0.246* | 0.284** |
|  | (0.072) | (0.136) | (0.142) | (0.120) | (0.131) | (0.118) |
| FE: Broad categ.-region-period | Y | Y | Y | Y | Y | Y |
| FE: Category-period | Y | Y | Y | Y | Y | Y |
| Observations | 3944 | 3944 | 3944 | 3944 | 3944 | 3944 |
| Pseudo-$R^2$ | 0.211 | 0.213 | 0.214 | 0.213 | 0.213 | 0.215 |
| BIC | 9527.5 | 9537.0 | 9539.4 | 9545.0 | 9544.5 | 9553.1 |

Standard errors are clustered by period and region. * $p < 0.1$, ** $p < 0.05$, *** $p < 0.01$

## 3.3. Main regression tables explaining exits of activities

Table S6 shows the results of the logistic regression model explaining exits of activities estimating the underlying expected number of immigrants, emigrants and locals with the "bins and balls" model. Columns 2-6 of Table S6 correspond to columns 6-10 of Table 1 in the main text.

As mentioned in the main text we find a significantly negative correlation between exits of a specific activity and a disproportionate inflow of famous individuals with knowledge in that activity. Also, the related knowledge of immigrants correlates negatively with the probability of future exits. These coefficients are robust to other specifications and period subsets.

Considering the control variables, we find a robust negative correlation between an activity's ubiquity and the probability of exit. More ubiquitous activities exhibit a lower probability of exit. Also, we find evidence that the probability of exit grows with the presence of the same specialization in geographically close regions ($\rho_{ik,t-1}^{M}$).

*Table S6. Main results of logistic regressions explaining the exit of activities*

| | \multicolumn{6}{c}{Dependent Variable: $Exit_{ik,t}$} | | | | | |
|---|---|---|---|---|---|---|
| | (1) | (2) | (3) | (4) | (5) | (6) |
| $M_{ik,t-1}^{immi}$ | | -0.603*** | -0.584*** | -0.591*** | -0.587*** | -0.571*** |
| | | (0.127) | (0.134) | (0.120) | (0.126) | (0.126) |
| $M_{ik,t-1}^{emi}$ | | 0.310 | 0.330 | 0.233 | 0.306 | 0.291 |
| | | (0.240) | (0.232) | (0.216) | (0.222) | (0.203) |
| $\omega_{ik,t-1}^{immi}$ | | | -0.067*** | | | -0.064*** |
| | | | (0.016) | | | (0.011) |
| $\omega_{ik,t-1}^{emi}$ | | | | -0.048 | | -0.025 |
| | | | | (0.038) | | (0.063) |
| $\omega_{ik,t-1}^{births}$ | | | | | -0.059*** | -0.034 |
| | | | | | (0.018) | (0.041) |
| $ubiquity_{k,t-1}$ | -0.053*** | -0.054*** | -0.055*** | -0.051*** | -0.051*** | -0.052*** |
| | (0.011) | (0.012) | (0.013) | (0.012) | (0.013) | (0.013) |
| $\rho_{ik,t-1}^{M}$ | 5.537*** | 5.684*** | 6.614*** | 5.127*** | 5.102*** | 5.967*** |
| | (1.273) | (1.250) | (1.453) | (1.184) | (1.180) | (1.209) |
| $\rho_{ik,t-1}^{\omega}$ | 0.118 | 0.142 | 0.145 | 0.157 | 0.152 | 0.156 |
| | (0.164) | (0.169) | (0.217) | (0.144) | (0.157) | (0.171) |
| $R_{ik,t-1}^{births}$ | 0.011 | 0.009 | -0.008 | 0.004 | 0.005 | -0.012 |
| | (0.019) | (0.019) | (0.020) | (0.016) | (0.018) | (0.018) |
| FE: Broad categ.-region-period | Y | Y | Y | Y | Y | Y |
| FE: Category-period | Y | Y | Y | Y | Y | Y |
| Observations | 1051 | 1051 | 1051 | 1051 | 1051 | 1051 |
| Pseudo-$R^2$ | 0.216 | 0.224 | 0.230 | 0.226 | 0.226 | 0.232 |
| BIC | 3616.9 | 3619.6 | 3618.0 | 3623.4 | 3623.3 | 3628.8 |

Standard errors are clustered by period and region. * $p < 0.1$, ** $p < 0.05$, *** $p < 0.01$

### 3.4. Robustness checks

In this section, we provide various robustness checks for our main results concerning the role of migrants in the historical geography of knowledge.

Specifically, we explore

- potential endogeneity concerns estimating the expected number of immigrants, emigrants and locals in a negative binomial regression model (Section 3.4.1)

- different regression model specifications for both entries and exits. Because of the highly restrictive fixed effects in our main specification, the number of observations is artificially reduced. To see whether our main findings also hold for larger sample sizes, we provide results for several less restrictive fixed effects specifications (Section 3.4.2)

- the fact that distance played a more pronounced role in knowledge diffusion in earlier periods. We take this into account by interacting the measures of knowledge diffusion across space for other reasons than migration with century-dummies (Section 3.4.3)

- excluding the 20th century from the sample, since our dataset is unbalanced with respect to time (Section 3.4.4)

- a different definition of entries and exits (Section 3.4.5)

- interaction terms to investigate the role of migration in unrelated diversification (Section 3.4.6)

- heterogenous effects across different broad categories (Section 3.4.7)

### 3.4.1. Estimating the expected number of immigrants, emigrants and locals

In our main specification, we defined the specialization matrices based on the concept of the Revealed Comparative Advantage. We say that a region is specialized in an activity, if the observed number of immigrants ($N_{ik,t}^{immi}$), emigrants ($N_{ik,t}^{emi}$) or locals ($N_{ik,t}^{births}$) is larger than the expected number of immigrants ($\widehat{N}_{ik,t}^{immi}$), emigrants ($\widehat{N}_{ik,t}^{emi}$) or locals ($\widehat{N}_{ik,t}^{births}$), respectively, given the size of the region and the ubiquity of the occupation (see Eq. 2 in the main text).

But based on that definition, our results shown in Table S5 and Table S6 may be subject to endogeneity. For instance, a region's local factors may change, affecting both the migration flows and the probability of giving birth to famous individuals in an activity. This could distort our estimates of whether a region is, in fact, specialized in an activity or experiences disproportionate immigration. To address these endogeneity concerns, we estimate the expected number of immigrants, emigrants and locals using not only the number of individuals in a region and the occupation's ubiquity, but also a region's specialization structure in the previous century and further unobserved factors specific to a region, activity and century.

Specifically, we estimate the following negative binomial regression models:

$$N_{ik,t}^{immi} = f(\alpha_0 + \alpha_1 N_{ik,t-1}^{immi} + \alpha_2 S_{ik,t-1}^{births} + \theta_{it} + \vartheta_{kt} + \varepsilon_{ik,t})$$

$$N_{ik,t}^{emi} = f(\beta_0 + \beta_1 N_{ik,t-1}^{emi} + \beta_2 S_{ik,t-1}^{births} + \theta_{it} + \vartheta_{kt} + \varepsilon_{ik,t}) \quad , \quad (S9)$$

$$N_{ik,t}^{births} = f(\gamma_0 + \gamma_1 N_{ik,t-1}^{births} + \gamma_2 S_{ik,t-1}^{births} + \theta_{it} + \vartheta_{kt} + \varepsilon_{ik,t})$$

where $S_{ik,t-1}^{births} = \frac{N_{ik}^{births}/\sum_k N_{ik}^{births}}{\sum_i N_{ik}^{births}/\sum_{i,k} N_{ik}^{births}}$, while $\theta_{it}$ and $\vartheta_{kt}$ denote fixed-effects accounting for unobserved factors specific to a region in a specific century and to an occupation in a specific century, respectively. Table S7 shows the results.

We use the fitted values of these regression models as the expected values in creating the specialization matrices in Eq. S2 ($\widehat{N}_{ik,t}^{immi}$, $\widehat{N}_{ik,t}^{emi}$, $\widehat{N}_{ik,t}^{births}$).

Using these adjacency matrices, we calculate the proximities between activities as well as the relatedness densities for immigrants, emigrants and locals. Figure S8 shows that the original measures of related knowledge based on the more naïve definition of the expected numbers and the relatedness densities based on the expected numbers retrieved from Eq. S9 are highly correlated with an $R^2$ of 0.9 to 0.95.

We then use these new measures of the knowledge of immigrants, emigrants and locals in the logistic regression models described in Eq. S8. The results for both entries (Table S8) and exits (Table S9) remain qualitatively unchanged for the knowledge of immigrants.

*Table S7. Negative binomial regression models to estimate the expected number of immigrants, emigrants and locals.*

|  | $N_{ik,t}^{immi}$ | $N_{ik,t}^{emi}$ | $N_{ik,t}^{births}$ |
|---|---|---|---|
| $N_{ik,t-1}^{immi}$ | 0.033** | | |
|  | (0.013) | | |
| $N_{ik,t-1}^{emi}$ | | 0.038*** | |
|  | | (0.011) | |
| $N_{ik,t-1}^{births}$ | | | 0.033*** |
|  | | | (0.009) |
| $S_{ik,t-1}^{births}$ | 0.045*** | 0.038*** | 0.040*** |
|  | (0.006) | (0.005) | (0.005) |
| Overdispersion parameter | 2.033*** | 3.453*** | 2.369*** |
|  | (0.163) | (0.294) | (0.138) |
| FE: period-region | X | X | X |
| FE: period-occupation | X | X | X |
| Num.Obs. | 39131 | 43651 | 43755 |
| Pseudo-$R^2$ | 0.341 | 0.292 | 0.282 |
| AIC | 30775.6 | 39170.4 | 50334.3 |
| BIC | 40173.4 | 49573.8 | 60775.3 |

* $p < 0.1$, ** $p < 0.05$, *** $p < 0.01$

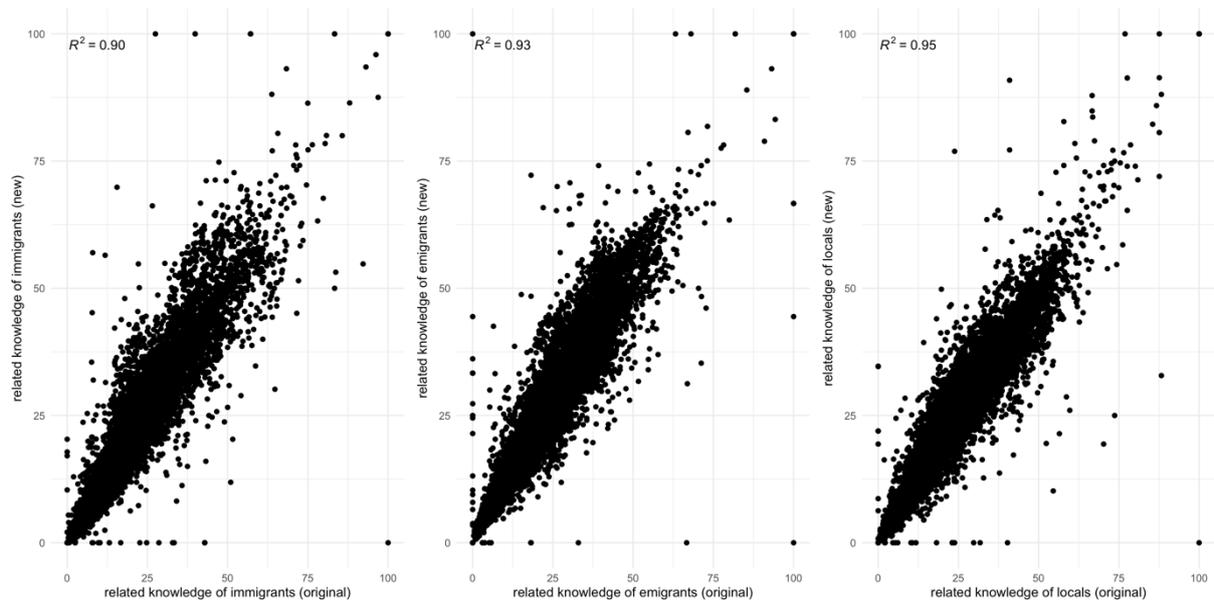

***Figure S8***. *Correlation between the relatedness densities based on the definition of expected number as in the Revealed Comparative Advantage (original) or the negative binomial regression (new).*

***Table S8***. *Logistic regression model explaining entries to new activities using the expected numbers of the model in Eq. S11*

|  | Dependent Variable: $Entry_{ik,t}$ | | | | | |
|---|---|---|---|---|---|---|
|  | (1) | (2) | (3) | (4) | (5) | (6) |
| $M^{immi}_{ik,t-1}$ |  | 0.295** | 0.254** | 0.295** | 0.293** | 0.251*** |
|  |  | (0.140) | (0.138) | (0.140) | (0.138) | (0.058) |
| $M^{emi}_{ik,t-1}$ |  | 0.347 | 0.355 | 0.347 | 0.334 | 0.316 |
|  |  | (0.372) | (0.364) | (0.374) | (0.372) | (0.325) |
| $\omega^{immi}_{ik,t-1}$ |  |  | 0.027** |  |  | 0.026*** |
|  |  |  | (0.013) |  |  | (0.008) |
| $\omega^{emi}_{ik,t-1}$ |  |  |  | 0.0003 |  | -0.026 |
|  |  |  |  | (0.014) |  | (0.019) |
| $\omega^{births}_{ik,t-1}$ |  |  |  |  | 0.028** | 0.044** |
|  |  |  |  |  | (0.015) | (0.016) |
| $ubiquity_{k,t-1}$ | 0.011 | 0.011 | 0.010 | 0.011 | 0.010 | 0.010*** |
|  | (0.008) | (0.007) | (0.007) | (0.007) | (0.007) | (0.003) |
| $\rho^{M}_{ik,t-1}$ | -0.261 | -0.289 | -0.311 | -0.286 | -0.294 | -0.312 |
|  | (0.916) | (0.932) | (0.948) | (0.934) | (0.932) | (0.582) |
| $\rho^{\omega}_{ik,t-1}$ | 0.096 | 0.091 | 0.075 | 0.096 | 0.081 | 0.071 |
|  | (0.059) | (0.059) | (0.063) | (0.060) | (0.056) | (0.047) |
| $R^{births}_{ik,t-1}$ | 0.332 | 0.236 | 0.295 | 0.228 | 0.246 | 0.284** |
|  | (0.262) | (0.271) | (0.271) | (0.274) | (0.273) | (0.118) |
| FE: Broad categ.-region-period | Y | Y | Y | Y | Y | Y |
| FE: Category-period | Y | Y | Y | Y | Y | Y |
| Observations | 4049 | 4049 | 4049 | 4049 | 4049 | 4049 |
| Pseudo-$R^2$ | 0.224 | 0.225 | 0.227 | 0.225 | 0.226 | 0.228 |
| BIC | 9796.7 | 9806.5 | 9807.7 | 9814.8 | 9810.7 | 9818.0 |

Standard errors are clustered by period and region. * $p < 0.1$, ** $p < 0.05$, *** $p < 0.01$

*Table S9. Logistic regression model explaining exits from existing areas of specialization using the expected numbers of the model in Eq. S11*

|  | \multicolumn{6}{c}{Dependent Variable: $Exit_{ik,t}$} |
|---|---|---|---|---|---|---|
|  | (1) | (2) | (3) | (4) | (5) | (6) |
| $M^{immi}_{ik,t-1}$ |  | -0.411* | -0.383* | -0.409* | -0.411* | -0.388** |
|  |  | (0.226) | (0.218) | (0.225) | (0.226) | (0.159) |
| $M^{emi}_{ik,t-1}$ |  | 0.851** | 0.864** | 0.822** | 0.828** | 0.884*** |
|  |  | (0.337) | (0.339) | (0.328) | (0.339) | (0.206) |
| $\omega^{immi}_{ik,t-1}$ |  |  | -0.042 |  |  | -0.040* |
|  |  |  | (0.028) |  |  | (0.022) |
| $\omega^{emi}_{ik,t-1}$ |  |  |  | -0.018 |  | 0.028 |
|  |  |  |  | (0.047) |  | (0.047) |
| $\omega^{births}_{ik,t-1}$ |  |  |  |  | -0.063 | -0.080** |
|  |  |  |  |  | (0.041) | (0.041) |
| $ubiquity_{k,t-1}$ | -0.052*** | -0.053*** | -0.054*** | -0.052*** | -0.050*** | -0.053*** |
|  | (0.012) | (0.012) | (0.012) | (0.012) | (0.012) | (0.009) |
| $\rho^{M}_{ik,t-1}$ | 4.793** | 4.550** | 4.886** | 4.450* | 4.207* | 4.621* |
|  | (2.310) | (2.254) | (2.277) | (2.331) | (2.236) | (2.139) |
| $\rho^{\omega}_{ik,t-1}$ | 0.209 | 0.241* | 0.262* | 0.243* | 0.247* | 0.265** |
|  | (0.128) | (0.133) | (0.133) | (0.132) | (0.127) | (0.072) |
| $R^{births}_{ik,t-1}$ | 0.028 | 0.022 | 0.012 | 0.022 | 0.020 | 0.010 |
|  | (0.039) | (0.038) | (0.035) | (0.037) | (0.036) | (0.024) |
| FE: Broad categ.-region-period | Y | Y | Y | Y | Y | Y |
| FE: Category-period | Y | Y | Y | Y | Y | Y |
| Observations | 1084 | 1084 | 1084 | 1084 | 1084 | 1084 |
| Pseudo-$R^2$ | 0.195 | 0.205 | 0.209 | 0.206 | 0.208 | 0.212 |
| BIC | 3804.0 | 3802.9 | 3805.0 | 3809.5 | 3805.3 | 3814.1 |

Standard errors are clustered by period and region. * $p < 0.1$, ** $p < 0.05$, *** $p < 0.01$

### 3.4.2. Model specifications

To control for unobserved factors that have an influence on both the probability of entering or exiting an activity and migration flows, we use fixed effects. For example, period fixed effects account for any unobserved heterogeneity that is time-specific, but independent of the location and the activity. Thus, this controls for the effect that entering a new activity may have become more easier over time, because of economic development and urbanization.

In the main results, we employ a highly restrictive fixed effects structure accounting for unobserved factors specific to a region, century and broad category as well as unobserved factors specific to an occupation category and century (see Eq. S8, Table S5 and Table S6).

Although this specification addresses endogeneity issues thoroughly, this highly restrictive fixed effects specification (more than 700 fixed effect coefficients) comes at a cost, too. That is, our sample size is artificially reduced, since observations with no changes in the dependent variable within a fixed effect category are removed. In this section, we provide several less restrictive fixed effects specifications to see how our results change.

Furthermore, a less restrictive fixed effects specification allows for including more control variables that previously were included in the fixed effects. That includes the number of occupations a location is specialized in (diversity, $\sum_k M^{births}_{ik}$), since the probability of entry or exit likely grows with the number of occupations already present in the respective location. Furthermore, we control for a region's population (Section 1.3) at the beginning of the century ($pop_{i,t}$), because we suspect a correlation between population size and the probability of entering or exiting an activity.

In the following, we control either for time-location and occupation category fixed effects (Table S10 and Table S14 for entries and exits, respectively), for time, location and occupation category fixed effects (Table S11 and Table S15 for entries and exits, respectively), for time and location fixed effects (Table S12 and Table S16 for entries and exits, respectively) or only for time fixed effects (Table S13 and Table S17 for entries and exits, respectively). Despite the different fixed effects, the main results for $M_{ik,t-1}^{immi}$ and $\omega_{ik,t-1}^{immi}$ remain unchanged.

*Table S10. Logistic regression model explaining entries, accounting for period-region and occupation category fixed effects*

|  | \multicolumn{6}{c}{Dependent Variable: $Entry_{ik,t}$} |
|---|---|---|---|---|---|---|
|  | (1) | (2) | (3) | (4) | (5) | (6) |
| $M_{ik,t-1}^{immi}$ |  | 0.389*** | 0.339*** | 0.392*** | 0.379*** | 0.326*** |
|  |  | (0.095) | (0.084) | (0.095) | (0.096) | (0.083) |
| $M_{ik,t-1}^{emi}$ |  | 0.033 | 0.011 | 0.026 | 0.031 | -0.017 |
|  |  | (0.357) | (0.369) | (0.362) | (0.353) | (0.373) |
| $\omega_{ik,t-1}^{immi}$ |  |  | 0.019** |  |  | 0.020** |
|  |  |  | (0.009) |  |  | (0.009) |
| $\omega_{ik,t-1}^{emi}$ |  |  |  | -0.007 |  | -0.024** |
|  |  |  |  | (0.009) |  | (0.010) |
| $\omega_{ik,t-1}^{births}$ |  |  |  |  | 0.013* | 0.029*** |
|  |  |  |  |  | (0.007) | (0.007) |
| $ubiquity_{k,t-1}$ | 0.007 | 0.007 | 0.006 | 0.007 | 0.007 | 0.007 |
|  | (0.005) | (0.005) | (0.004) | (0.004) | (0.005) | (0.004) |
| $\rho_{ik,t-1}^{M}$ | 0.652 | 0.588 | 0.604 | 0.592 | 0.597 | 0.620 |
|  | (0.721) | (0.729) | (0.716) | (0.731) | (0.730) | (0.719) |
| $\rho_{ik,t-1}^{\omega}$ | 0.034 | 0.034 | 0.025 | 0.039 | 0.021 | 0.015 |
|  | (0.031) | (0.031) | (0.031) | (0.031) | (0.031) | (0.032) |
| $R_{ik,t-1}^{births}$ | 0.525** | 0.463* | 0.493* | 0.458* | 0.471* | 0.496* |
|  | (0.249) | (0.265) | (0.273) | (0.268) | (0.258) | (0.272) |
| FE: period-region | Y | Y | Y | Y | Y | Y |
| FE: occu. category | Y | Y | Y | Y | Y | Y |
| Observations | 6165 | 6165 | 6165 | 6165 | 6165 | 6165 |
| Pseudo-$R^2$ | 0.131 | 0.134 | 0.136 | 0.135 | 0.135 | 0.137 |
| AIC | 5782.5 | 5768.7 | 5762.6 | 5770.1 | 5768.3 | 5758.7 |
| BIC | 7437.3 | 7437.0 | 7437.5 | 7445.0 | 7443.3 | 7447.1 |

Standard errors are clustered by region and period. * $p < 0.1$, ** $p < 0.05$, *** $p < 0.01$

**Table S11**. *Logistic regression model explaining entries, accounting for period-, location- and occupation category-fixed effects*

|  | Dependent Variable: $Entry_{ik,t}$ | | | | | |
|---|---|---|---|---|---|---|
|  | (1) | (2) | (3) | (4) | (5) | (6) |
| $M^{immi}_{ik,t-1}$ |  | 0.380*** | 0.335*** | 0.381*** | 0.375*** | 0.330*** |
|  |  | (0.073) | (0.066) | (0.074) | (0.074) | (0.066) |
| $M^{emi}_{ik,t-1}$ |  | 0.012 | -0.027 | 0.015 | 0.010 | -0.017 |
|  |  | (0.235) | (0.228) | (0.231) | (0.234) | (0.224) |
| $\omega^{immi}_{ik,t-1}$ |  |  | 0.011*** |  |  | 0.012*** |
|  |  |  | (0.003) |  |  | (0.002) |
| $\omega^{emi}_{ik,t-1}$ |  |  |  | -0.001 |  | -0.008*** |
|  |  |  |  | (0.004) |  | (0.002) |
| $\omega^{births}_{ik,t-1}$ |  |  |  |  | 0.006 | 0.011** |
|  |  |  |  |  | (0.006) | (0.006) |
| $diversity_{i,t-1}$ | -0.008 | -0.007 | -0.009 | -0.005 | -0.021 | -0.022 |
|  | (0.020) | (0.020) | (0.020) | (0.017) | (0.020) | (0.022) |
| $ubiquity_{k,t-1}$ | 0.008*** | 0.007*** | 0.007*** | 0.007*** | 0.007*** | 0.007*** |
|  | (0.001) | (0.001) | (0.001) | (0.001) | (0.001) | (0.001) |
| $\rho^{M}_{ik,t-1}$ | 0.622*** | 0.562*** | 0.590*** | 0.562*** | 0.568*** | 0.598*** |
|  | (0.207) | (0.159) | (0.122) | (0.161) | (0.175) | (0.155) |
| $\rho^{\omega}_{ik,t-1}$ | 0.029 | 0.029 | 0.021 | 0.030 | 0.023 | 0.016 |
|  | (0.025) | (0.024) | (0.024) | (0.024) | (0.024) | (0.024) |
| $R^{births}_{ik,t-1}$ | 0.533*** | 0.478** | 0.493*** | 0.477** | 0.483*** | 0.499*** |
|  | (0.124) | (0.187) | (0.178) | (0.185) | (0.182) | (0.178) |
| $\log(pop_{i,t})$ | 0.368*** | 0.351*** | 0.249*** | 0.352*** | 0.349*** | 0.249*** |
|  | (0.067) | (0.074) | (0.089) | (0.074) | (0.069) | (0.083) |
| FE: period | Y | Y | Y | Y | Y | Y |
| FE: region | Y | Y | Y | Y | Y | Y |
| FE: occu. category | Y | Y | Y | Y | Y | Y |
| Observations | 6180 | 6180 | 6180 | 6180 | 6180 | 6180 |
| Pseudo-$R^2$ | 0.121 | 0.124 | 0.124 | 0.124 | 0.124 | 0.125 |
| AIC | 5694.6 | 5680.9 | 5677.5 | 5682.8 | 5682.1 | 5679.4 |
| BIC | 6811.6 | 6811.4 | 6814.7 | 6820.0 | 6819.3 | 6830.0 |

Standard errors are clustered by period and region. * $p < 0.1$, ** $p < 0.05$, *** $p < 0.01$

**Table S12**. Logistic regression model explaining entries, accounting for period- and location-fixed effects

|  | Dependent Variable: $Entry_{ik,t}$ | | | | | |
|---|---|---|---|---|---|---|
|  | (1) | (2) | (3) | (4) | (5) | (6) |
| $M^{immi}_{ik,t-1}$ |  | 0.367*** | 0.324*** | 0.368*** | 0.361*** | 0.317*** |
|  |  | (0.086) | (0.076) | (0.087) | (0.087) | (0.076) |
| $M^{emi}_{ik,t-1}$ |  | 0.207 | 0.174 | 0.208 | 0.204 | 0.182 |
|  |  | (0.223) | (0.220) | (0.219) | (0.221) | (0.215) |
| $\omega^{immi}_{ik,t-1}$ |  |  | 0.010*** |  |  | 0.011*** |
|  |  |  | (0.003) |  |  | (0.003) |
| $\omega^{emi}_{ik,t-1}$ |  |  |  | -0.001 |  | -0.008*** |
|  |  |  |  | (0.003) |  | (0.002) |
| $\omega^{births}_{ik,t-1}$ |  |  |  |  | 0.008 | 0.013* |
|  |  |  |  |  | (0.007) | (0.007) |
| $diversity_{i,t-1}$ | -0.012 | -0.011 | -0.013 | -0.010 | -0.028 | -0.030 |
|  | (0.019) | (0.019) | (0.019) | (0.015) | (0.019) | (0.020) |
| $ubiquity_{k,t-1}$ | 0.009*** | 0.009*** | 0.009*** | 0.009*** | 0.009*** | 0.009*** |
|  | (0.003) | (0.003) | (0.003) | (0.003) | (0.003) | (0.003) |
| $\rho^{M}_{ik,t-1}$ | 0.290 | 0.245* | 0.255* | 0.245* | 0.259* | 0.273 |
|  | (0.185) | (0.128) | (0.152) | (0.128) | (0.154) | (0.186) |
| $\rho^{\omega}_{ik,t-1}$ | 0.029 | 0.026 | 0.020 | 0.027 | 0.020 | 0.013 |
|  | (0.045) | (0.043) | (0.044) | (0.043) | (0.040) | (0.041) |
| $R^{births}_{ik,t-1}$ | 0.298*** | 0.161 | 0.171 | 0.160 | 0.170 | 0.184 |
|  | (0.114) | (0.181) | (0.173) | (0.180) | (0.171) | (0.170) |
| $\log(pop_{i,t})$ | 0.363*** | 0.350*** | 0.257*** | 0.352*** | 0.349*** | 0.258*** |
|  | (0.055) | (0.061) | (0.074) | (0.061) | (0.057) | (0.068) |
| FE: period | Y | Y | Y | Y | Y | Y |
| FE: region | Y | Y | Y | Y | Y | Y |
| Observations | 6180 | 6180 | 6180 | 6180 | 6180 | 6180 |
| Pseudo-$R^2$ | 0.092 | 0.095 | 0.096 | 0.095 | 0.095 | 0.096 |
| AIC | 5819.7 | 5806.0 | 5803.1 | 5808.0 | 5806.7 | 5804.3 |
| BIC | 6775.3 | 6775.0 | 6778.8 | 6783.7 | 6782.4 | 6793.5 |

Standard errors are clustered by region and period. * $p < 0.1$, ** $p < 0.05$, *** $p < 0.01$

***Table S13.*** *Logistic regression model explaining entries, accounting for period-fixed effects*

| | Dependent Variable: $Entry_{ik,t}$ | | | | | |
|---|---|---|---|---|---|---|
| | (1) | (2) | (3) | (4) | (5) | (6) |
| $M^{immi}_{ik,t-1}$ | | 0.406*** | 0.327*** | 0.404*** | 0.403*** | 0.322*** |
| | | (0.072) | (0.076) | (0.072) | (0.073) | (0.077) |
| $M^{emi}_{ik,t-1}$ | | 0.195 | 0.153 | 0.189 | 0.194 | 0.147 |
| | | (0.175) | (0.171) | (0.176) | (0.173) | (0.169) |
| $\omega^{immi}_{ik,t-1}$ | | | 0.011*** | | | 0.011*** |
| | | | (0.002) | | | (0.003) |
| $\omega^{emi}_{ik,t-1}$ | | | | 0.004 | | 0.003 |
| | | | | (0.002) | | (0.003) |
| $\omega^{births}_{ik,t-1}$ | | | | | 0.005 | 0.004 |
| | | | | | (0.005) | (0.006) |
| $diversity_{i,t-1}$ | 0.041*** | 0.037*** | 0.025*** | 0.030*** | 0.028*** | 0.012 |
| | (0.004) | (0.005) | (0.007) | (0.006) | (0.011) | (0.015) |
| $ubiquity_{k,t-1}$ | 0.008*** | 0.008*** | 0.008*** | 0.007*** | 0.007*** | 0.007*** |
| | (0.003) | (0.002) | (0.002) | (0.003) | (0.003) | (0.003) |
| $\rho^M_{ik,t-1}$ | 0.235 | 0.186 | 0.208 | 0.188 | 0.192 | 0.215 |
| | (0.337) | (0.323) | (0.329) | (0.325) | (0.328) | (0.336) |
| $\rho^\omega_{ik,t-1}$ | 0.040 | 0.037 | 0.029 | 0.035 | 0.034 | 0.025 |
| | (0.039) | (0.038) | (0.035) | (0.037) | (0.035) | (0.032) |
| $R^{births}_{ik,t-1}$ | 0.366*** | 0.224 | 0.219 | 0.222 | 0.230 | 0.223 |
| | (0.111) | (0.163) | (0.157) | (0.160) | (0.156) | (0.148) |
| $\log(pop_{i,t})$ | 0.164*** | 0.158** | 0.130* | 0.160** | 0.161** | 0.134* |
| | (0.062) | (0.067) | (0.070) | (0.067) | (0.069) | (0.071) |
| FE: period | Y | Y | Y | Y | Y | Y |
| Observations | 6180 | 6180 | 6180 | 6180 | 6180 | 6180 |
| Pseudo-$R^2$ | 0.070 | 0.073 | 0.075 | 0.073 | 0.073 | 0.075 |
| AIC | 5697.9 | 5679.1 | 5670.1 | 5680.3 | 5680.5 | 5672.9 |
| BIC | 5778.6 | 5773.3 | 5771.1 | 5781.3 | 5781.4 | 5787.3 |

Standard errors are clustered by region and period. * $p < 0.1$, ** $p < 0.05$, *** $p < 0.01$

***Table S14.*** *Logistic regression model explaining exits, accounting for period-region and occupation category fixed effects*

| | Dependent Variable: $Exit_{ik,t}$ | | | | | |
|---|---|---|---|---|---|---|
| | (1) | (2) | (3) | (4) | (5) | (6) |
| $M^{immi}_{ik,t-1}$ | | -0.565*** | -0.523*** | -0.563*** | -0.553*** | -0.514*** |
| | | (0.155) | (0.148) | (0.155) | (0.157) | (0.150) |
| $M^{emi}_{ik,t-1}$ | | 0.142 | 0.148 | 0.129 | 0.151 | 0.157 |
| | | (0.168) | (0.173) | (0.169) | (0.169) | (0.174) |
| $\omega^{immi}_{ik,t-1}$ | | | -0.037*** | | | -0.036*** |
| | | | (0.014) | | | (0.014) |
| $\omega^{emi}_{ik,t-1}$ | | | | -0.016 | | 0.000 |
| | | | | (0.012) | | (0.017) |
| $\omega^{births}_{ik,t-1}$ | | | | | -0.040*** | -0.038* |
| | | | | | (0.015) | (0.021) |
| $ubiquity_{k,t-1}$ | -0.027*** | -0.026*** | -0.026*** | -0.025*** | -0.024*** | -0.024*** |
| | (0.007) | (0.006) | (0.006) | (0.007) | (0.007) | (0.007) |
| $\rho^M_{ik,t-1}$ | 1.371 | 1.427* | 1.685** | 1.297 | 1.122 | 1.399* |
| | (0.915) | (0.848) | (0.807) | (0.801) | (0.851) | (0.796) |
| $\rho^\omega_{ik,t-1}$ | 0.033 | 0.040 | 0.037 | 0.037 | 0.039 | 0.037 |
| | (0.049) | (0.052) | (0.057) | (0.051) | (0.052) | (0.056) |
| $R^{births}_{ik,t-1}$ | -0.021 | -0.019 | -0.026 | -0.020 | -0.022 | -0.029 |
| | (0.022) | (0.024) | (0.026) | (0.024) | (0.024) | (0.026) |
| FE: period-region | Y | Y | Y | Y | Y | Y |
| FE: occu. category | Y | Y | Y | Y | Y | Y |
| Observations | 1989 | 1989 | 1989 | 1989 | 1989 | 1989 |
| Pseudo-$R^2$ | 0.159 | 0.168 | 0.172 | 0.168 | 0.170 | 0.173 |
| AIC | 2665.7 | 2646.5 | 2637.9 | 2647.0 | 2643.1 | 2637.3 |
| BIC | 3969.4 | 3961.4 | 3958.5 | 3967.5 | 3963.6 | 3969.0 |

Standard errors are clustered by region and period. * $p < 0.1$, ** $p < 0.05$, *** $p < 0.01$

*Table S15*. Logistic regression model explaining exits, accounting for period-, location- and occupation category-fixed effects

| | | | Dependent Variable: $Exit_{ik,t}$ | | | |
|---|---|---|---|---|---|---|
| | (1) | (2) | (3) | (4) | (5) | (6) |
| $M_{ik,t-1}^{immi}$ | | -0.509*** | -0.464*** | -0.509*** | -0.510*** | -0.461*** |
| | | (0.051) | (0.064) | (0.061) | (0.051) | (0.072) |
| $M_{ik,t-1}^{emi}$ | | 0.131 | 0.123 | 0.125 | 0.131 | 0.114 |
| | | (0.177) | (0.180) | (0.185) | (0.177) | (0.190) |
| $\omega_{ik,t-1}^{immi}$ | | | -0.014* | | | -0.014** |
| | | | (0.008) | | | (0.007) |
| $\omega_{ik,t-1}^{emi}$ | | | | 0.005 | | 0.008 |
| | | | | (0.015) | | (0.019) |
| $\omega_{ik,t-1}^{births}$ | | | | | 0.001 | -0.002 |
| | | | | | (0.006) | (0.012) |
| $diversity_{i,t-1}$ | -0.042 | -0.044 | -0.039 | -0.051 | -0.046 | -0.045 |
| | (0.029) | (0.032) | (0.033) | (0.044) | (0.036) | (0.039) |
| $ubiquity_{k,t-1}$ | -0.027*** | -0.026*** | -0.025*** | -0.026*** | -0.026*** | -0.026*** |
| | (0.006) | (0.006) | (0.006) | (0.005) | (0.006) | (0.005) |
| $\rho_{ik,t-1}^{M}$ | 1.399* | 1.405 | 1.417 | 1.459* | 1.417 | 1.473* |
| | (0.836) | (0.873) | (0.882) | (0.844) | (0.896) | (0.891) |
| $\rho_{ik,t-1}^{\omega}$ | 0.038 | 0.042 | 0.040 | 0.043 | 0.043 | 0.042 |
| | (0.050) | (0.053) | (0.052) | (0.050) | (0.054) | (0.051) |
| $R_{ik,t-1}^{births}$ | -0.023*** | -0.022*** | -0.026*** | -0.021*** | -0.022*** | -0.025*** |
| | (0.006) | (0.007) | (0.008) | (0.006) | (0.007) | (0.007) |
| $\log(pop_{i,t})$ | -0.372 | -0.340 | -0.265 | -0.357 | -0.343 | -0.283 |
| | (0.277) | (0.272) | (0.272) | (0.276) | (0.278) | (0.280) |
| FE: period | Y | Y | Y | Y | Y | Y |
| FE: region | Y | Y | Y | Y | Y | Y |
| FE: occu. category | Y | Y | Y | Y | Y | Y |
| Observations | 2017 | 2017 | 2017 | 2017 | 2017 | 2017 |
| Pseudo-$R^2$ | 0.138 | 0.145 | 0.147 | 0.146 | 0.145 | 0.147 |
| AIC | 2624.5 | 2607.6 | 2605.7 | 2609.2 | 2609.6 | 2608.9 |
| BIC | 3533.3 | 3527.5 | 3531.2 | 3534.7 | 3535.1 | 3545.6 |

Standard errors are clustered by region and period. * $p < 0.1$, ** $p < 0.05$, *** $p < 0.01$

*Table S16*. Logistic regression model explaining exits, accounting for period- and location-fixed effects

| | | | Dependent Variable: $Exit_{ik,t}$ | | | |
|---|---|---|---|---|---|---|
| | (1) | (2) | (3) | (4) | (5) | (6) |
| $M_{ik,t-1}^{immi}$ | | -0.520*** | -0.464*** | -0.520*** | -0.521*** | -0.460*** |
| | | (0.034) | (0.033) | (0.041) | (0.034) | (0.042) |
| $M_{ik,t-1}^{emi}$ | | 0.060 | 0.053 | 0.052 | 0.060 | 0.042 |
| | | (0.182) | (0.185) | (0.191) | (0.182) | (0.195) |
| $\omega_{ik,t-1}^{immi}$ | | | -0.016*** | | | -0.017*** |
| | | | (0.006) | | | (0.006) |
| $\omega_{ik,t-1}^{emi}$ | | | | 0.007 | | 0.010 |
| | | | | (0.012) | | (0.012) |
| $\omega_{ik,t-1}^{births}$ | | | | | 0.003 | -0.002 |
| | | | | | (0.006) | (0.007) |
| $diversity_{i,t-1}$ | -0.036 | -0.037 | -0.031 | -0.047 | -0.042 | -0.041 |
| | (0.027) | (0.028) | (0.027) | (0.039) | (0.029) | (0.030) |
| $ubiquity_{k,t-1}$ | -0.020*** | -0.019*** | -0.018*** | -0.019*** | -0.019*** | -0.019*** |
| | (0.005) | (0.005) | (0.005) | (0.005) | (0.005) | (0.005) |
| $\rho_{ik,t-1}^{M}$ | 1.277 | 1.305 | 1.317 | 1.384 | 1.332 | 1.404 |
| | (0.891) | (0.956) | (1.007) | (0.923) | (0.976) | (1.009) |
| $\rho_{ik,t-1}^{\omega}$ | 0.027 | 0.032 | 0.031 | 0.033 | 0.032 | 0.033 |
| | (0.045) | (0.049) | (0.049) | (0.047) | (0.049) | (0.048) |
| $R_{ik,t-1}^{births}$ | -0.012** | -0.011* | -0.015*** | -0.010* | -0.011 | -0.014** |
| | (0.005) | (0.006) | (0.005) | (0.006) | (0.007) | (0.006) |
| $\log(pop_{i,t})$ | -0.316 | -0.284 | -0.196 | -0.308 | -0.291 | -0.221 |
| | (0.257) | (0.251) | (0.249) | (0.252) | (0.256) | (0.254) |
| FE: period | Y | Y | Y | Y | Y | Y |
| FE: region | Y | Y | Y | Y | Y | Y |
| Observations | 2023 | 2023 | 2023 | 2023 | 2023 | 2023 |
| Pseudo-$R^2$ | 0.102 | 0.111 | 0.113 | 0.111 | 0.111 | 0.114 |
| AIC | 2677.9 | 2659.0 | 2655.1 | 2660.1 | 2661.0 | 2657.5 |
| BIC | 3458.0 | 3450.4 | 3452.0 | 3457.1 | 3457.9 | 3465.7 |

Standard errors are clustered by region and period. * $p < 0.1$, ** $p < 0.05$, *** $p < 0.01$

*Table S17. Logistic regression model explaining exits, accounting for period-fixed effects*

|  | \multicolumn{6}{c}{Dependent Variable: $Exit_{ik,t}$} | | | | | |
| --- | --- | --- | --- | --- | --- | --- |
|  | (1) | (2) | (3) | (4) | (5) | (6) |
| $M_{ik,t-1}^{immi}$ |  | -0.549*** | -0.469*** | -0.541*** | -0.550*** | -0.464*** |
|  |  | (0.043) | (0.047) | (0.047) | (0.042) | (0.054) |
| $M_{ik,t-1}^{emi}$ |  | 0.159 | 0.131 | 0.121 | 0.158 | 0.097 |
|  |  | (0.181) | (0.177) | (0.189) | (0.184) | (0.195) |
| $\omega_{ik,t-1}^{immi}$ |  |  | -0.012*** |  |  | -0.012*** |
|  |  |  | (0.003) |  |  | (0.003) |
| $\omega_{ik,t-1}^{emi}$ |  |  |  | 0.011 |  | 0.009 |
|  |  |  |  | (0.008) |  | (0.009) |
| $\omega_{ik,t-1}^{births}$ |  |  |  |  | 0.011 | 0.004 |
|  |  |  |  |  | (0.008) | (0.008) |
| $diversity_{i,t-1}$ | -0.043*** | -0.039*** | -0.024*** | -0.058*** | -0.063*** | -0.051** |
|  | (0.006) | (0.004) | (0.006) | (0.017) | (0.017) | (0.020) |
| $ubiquity_{k,t-1}$ | -0.014*** | -0.013*** | -0.013*** | -0.014*** | -0.014*** | -0.014*** |
|  | (0.003) | (0.004) | (0.004) | (0.003) | (0.004) | (0.003) |
| $\rho_{ik,t-1}^{M}$ | 1.035 | 1.056 | 1.022 | 1.188 | 1.176 | 1.183 |
|  | (0.782) | (0.872) | (0.904) | (0.802) | (0.838) | (0.837) |
| $\rho_{ik,t-1}^{\omega}$ | -0.013 | -0.006 | -0.001 | -0.006 | -0.005 | -0.001 |
|  | (0.025) | (0.025) | (0.027) | (0.025) | (0.025) | (0.027) |
| $R_{ik,t-1}^{births}$ | -0.007 | -0.008 | -0.010** | -0.006 | -0.007 | -0.008 |
|  | (0.006) | (0.005) | (0.005) | (0.005) | (0.005) | (0.005) |
| $\log(pop_{i,t})$ | -0.265*** | -0.247*** | -0.218*** | -0.246*** | -0.241*** | -0.215*** |
|  | (0.019) | (0.022) | (0.031) | (0.023) | (0.019) | (0.031) |
| FE: period | Y | Y | Y | Y | Y | Y |
| Observations | 2045 | 2045 | 2045 | 2045 | 2045 | 2045 |
| Pseudo-R² | 0.059 | 0.070 | 0.073 | 0.072 | 0.071 | 0.074 |
| AIC | 2558.5 | 2531.7 | 2526.2 | 2530.7 | 2532.3 | 2527.1 |
| BIC | 2626.0 | 2610.5 | 2610.6 | 2615.1 | 2616.6 | 2622.7 |

Standard errors are clustered by region and period. * $p < 0.1$, ** $p < 0.05$, *** $p < 0.01$

### 3.4.3. Century-specific distance measures

Furthermore, we acknowledge that distances across space, if looking at such long periods, are not constant over time, but decrease due to improvements in the infrastructure or technological progress. Hence, we interact our measures of spatial proximity, $\rho_{ik,t-1}^{M}$ and $\rho_{ik,t-1}^{\omega}$ (see Section 2.4), with dummies indicating the different centuries to alleviate concerns that our results are subject to omitted variable bias. Table S18 and Table S19 show that the results remain unchanged for both entries and exits, respectively.

***Table S18***. *Logistic regression model explaining entries to new activities, interacting measures of spatial proximity with period fixed-effects.*

| | \multicolumn{6}{c}{Dependent Variable: $Entry_{ik,t}$} | | | | | |
|---|---|---|---|---|---|---|
| | (1) | (2) | (3) | (4) | (5) | (6) |
| $M_{ik,t-1}^{immigrants}$ | | 0.322*** | 0.289*** | 0.324** | 0.318*** | 0.284** |
| | | (0.120) | (0.108) | (0.127) | (0.120) | (0.121) |
| $M_{ik,t-1}^{emigrants}$ | | 0.163 | 0.083 | 0.144 | 0.171 | 0.044 |
| | | (0.469) | (0.464) | (0.503) | (0.493) | (0.570) |
| $\omega_{ik,t-1}^{immigrants}$ | | | 0.031*** | | | 0.032** |
| | | | (0.010) | | | (0.015) |
| $\omega_{ik,t-1}^{emigrants}$ | | | | -0.010 | | -0.030 |
| | | | | (0.019) | | (0.029) |
| $\omega_{ik,t-1}^{births}$ | | | | | 0.010 | 0.029 |
| | | | | | (0.021) | (0.030) |
| $ubiquity_{k,t-1}$ | 0.003 | 0.003 | 0.001 | 0.003 | 0.002 | 0.001 |
| | (0.004) | (0.004) | (0.004) | (0.004) | (0.003) | (0.005) |
| $\rho_{ik,t-1}^{M}$ | 0.696** | 0.744** | 0.751** | 0.748** | 0.772** | 0.848*** |
| | (0.314) | (0.338) | (0.326) | (0.349) | (0.334) | (0.308) |
| $\rho_{ik,t-1}^{\omega}$ | 0.146*** | 0.142*** | 0.131*** | 0.152*** | 0.131*** | 0.128*** |
| | (0.051) | (0.035) | (0.032) | (0.045) | (0.042) | (0.045) |
| $R_{ik,t-1}^{births}$ | 0.379 | 0.268 | 0.344 | 0.258 | 0.271 | 0.325 |
| | (0.302) | (0.345) | (0.393) | (0.384) | (0.324) | (0.438) |
| FE: Broad categ.-region-period | Y | Y | Y | Y | Y | Y |
| FE: Category-period | Y | Y | Y | Y | Y | Y |
| $\rho_{ik,t-1}^{M}$ * period | Y | Y | Y | Y | Y | Y |
| $\rho_{ik,t-1}^{\omega}$ * period | Y | Y | Y | Y | Y | Y |
| Observations | 3944 | 3944 | 3944 | 3944 | 3944 | 3944 |
| Pseudo-$R^2$ | 0.216 | 0.218 | 0.219 | 0.218 | 0.218 | 0.220 |
| BIC | 9585.9 | 9595.9 | 9597.1 | 9603.6 | 9603.6 | 9609.8 |

Standard errors are clustered by region and period. * $p < 0.1$, ** $p < 0.05$, *** $p < 0.01$

***Table S19***. *Logistic regression model explaining exits from existing areas of specializations, interacting measures of spatial proximity with period fixed-effects.*

| | \multicolumn{6}{c}{Dependent Variable: $Entry_{ik,t}$} | | | | | |
|---|---|---|---|---|---|---|
| | (1) | (2) | (3) | (4) | (5) | (6) |
| $M_{ik,t-1}^{immigrants}$ | | -0.611*** | -0.592*** | -0.611*** | -0.594*** | -0.596*** |
| | | (0.094) | (0.113) | (0.133) | (0.115) | (0.171) |
| $M_{ik,t-1}^{emigrants}$ | | 0.241 | 0.246 | 0.240 | 0.249 | 0.319 |
| | | (0.325) | (0.326) | (0.328) | (0.470) | (0.390) |
| $\omega_{ik,t-1}^{immigrants}$ | | | -0.084*** | | | -0.085*** |
| | | | (0.013) | | | (0.028) |
| $\omega_{ik,t-1}^{emigrants}$ | | | | -0.0008 | | 0.055 |
| | | | | (0.018) | | (0.041) |
| $\omega_{ik,t-1}^{births}$ | | | | | -0.041** | -0.061 |
| | | | | | (0.020) | (0.049) |
| $ubiquity_{k,t-1}$ | -0.076*** | -0.079*** | -0.078*** | -0.079*** | -0.078*** | -0.081*** |
| | (0.014) | (0.016) | (0.016) | (0.015) | (0.015) | (0.015) |
| $\rho_{ik,t-1}^{M}$ | 6.551** | 7.007** | 6.991** | 7.001** | 7.025*** | 7.446*** |
| | (2.480) | (2.685) | (2.863) | (2.785) | (2.667) | (2.705) |
| $\rho_{ik,t-1}^{\omega}$ | 0.394*** | 0.433*** | 0.463*** | 0.432*** | 0.434*** | 0.482*** |
| | (0.074) | (0.078) | (0.091) | (0.076) | (0.084) | (0.098) |
| $R_{ik,t-1}^{births}$ | 0.032 | 0.032 | 0.015 | 0.032 | 0.027 | 0.018 |
| | (0.026) | (0.041) | (0.023) | (0.024) | (0.027) | (0.021) |
| FE: Broad categ.-region-period | Y | Y | Y | Y | Y | Y |
| FE: Category-period | Y | Y | Y | Y | Y | Y |
| $\rho_{ik,t-1}^{M}$ * period | Y | Y | Y | Y | Y | Y |
| $\rho_{ik,t-1}^{\omega}$ * period | Y | Y | Y | Y | Y | Y |
| Observations | 1051 | 1051 | 1051 | 1051 | 1051 | 1051 |
| Pseudo-$R^2$ | 0.240 | 0.248 | 0.256 | 0.248 | 0.249 | 0.258 |
| BIC | 3644.2 | 3647.6 | 3642.1 | 3654.5 | 3653.2 | 3654.0 |

Standard errors are clustered by region and period. * $p < 0.1$, ** $p < 0.05$, *** $p < 0.01$

### 3.4.4. Excluding the 20th century, and exploring heterogeneous effects across time

We established in Table S2 that our dataset is unbalanced with respect to the different centuries. The majority of famous individuals in our dataset are born in the 19th and 20th century. Hence, observations of relatedness densities and entries or exits are also unbalanced. To check the robustness of our results, we run the logistic regression models excluding the 20th century. Table S20 shows the results explaining entries to new activities, Table S21 for exits of activities. For both entries and exits we find that our main results remain unchanged. That is, $M_{ik,t-1}^{immi}$ and $\omega_{ik,t-1}^{immi}$ correlate positively with future entries and negatively with future exits.

Also, we provide the results for including only the 20th century in Table S22 (entries) and Table S23 (exits). Comparing the coefficients for the 11th to 19th century and the 20th century, we find that the related knowledge of immigrants plays a more significant role in the 20th century than before. Spillovers of migrants within the same activity, however, play a smaller role in the 20th century. Exploring differences across time periods further may be an interesting avenue for future research, since the cost of migration changed substantially over the past centuries.

*Table S20. Logistic regression model explaining entries to new activities, subsample for 11th to 19th century.*

|  | Dependent Variable: $Entry_{ik,t}$ | | | | |
|---|---|---|---|---|---|
|  | (1) | (2) | (3) | (4) | (5) |
| $M_{ik,t-1}^{immi}$ | 0.487** | 0.461* | 0.461** | 0.461** | 0.447* |
|  | (0.168) | (0.197) | (0.175) | (0.181) | (0.230) |
| $M_{ik,t-1}^{emi}$ | -0.064 | -0.092 | -0.160 | -0.188 | -0.389 |
|  | (0.395) | (0.398) | (0.461) | (0.505) | (0.886) |
| $\omega_{ik,t-1}^{immi}$ | 0.009** | 0.015* | 0.014*** | 0.018** | 0.026* |
|  | (0.002) | (0.007) | (0.005) | (0.008) | (0.015) |
| $\omega_{ik,t-1}^{emi}$ | 0.0006 | -0.012 | -0.010 | -0.019** | -0.023 |
|  | (0.004) | (0.008) | (0.007) | (0.009) | (0.022) |
| $\omega_{ik,t-1}^{births}$ | 0.008 | 0.016 | 0.019** | 0.028*** | 0.022 |
|  | (0.009) | (0.009) | (0.008) | (0.010) | (0.019) |
| $diversity_{i,t-1}$ | -0.034 | -0.146** | -0.164*** |  |  |
|  | (0.034) | (0.039) | (0.050) |  |  |
| $ubiquity_{k,t-1}$ | -0.006 | 0.001 | 0.003 | 0.002 | 0.004 |
|  | (0.003) | (0.004) | (0.011) | (0.012) | (0.019) |
| $\rho_{ik,t-1}^{M}$ | 1.125* | 0.901* | 0.564 | 0.637 | 0.083 |
|  | (0.437) | (0.373) | (0.737) | (0.761) | (1.212) |
| $\rho_{ik,t-1}^{\omega}$ | -0.010 | -0.032** | -0.017 | -0.018 | 0.036 |
|  | (0.019) | (0.007) | (0.034) | (0.039) | (0.068) |
| $R_{ik,t-1}^{births}$ | 0.585 | 0.546 | 0.773** | 0.802** | 0.719 |
|  | (0.448) | (0.550) | (0.342) | (0.376) | (0.744) |
| $\log(pop_{i,t})$ | -0.024 | 0.233 | 0.238 |  |  |
|  | (0.012) | (0.259) | (0.230) |  |  |
| FE: period | Y | Y | Y |  |  |
| FE: region |  | Y | Y |  |  |
| FE: category |  |  | Y | Y |  |
| FE: period-region |  |  |  | Y |  |
| FE: region-period-broad category |  |  |  |  | Y |
| FE: category-period |  |  |  |  | Y |
| Observations | 1397 | 1386 | 1386 | 1382 | 883 |
| Pseudo-R² | 0.028 | 0.058 | 0.078 | 0.089 | 0.174 |
| BIC | 1803.6 | 2126.3 | 2215.5 | 2409.6 | 2601.6 |

Standard errors are clustered by region. * p < 0.1, ** p < 0.05, *** p < 0.01

**Table S21**. *Logistic regression model explaining exits of activities, subsample for 11$^{th}$ to 19$^{th}$ century.*

|  | Dependent Variable: $Exit_{ik,t}$ | | | | |
|---|---|---|---|---|---|
|  | (1) | (2) | (3) | (4) | (5) |
| $M^{immi}_{ik,t-1}$ | -0.431** | -0.448** | -0.445** | -0.548** | -0.635 |
|  | (0.169) | (0.191) | (0.213) | (0.223) | (0.689) |
| $M^{emi}_{ik,t-1}$ | -0.233 | -0.220 | -0.115 | -0.225 | -0.284 |
|  | (0.253) | (0.287) | (0.276) | (0.299) | (0.791) |
| $\omega^{immi}_{ik,t-1}$ | -0.431** | -0.448** | -0.445** | -0.548** | -0.635 |
|  | (0.008) | (0.012) | (0.012) | (0.017) | (0.038) |
| $\omega^{emi}_{ik,t-1}$ | 0.010 | 0.015 | 0.015 | -0.005 | -0.076 |
|  | (0.008) | (0.015) | (0.015) | (0.022) | (0.082) |
| $\omega^{births}_{ik,t-1}$ | -0.016 | -0.028 | -0.027 | -0.031 | 0.018 |
|  | (0.013) | (0.018) | (0.020) | (0.027) | (0.077) |
| $diversity_{i,t-1}$ | 0.070 | 0.336** | 0.310** |  |  |
|  | (0.066) | (0.138) | (0.138) |  |  |
| $ubiquity_{k,t-1}$ | 0.007 | 0.011 | 0.010 | 0.024 | 0.047 |
|  | (0.018) | (0.021) | (0.026) | (0.030) | (0.064) |
| $\rho^{M}_{ik,t-1}$ | 0.659 | 0.423 | 0.464 | 0.151 | 1.074 |
|  | (1.387) | (1.564) | (1.717) | (1.831) | (3.816) |
| $\rho^{\omega}_{ik,t-1}$ | -0.025 | -0.050 | -0.047 | -0.056 | -0.133 |
|  | (0.029) | (0.040) | (0.047) | (0.066) | (0.141) |
| $R^{births}_{ik,t-1}$ | 0.001 | -0.005 | -0.012 | -0.002 | -0.023 |
|  | (0.037) | (0.057) | (0.063) | (0.067) | (0.119) |
| $\log(pop_{i,t})$ | -0.141 | -1.478** | -1.513*** |  |  |
|  | (0.107) | (0.572) | (0.531) |  |  |
| FE: period | Y | Y | Y |  |  |
| FE: region |  | Y | Y |  |  |
| FE: category |  |  | Y | Y |  |
| FE: period-region |  |  |  | Y |  |
| FE: region-period-broad category |  |  |  |  | Y |
| FE: category-period |  |  |  |  | Y |
| Observations | 556 | 556 | 550 | 522 | 253 |
| Pseudo-R$^2$ | 0.033 | 0.094 | 0.111 | 0.137 | 0.266 |
| BIC | 846.3 | 1140.5 | 1213.8 | 1281.5 | 960.2 |

Standard errors are clustered by region. * $p < 0.1$, ** $p < 0.05$, *** $p < 0.01$

**Table S22.** *Logistic regression model explaining entries to activities, subsample for 20*[th] *century.*

| | Dependent Variable: $Entry_{ik,t}$ | | | |
|---|---|---|---|---|
| | (1) | (2) | (3) | (4) |
| $M_{ik,t-1}^{immi}$ | 0.232** | 0.237** | 0.288** | 0.257 |
| | (0.103) | (0.111) | (0.113) | (0.158) |
| $M_{ik,t-1}^{emi}$ | 0.248 | 0.188 | -0.042 | 0.139 |
| | (0.267) | (0.312) | (0.326) | (0.423) |
| $\omega_{ik,t-1}^{immi}$ | 0.016*** | 0.020* | 0.027** | 0.037** |
| | (0.004) | (0.011) | (0.012) | (0.017) |
| $\omega_{ik,t-1}^{emi}$ | 0.010 | -0.046 | -0.044 | -0.023 |
| | (0.013) | (0.030) | (0.032) | (0.049) |
| $\omega_{ik,t-1}^{births}$ | 0.012 | 0.046 | 0.045 | 0.025 |
| | (0.017) | (0.029) | (0.032) | (0.046) |
| $diversity_{i,t-1}$ | -0.026 | | | |
| | (0.026) | | | |
| $ubiquity_{k,t-1}$ | -0.006 | -0.013** | -0.003 | 0.001 |
| | (0.005) | (0.006) | (0.006) | (0.008) |
| $\rho_{ik,t-1}^{M}$ | 1.375 | 2.380** | 1.491 | 0.863 |
| | (1.050) | (1.178) | (1.221) | (1.714) |
| $\rho_{ik,t-1}^{\omega}$ | 0.113*** | 0.180*** | 0.113** | 0.122* |
| | (0.033) | (0.043) | (0.049) | (0.065) |
| $R_{ik,t-1}^{births}$ | 0.300 | 0.314 | 0.498* | 0.246 |
| | (0.226) | (0.260) | (0.276) | (0.372) |
| $\log(pop_{i,t})$ | 0.206*** | | | |
| | (0.042) | | | |
| FE: region | | Y | Y | |
| FE: category | | | Y | Y |
| FE: region-broad category | | | | Y |
| Observations | 4783 | 4783 | 4783 | 3061 |
| Pseudo-R² | 0.072 | 0.100 | 0.148 | 0.213 |
| BIC | 4011.4 | 4953.5 | 4946.0 | 6534.4 |

Standard errors are clustered by region. * p < 0.1, ** p < 0.05, *** p < 0.01

**Table S23.** *Logistic regression model explaining exits of activities, subsample for 20*[th] *century.*

| | Dependent Variable: $Exit_{ik,t}$ | | | |
|---|---|---|---|---|
| | (1) | (2) | (3) | (4) |
| $M_{ik,t-1}^{immi}$ | -0.487*** | -0.511*** | -0.540*** | -0.625** |
| | (0.145) | (0.160) | (0.172) | (0.297) |
| $M_{ik,t-1}^{emi}$ | 0.277 | 0.270 | 0.347 | 0.465 |
| | (0.202) | (0.228) | (0.249) | (0.418) |
| $\omega_{ik,t-1}^{immi}$ | -0.012** | -0.051*** | -0.058** | -0.098** |
| | (0.006) | (0.018) | (0.023) | (0.046) |
| $\omega_{ik,t-1}^{emi}$ | 0.011 | 0.034 | 0.020 | 0.079 |
| | (0.016) | (0.031) | (0.033) | (0.063) |
| $\omega_{ik,t-1}^{births}$ | 0.019 | -0.038 | -0.027 | -0.104 |
| | (0.018) | (0.037) | (0.039) | (0.075) |
| $diversity_{i,t-1}$ | -0.082** | | | |
| | (0.041) | | | |
| $ubiquity_{k,t-1}$ | -0.014* | -0.030*** | -0.043*** | -0.091*** |
| | (0.007) | (0.008) | (0.010) | (0.023) |
| $\rho_{ik,t-1}^{M}$ | 0.390 | 1.931 | 1.855 | 9.130* |
| | (1.510) | (1.931) | (2.169) | (4.940) |
| $\rho_{ik,t-1}^{\omega}$ | 0.021 | 0.184*** | 0.256*** | 0.555*** |
| | (0.048) | (0.060) | (0.079) | (0.165) |
| $R_{ik,t-1}^{births}$ | -0.002 | 0.000 | -0.012 | 0.033 |
| | (0.016) | (0.021) | (0.024) | (0.044) |
| $\log(pop_{i,t})$ | -0.240** | | | |
| | (0.093) | | | |
| FE: region | | Y | Y | |
| FE: category | | | Y | Y |
| FE: region-broad category | | | | Y |
| Observations | 1489 | 1467 | 1461 | 798 |
| Pseudo-R² | 0.065 | 0.131 | 0.181 | 0.233 |
| BIC | 1824.5 | 2575.8 | 2640.6 | 2457.8 |

Standard errors are clustered by region. * p < 0.1, ** p < 0.05, *** p < 0.01

### 3.4.5. Redefining entries and exits

In the main results, we define entries based on births in a location in the coming century. That is, $Entry_{ik,t} = 1$ if $M_{ik,t-1}^{births} = 0$ and $M_{ik,t}^{births} = 1$ (see Eq. S2). Similarly, we defined $Exit_{ik,t} = 1$ if $M_{ik,t-1}^{births} = 1$ and $M_{ik,t}^{births} = 0$

To check the robustness of our results, we apply a different definition of entries and exits. Instead of defining the entry to a new activity by developing a new specialization, we can define entry as a location exhibiting births of famous individuals with a certain occupation for the first time. Specifically, let us refer to this definition of entry as $Entry2_{ik,t}$ and let it be defined as $Entry2_{ik,t} = 1$ if $N_{ik,t-1}^{births} = 0$ and $N_{ik,t}^{births} > 0$. Similarly, we can define exits as $Exit2_{ik,t} = 1$ if $N_{ik,t-1}^{births} > 0$ and $N_{ik,t}^{births} = 0$.

Table S24 shows the results of the logistic regression model for this definition of entry, for different specifications of fixed effects. As can be seen, the results are robust, that is, $M_{ik,t-1}^{immi}$ and $\omega_{ik,t-1}^{immi}$ correlate positively with future entries. The same holds for redefining exits as last births of famous individuals with a certain activity (

Table S25). While $M_{ik,t-1}^{immi}$ correlates significantly for all fixed effects, the diffusion of related knowledge is – in contrast to the main findings defining entries and exits as gaining or losing a specialization – not significant for the most restrictive specifications.

***Table S24***. *Regression results explaining entries to new activities, redefining entries as first births of famous individuals with occupation k in location i*

|  | Dependent Variable: $Entry2_{ik,t}$ | | | | |
|---|---|---|---|---|---|
|  | (1) | (2) | (3) | (4) | (5) |
| $M^{immi}_{ik,t-1}$ | 0.268** | 0.271** | 0.308*** | 0.346*** | 0.376** |
|  | (0.125) | (0.125) | (0.099) | (0.118) | (0.154) |
| $\omega^{immi}_{ik,t-1}$ | 0.021*** | 0.021** | 0.024*** | 0.028*** | 0.011 |
|  | (0.004) | (0.006) | (0.004) | (0.005) | (0.009) |
| $\omega^{emi}_{ik,t-1}$ | 0.015*** | 0.006 | 0.004 | -0.012 | 0.011 |
|  | (0.002) | (0.007) | (0.007) | (0.023) | (0.029) |
| $\omega^{births}_{ik,t-1}$ | -0.009 | 0.002 | 0.0002 | 0.008 | -0.017 |
|  | (0.008) | (0.006) | (0.006) | (0.010) | (0.015) |
| $diversity_{i,t-1}$ | 0.005 | -0.077** | -0.059 |  |  |
|  | (0.023) | (0.035) | (0.036) |  |  |
| $ubiquity_{k,t-1}$ | 0.003 | 0.004 | 0.00001 | -0.001 | -0.010 |
|  | (0.004) | (0.006) | (0.006) | (0.007) | (0.009) |
| $\rho^{M}_{ik,t-1}$ | 2.175*** | 2.320*** | 2.830** | 3.065*** | 3.949** |
|  | (0.410) | (0.395) | (1.109) | (1.165) | (1.751) |
| $\rho^{\omega}_{ik,t-1}$ | 0.047 | 0.041 | 0.075** | 0.090** | 0.228** |
|  | (0.043) | (0.067) | (0.036) | (0.040) | (0.065) |
| $\log(pop_{i,t})$ | 0.279*** | 0.197 | 0.237 |  |  |
|  | (0.069) | (0.166) | (0.199) |  |  |
| FE: period | Y | Y | Y |  |  |
| FE: region |  | Y | Y |  |  |
| FE: category |  |  | Y | Y |  |
| FE: period-region |  |  |  | Y |  |
| FE: region-period-broad category |  |  |  |  | Y |
| FE: category-period |  |  |  |  | Y |
| Observations | 5569 | 5539 | 5539 | 5487 | 3536 |
| Pseudo-R2 | 0.168 | 0.225 | 0.317 | 0.320 | 0.384 |
| BIC | 3181.3 | 3998.6 | 3805.8 | 4215.2 | 4630.0 |

* p < 0.1, ** p < 0.05, *** p < 0.01

***Table S25***. *Regression results explaining exits to new activities, redefining exits as last births of famous individuals with occupation k in location i*

|  | Dependent Variable: $Exit2_{ik,t}$ | | | | |
|---|---|---|---|---|---|
|  | (1) | (2) | (3) | (4) | (5) |
| $M^{immi}_{ik,t-1}$ | -0.360*** | -0.380*** | -0.416*** | -0.459*** | -0.336** |
|  | (0.061) | (0.066) | (0.086) | (0.093) | (0.141) |
| $M^{emi}_{ik,t-1}$ | 0.216 | 0.147 | 0.074 | 0.083 | 0.367 |
|  | (0.133) | (0.119) | (0.180) | (0.202) | (0.305) |
| $\omega^{immi}_{ik,t-1}$ | -0.017** | -0.026*** | -0.025*** | -0.025 | -0.058 |
|  | (0.007) | (0.010) | (0.006) | (0.017) | (0.037) |
| $\omega^{emi}_{ik,t-1}$ | 0.010 | 0.015* | 0.012* | 0.023 | 0.029 |
|  | (0.009) | (0.008) | (0.007) | (0.016) | (0.024) |
| $\omega^{births}_{ik,t-1}$ | 0.007 | -0.016** | -0.018*** | -0.037*** | 0.010 |
|  | (0.010) | (0.007) | (0.007) | (0.012) | (0.046) |
| $diversity_{i,t-1}$ | -0.067 | 0.071* | 0.082* |  |  |
|  | (0.038) | (0.031) | (0.033) |  |  |
| $ubiquity_{k,t-1}$ | -0.015*** | -0.019** | -0.030** | -0.029** | -0.042** |
|  | (0.003) | (0.006) | (0.008) | (0.010) | (0.013) |
| $\rho^{M}_{ik,t-1}$ | -1.866** | -2.428** | -1.373** | -1.919** | -3.378 |
|  | (0.586) | (1.019) | (0.590) | (0.601) | (2.349) |
| $\rho^{\omega}_{ik,t-1}$ | 0.041 | 0.113* | 0.096 | 0.098 | 0.060 |
|  | (0.042) | (0.059) | (0.072) | (0.094) | (0.169) |
| $\log(pop_{i,t})$ | -0.358*** | -0.211 | -0.275 |  |  |
|  | (0.036) | (0.345) | (0.360) |  |  |
| FE: period | Y | Y | Y |  |  |
| FE: region |  | Y | Y |  |  |
| FE: category |  |  | Y | Y |  |
| FE: period-region |  |  |  | Y |  |
| FE: region-period-broad category |  |  |  |  | Y |
| FE: category-period |  |  |  |  | Y |
| Observations | 2656 | 2656 | 2628 | 2521 | 1488 |
| Pseudo-R² | 0.168 | 0.225 | 0.317 | 0.320 | 0.384 |
| BIC | 3181.3 | 3998.6 | 3805.8 | 4215.2 | 4630.0 |

* p < 0.1, ** p < 0.05, *** p < 0.01

### 3.4.6. Interaction terms

Following the literature on the role of migration in unrelated diversification (Elekes et al., 2019; Miguelez & Morrison, 2022; Neffke et al., 2018), we add interaction terms between various relatedness densities to the main specification of column 6 of Table S5. For example, a significantly negative interaction term between $\omega_{ik}^{immi}$ and $\omega_{ik}^{births}$ would indicate that the related knowledge of immigrants and those of individuals born in a location are substitutes to each other. Put differently, if $\omega_{ik}^{births}$ is high, the correlation of $\omega_{ik}^{immi}$ with the probability of entry decreases. As Table S26 shows, the interaction term between $\omega_{ik}^{immi}$ and $\omega_{ik}^{births}$ is indeed significantly negative across all fixed-effects specifications. However, quantitatively, the coefficient is very small compared to the overall coefficient $\omega_{ik}^{immi}$. Thus, we cannot conclude that migration contributes substantially to unrelated diversification.

*Table S26. Regression results explaining entries to new activities, including interaction terms*

| | (1) | (2) | (3) | (4) | (5) | (6) | (7) | (8) | (9) |
|---|---|---|---|---|---|---|---|---|---|
| | \multicolumn{9}{c}{Dependent Variable: $Entry_{ik,t}$} | | | | | | | | |
| $M_{ik,t-1}^{immi}$ | 0.320*** | 0.319*** | 0.319*** | 0.317*** | 0.316*** | 0.316*** | 0.330*** | 0.329*** | 0.330*** |
| | (0.077) | (0.077) | (0.076) | (0.074) | (0.074) | (0.075) | (0.064) | (0.063) | (0.066) |
| $M_{ik,t-1}^{emi}$ | 0.155 | 0.137 | 0.118 | 0.186 | 0.168 | 0.152 | -0.012 | -0.024 | -0.035 |
| | (0.163) | (0.164) | (0.164) | (0.210) | (0.204) | (0.212) | (0.222) | (0.224) | (0.228) |
| $\omega_{ik,t-1}^{immi}$ | 0.021*** | 0.030*** | 0.011*** | 0.025*** | 0.045*** | 0.010*** | 0.024*** | 0.045*** | 0.011*** |
| | (0.003) | (0.005) | (0.003) | (0.008) | (0.010) | (0.003) | (0.007) | (0.010) | (0.003) |
| $\omega_{ik,t-1}^{emi}$ | 0.013*** | 0.001 | 0.015 | 0.009 | -0.012*** | 0.007 | 0.007* | -0.012*** | 0.003 |
| | (0.005) | (0.004) | (0.009) | (0.007) | (0.004) | (0.014) | (0.003) | (0.005) | (0.009) |
| $\omega_{ik,t-1}^{births}$ | 0.003 | 0.024*** | 0.019 | 0.011 | 0.050*** | 0.029 | 0.009 | 0.047*** | 0.023* |
| | (0.005) | (0.004) | (0.012) | (0.007) | (0.013) | (0.020) | (0.006) | (0.010) | (0.012) |
| $\omega_{ik,t-1}^{immi} * \omega_{ik,t-1}^{emi}$ | -0.000*** | | | -0.000* | | | -0.000** | | |
| | (0.000) | | | (0.000) | | | (0.000) | | |
| $\omega_{ik,t-1}^{immi} * \omega_{ik,t-1}^{births}$ | | -0.001*** | | | -0.001*** | | | -0.001*** | |
| | | (0.000) | | | (0.000) | | | (0.000) | |
| $\omega_{ik,t-1}^{emi} * \omega_{ik,t-1}^{births}$ | | | 0.000 | | | 0.000 | | | 0.000 |
| | | | (0.000) | | | (0.000) | | | (0.000) |
| $diversity_{i,t-1}$ | 0.008 | 0.004 | 0.003 | -0.030 | -0.025 | -0.032 | -0.022 | -0.017 | -0.024 |
| | (0.013) | (0.012) | (0.018) | (0.021) | (0.024) | (0.022) | (0.022) | (0.025) | (0.022) |
| $ubiquity_{k,t-1}$ | 0.006** | 0.006** | 0.007** | 0.008** | 0.007** | 0.008** | 0.006*** | 0.006*** | 0.007*** |
| | (0.003) | (0.003) | (0.003) | (0.004) | (0.004) | (0.004) | (0.001) | (0.001) | (0.001) |
| $\rho_{ik,t-1}^{M}$ | 0.232 | 0.163 | 0.096 | 0.302 | 0.226*** | 0.163*** | 0.621*** | 0.545*** | 0.514*** |
| | (0.340) | (0.292) | (0.220) | (0.193) | (0.086) | (0.054) | (0.146) | (0.053) | (0.146) |
| $\rho_{ik,t-1}^{\omega}$ | 0.026 | 0.027 | 0.028 | 0.013 | 0.012 | 0.016 | 0.016 | 0.015 | 0.018 |
| | (0.033) | (0.033) | (0.032) | (0.041) | (0.040) | (0.041) | (0.024) | (0.023) | (0.024) |
| $R_{ik,t-1}^{births}$ | 0.255* | 0.288* | 0.275* | 0.222 | 0.259 | 0.234 | 0.531*** | 0.557*** | 0.530*** |
| | (0.152) | (0.155) | (0.147) | (0.169) | (0.169) | (0.170) | (0.185) | (0.188) | (0.184) |
| $\log(pop_{i,t})$ | 0.138* | 0.144** | 0.143* | 0.249*** | 0.222*** | 0.265*** | 0.242*** | 0.218** | 0.255*** |
| | (0.072) | (0.070) | (0.078) | (0.068) | (0.081) | (0.079) | (0.081) | (0.089) | (0.087) |
| FE: period | Y | Y | Y | Y | Y | Y | Y | Y | Y |
| FE: region | | | | Y | Y | Y | Y | Y | Y |
| FE: occu. category | | | | | | | Y | Y | Y |
| Observations | 6180 | 6180 | 6180 | 6180 | 6180 | 6180 | 6180 | 6180 | 6180 |
| Pseudo-$R^2$ | 0.076 | 0.077 | 0.076 | 0.097 | 0.099 | 0.097 | 0.125 | 0.127 | 0.125 |
| AIC | 5671.9 | 5666.3 | 5670.7 | 5802.4 | 5789.8 | 5802.3 | 5678.6 | 5666.8 | 5679.5 |
| BIC | 5793.0 | 5787.5 | 5791.8 | 6798.3 | 6785.7 | 6798.2 | 6836.0 | 6824.2 | 6836.9 |

Standard errors are clustered by period and region. * $p < 0.1$, ** $p < 0.05$, *** $p < 0.01$

### 3.4.7. Heterogenous effects across activities

The effects of knowledge spillovers may differ across activities. The emergence of artists or scientists may be more demand-driven than other categories, since it is well known that wealthy patrons supported the arts and sciences. The same may apply to sports persons. The emergence of politicians may be less related to the presence of artists or sports persons but is affected by political decisions. Businessmen may profit from spillovers across the sciences and institutions.

To explore these heterogenous effects, we split our dataset into six different, highly aggregated occupational categories:

(1) "Arts",

(2) "Humanities",

(3) "Sciences",

(4) "Business & Technology" (this category includes the category 'Business & Law' as well as engineers and inventors),

(5) "Sports", and

(6) "Public Figures & Institutions".

This distinction loosely follows the full taxonomy (Table S1).

For each of these categories, we run logistic regression models for entries and exits to estimate whether the role of immigrants, emigrants and locals differs across these categories.

The results are shown in Table S27 and Table S28**Fehler! Verweisquelle konnte nicht gefunden werden.**.

For entries (Table S27), we find that the knowledge of immigrants in the same activity ($M_{ik,t-1}^{immi}$) correlates with future entries in "Sciences" as well as for "Public Figures & Institutions". That is, the probability of entering e.g. physics or politics increases with a disproportionate inflow of physicists or politicians. The related knowledge of immigrants ($\omega_{ik,t-1}^{immi}$) correlates positively with entries in "Sports" and "Humanities". That is, the emergence of sportsmen, writers or philosophers correlates with the immigration of individuals in related activities.

For exits (Table S28), we find that the probability of exit decreases with the knowledge of immigrants in the same activity ($M_{ik,t-1}^{immi}$) in "Arts" and "Public Figures & Institutions". A larger than expected inflow of e.g. painters decreases the probability of exiting painting. The related knowledge of immigrants ($\omega_{ik,t-1}^{immi}$) correlates negatively with exits in "Business &

Technology". The immigration of individuals with related activities decreases the probability of exit occupations in business or technology.

These models help understand the mechanisms behind our findings. The number of observations, however, is more limited in these models than in our full model and, hence, we cannot be as restrictive with fixed effects. More comprehensive data would be required to analyze the heterogeneity across different fields in more detail and to provide more robust evidence on the specific mechanisms. Exploring the heterogeneity across disciplines further may be an interesting avenue for future research.

*Table S27. Heterogeneous effects for entries across occupation categories*

| | Dependent Variable: $Entry_{ik,t}$ | | | | | |
|---|---|---|---|---|---|---|
| | Arts | Humanities | Sciences | Business & Technology | Sports | Public Figures & Institutions |
| | (1) | (2) | (3) | (4) | (5) | (6) |
| $M^{immi}_{ik,t-1}$ | 0.077 | 0.304 | 0.438** | -0.665 | -0.158 | 0.576*** |
| | (0.151) | (0.393) | (0.125) | (0.545) | (0.563) | (0.094) |
| $M^{emi}_{ik,t-1}$ | 0.035 | -0.343 | -0.851 | -1.227 | | 0.625 |
| | (0.714) | (0.543) | (0.491) | (1.652) | | (0.770) |
| $\omega^{immi}_{ik,t-1}$ | 0.008 | 0.031** | 0.015 | -0.020 | 0.163*** | 0.020 |
| | (0.019) | (0.012) | (0.012) | (0.019) | (0.060) | (0.014) |
| $\omega^{emi}_{ik,t-1}$ | -0.004 | -0.011 | 0.004 | -0.039 | 0.043 | -0.009 |
| | (0.031) | (0.020) | (0.022) | (0.029) | (0.094) | (0.022) |
| $\omega^{births}_{ik,t-1}$ | 0.038 | 0.005 | 0.007 | -0.015 | -0.061 | 0.016 |
| | (0.025) | (0.025) | (0.033) | (0.021) | (0.103) | (0.032) |
| $diversity_{i,t-1}$ | -0.058 | -0.051 | -0.011 | 0.188 | | -0.128 |
| | (0.050) | (0.079) | (0.065) | (0.103) | | (0.084) |
| $ubiquity_{k,t-1}$ | -0.001 | 0.009 | 0.024** | 0.173*** | 0.041 | 0.007 |
| | (0.009) | (0.009) | (0.006) | (0.027) | (0.036) | (0.008) |
| $\rho^{M}_{ik,t-1}$ | -0.654 | 3.744** | -1.147 | 2.131 | -2.384 | 1.170 |
| | (1.592) | (1.641) | (1.067) | (2.097) | (2.142) | (1.717) |
| $\rho^{\omega}_{ik,t-1}$ | 0.131 | -0.198*** | 0.068 | -0.446** | -0.082 | 0.034 |
| | (0.082) | (0.068) | (0.044) | (0.102) | (0.238) | (0.033) |
| $R^{births}_{ik,t-1}$ | 0.188 | 0.134 | 0.383 | 1.625 | 2.569 | 1.251*** |
| | (0.297) | (0.842) | (0.193) | (2.136) | (2.993) | (0.286) |
| $\log(pop_{i,t})$ | 0.277 | 0.340 | 0.808** | 0.196 | | 0.700 |
| | (0.191) | (0.227) | (0.245) | (0.948) | | (0.421) |
| FE: period | Y | Y | Y | Y | Y | Y |
| FE: region | Y | Y | Y | Y | Y | Y |
| FE: category | Y | Y | Y | Y | Y | Y |
| Num.Obs. | 1502 | 515 | 1094 | 288 | 521 | 939 |
| Pseudo-R2 | 0.194 | 0.199 | 0.179 | 0.248 | 0.210 | 0.246 |
| BIC | 2316.1 | 1146.5 | 1807.0 | 665.1 | 1044.0 | 1581.4 |

Standard errors are clustered by period and region. * $p < 0.1$, ** $p < 0.05$, *** $p < 0.01$

*Table S28. Heterogeneous effects for exits across occupation categories*

|  | Dependent Variable: $Exit_{ik,t}$ | | | | | |
| --- | --- | --- | --- | --- | --- | --- |
|  | Arts | Humanities | Sciences | Business & Technology | Sports | Public Figures & Institutions |
|  | (1) | (2) | (3) | (4) | (5) | (6) |
| $M^{immi}_{ik,t-1}$ | -1.348** | -0.929 | 0.031 | -3.827 | -1.335 | -0.218* |
|  | (0.426) | (0.762) | (0.211) | (3.747) | (1.139) | (0.083) |
| $M^{emi}_{ik,t-1}$ | -0.214 | -1.115 | 0.211 | 4.889* |  | -0.533 |
|  | (0.243) | (0.816) | (0.243) | (2.349) |  | (0.615) |
| $\omega^{immi}_{ik,t-1}$ | -0.026 | 0.071 | -0.007 | -0.811** | -0.610 | -0.006 |
|  | (0.023) | (0.046) | (0.020) | (0.363) | (0.455) | (0.011) |
| $\omega^{emi}_{ik,t-1}$ | -0.031 | -0.023 | 0.019 | -0.622 |  | 0.002 |
|  | (0.015) | (0.036) | (0.049) | (0.479) |  | (0.031) |
| $\omega^{births}_{ik,t-1}$ | -0.017 | 0.031 | -0.028 | 0.486 | 0.406 | 0.042 |
|  | (0.047) | (0.037) | (0.057) | (0.555) | (0.291) | (0.023) |
| $diversity_{i,t-1}$ | 0.231* | -0.458* | -0.084 | -2.513** |  | -0.233 |
|  | (0.085) | (0.259) | (0.121) | (1.060) |  | (0.133) |
| $ubiquity_{k,t-1}$ | -0.030 | -0.052*** | -0.027 | -0.102 | -0.886** | -0.014 |
|  | (0.036) | (0.018) | (0.017) | (0.300) | (0.341) | (0.011) |
| $\rho^M_{ik,t-1}$ | 2.116 | 4.253 | 3.286 | 68.536 | 70.545 | 0.622 |
|  | (7.694) | (4.700) | (1.690) | (45.770) | (48.115) | (2.932) |
| $\rho^\omega_{ik,t-1}$ | 0.203 | 0.083 | -0.052 | 0.508 | 4.863* | -0.285*** |
|  | (0.169) | (0.187) | (0.151) | (2.412) | (2.386) | (0.043) |
| $R^{births}_{ik,t-1}$ | 0.017 | -0.089 | 0.005 | 0.118 | 0.490** | -0.212* |
|  | (0.045) | (0.088) | (0.030) | (0.195) | (0.204) | (0.084) |
| $\log(pop_{i,t})$ | -0.505 | 2.274 | -0.891 | 2.534 |  | -0.772 |
|  | (0.646) | (1.643) | (0.766) | (6.673) |  | (0.398) |
| FE: period | Y | Y | Y | Y | Y | Y |
| FE: region | Y | Y | Y | Y | Y | Y |
| FE: category | Y | Y | Y |  | Y | Y |
| Num.Obs. | 360 | 159 | 435 | 63 | 46 | 323 |
| Pseudo-R2 | 0.239 | 0.253 | 0.201 | 0.662 | 0.592 | 0.209 |
| BIC | 884.8 | 494.0 | 1003.9 | 170.3 | 117.7 | 862.0 |

Standard errors are clustered by period and region. We cannot include category fixed-effects in column (4), since the maximum likelihood estimator does not converge if they are included. * p < 0.1, ** p < 0.05, *** p < 0.01

### 3.4.8. Heterogenous effects across city size

It may be that the patterns we discover here vary for cities of different sizes. Hence, we explore this potential heterogeneity by splitting the sample into small cities (population levels by Buringh (2021) below the median of the respective century) and large cities (population levels above the median of the respective century).

Table S29 shows the results for entries and exits for both, small and large cities. Indeed, the correlational patterns we uncover in this study vary across sizes. The knowledge of immigrants in the same activity and in related activities is significantly more relevant for large cities with respect to both, entries and exits. In contrast, it can be seen that the knowledge of emigrants is a highly relevant factor in predicting future exits of activities for small cities.

*Table S29. Heterogeneous effects across city size for entries and exits*

|  | Dependent Variable: $Entry_{ik,t}$ | | Dependent Variable: $Exit_{ik,t}$ | |
|---|---|---|---|---|
|  | Small cities (1) | Large cities (2) | Small cities (3) | Large cities (4) |
| $M^{immi}_{ik,t-1}$ | 0.042 | 0.310*** | -1.065* | -0.576*** |
|  | (0.242) | (0.085) | (0.607) | (0.172) |
| $M^{emi}_{ik,t-1}$ | -0.581 | -0.019 | 2.519*** | 0.005 |
|  | (0.323) | (0.350) | (0.219) | (0.228) |
| $\omega^{immi}_{ik,t-1}$ | -0.003 | 0.030*** | -0.122 | -0.052*** |
|  | (0.038) | (0.006) | (0.089) | (0.008) |
| $\omega^{emi}_{ik,t-1}$ | 0.066 | -0.027 | 0.214 | -0.043 |
|  | (0.057) | (0.021) | (0.275) | (0.064) |
| $\omega^{births}_{ik,t-1}$ | -0.090 | 0.036** | -0.080 | -0.025 |
|  | (0.061) | (0.016) | (0.063) | (0.044) |
| $ubiquity_{k,t-1}$ | 0.008 | 0.006** | -0.072** | -0.051*** |
|  | (0.008) | (0.003) | (0.023) | (0.013) |
| $\rho^{M}_{ik,t-1}$ | -2.128 | -0.140 | 6.702*** | 5.876*** |
|  | (2.091) | (0.663) | (0.115) | (1.294) |
| $\rho^{\omega}_{ik,t-1}$ | 0.208** | 0.073* | 0.087 | 0.153 |
|  | (0.049) | (0.044) | (0.344) | (0.174) |
| $R^{births}_{ik,t-1}$ | 2.741*** | 0.096 | -0.071 | 0.018 |
|  | (0.550) | (0.241) | (0.054) | (0.035) |
| FE: region-period-broad category | Y | Y | Y | Y |
| FE: category-period | Y | Y | Y | Y |
| Num.Obs. | 906 | 2994 | 142 | 899 |
| Pseudo-R² | 0.254 | 0.215 | 0.285 | 0.246 |

The number of observations is not equal for small and large cities in spite of splitting the sample at the median of population levels, because smaller cities experience less entries of new and exits of existing activities than large cities do. Hence, many observations for small cities are be removed due to no variety within the fixed-effects structure. Standard errors are clustered by period and region.
* $p < 0.1$, ** $p < 0.05$, *** $p < 0.01$

### 3.4.9. Marginal effects after decomposing RCA values

Taking the ratio of the observed and expected number throughout the study, e.g. by considering the Revealed Comparative Advantage / Location Quotient, is first and foremost a control of size. This makes it possible to sensibly compare Paris and London with East Wales and Lower Austria. However, these models are also opaque, not telling us whether our results are driven by changes in the observed or expected number (or both).

In this chapter, we provide results for entries and exits while including all terms of the original ratio. That is, we include the terms $\sum_k N^{births}_{ik,t}$ and $\sum_i N^{births}_{ik,t}$ from the dependent variables $Entry_{ik,t}$ and $Entry_{ik,t}$. We further decompose the terms $M^{immi}_{ik,t-1}$ and $M^{emi}_{ik,t-1}$ into $N^{immi}_{ik,t-1}$, $\sum_k N^{immi}_{ik,t-1}$, $\sum_i N^{immi}_{ik,t-1}$ and $N^{emi}_{ik,t-1}$, $\sum_k N^{emi}_{ik,t-1}$, $\sum_i N^{emi}_{ik,t-1}$, respectively. The coefficients of the observed values, i.e. $N^{immi}_{ik,t-1}$ and $N^{emi}_{ik,t-1}$, can then more directly be interpreted as the marginal effects of one additional immigrant or emigrant with a specific occupation. To reduce skew, all these terms enter transformed using the inverted hyperbolic sine function (denoted by *asinh*). We are using this instead of a log transformation, because we have observations with zeros.

Table S30 and Table S31 show the results for entries and exits, respectively, for various fixed effects specifications. The number of immigrants with a specific occupation ($N^{immi}_{ik,t-1}$) correlates positively with future entries and negatively with future exits, confirming our main results with composite indices. Using columns (4), we can assess the average marginal effects of one additional immigrant to a region with a certain occupation. We calculate the average marginal effect by comparing the predicted values of the model in column (4) to the predicted values if $N_{ik,t-1}^{immi}$ was increased by 1 for each observation (before using the inverted hyperbolic sine transformation). We find that this average marginal effect amounts to 1.68 percentage points for entries and -5.04 percentage points for exits.

We can also interpret the results as elasticities. We can directly use the coefficients of the model in column (4) to find that a 1 percent increase in $N_{ik,t-1}^{immi}$ translates into an increase of the odds ratio to enter by 0.16 percent ($\exp(0.16 * 0.01)$) and a decrease of the odds ratio to exit by 0.413 percent ($\exp(-0.414 * 0.01)$). To calculate elasticities with respect to the probability of entry or exit, we calculate the average marginal effect by comparing the predicted values of the model in column (4) to the predicted values if $N_{ik,t-1}^{immi}$ was increased by 1% for each observation (before using the inverted hyperbolic sine transformation). We find that a 1% increase in $N_{ik,t-1}^{immi}$ increases the probability of entry by 0.0053% and reduces the probability of exit by 0.033%.

Table S32 and Table S33 show the results for our second definition of entries and exits using the first and last births of individuals with a certain activity in a region (see SM chapter 3.4.5). Using again columns (4), we find an average marginal effect of one additional immigrant in a region with a certain activity of 3.38 percentage points for entries and -4.23 percentage points for exits.

Also, in all these regressions the results for the related knowledge of immigrants, emigrants and locals remain virtually unchanged compared to our original results.

*Table S30. Logistic regression models explaining entries, decomposing RCA values*

| | Dependent Variable: $Entry_{ik,t}$ | | | | |
|---|---|---|---|---|---|
| | (1) | (2) | (3) | (4) | (5) |
| asinh ($N_{ik,t-1}^{immi}$) | 0.127* | 0.129* | 0.140** | 0.160** | 0.040 |
| | (0.070) | (0.068) | (0.069) | (0.077) | (0.078) |
| asinh ($\sum_k N_{ik,t-1}^{immi}$) | -0.087** | -0.135 | -0.131 | | |
| | (0.041) | (0.141) | (0.137) | | |
| asinh ($\sum_i N_{ik,t-1}^{immi}$) | 0.108 | 0.081 | -0.277 | -0.358 | -0.375 |
| | (0.660) | (0.667) | (0.544) | (0.581) | (0.610) |
| asinh ($N_{ik,t-1}^{emi}$) | -0.151** | -0.166** | 0.001 | -0.042 | 0.050 |
| | (0.062) | (0.053) | (0.028) | (0.051) | (0.047) |
| asinh ($\sum_k N_{ik,t-1}^{emi}$) | 0.010 | 0.002 | -0.011 | | |
| | (0.107) | (0.185) | (0.187) | | |
| asinh ($\sum_i N_{ik,t-1}^{emi}$) | -0.550 | -0.517 | -0.079 | -0.034 | 0.053 |
| | (0.569) | (0.580) | (0.515) | (0.541) | (0.623) |
| asinh ($\sum_k N_{ik,t}^{births}$) | 0.349*** | 0.305** | 0.299* | | |
| | (0.041) | (0.144) | (0.160) | | |
| asinh ($\sum_i N_{ik,t}^{births}$) | 0.546*** | 0.554*** | 0.582*** | 0.591*** | 0.659*** |
| | (0.035) | (0.038) | (0.048) | (0.052) | (0.084) |
| $\omega_{ik,t-1}^{immi}$ | 0.012*** | 0.014*** | 0.014*** | 0.020** | 0.033** |
| | (0.003) | (0.003) | (0.003) | (0.007) | (0.013) |
| $\omega_{ik,t-1}^{emi}$ | 0.007*** | -0.002 | -0.002 | -0.010 | -0.014 |
| | (0.001) | (0.003) | (0.004) | (0.011) | (0.022) |
| $\omega_{ik,t-1}^{births}$ | 0.002 | 0.009 | 0.010 | 0.022** | 0.016 |
| | (0.003) | (0.006) | (0.006) | (0.009) | (0.015) |
| $diversity_{i,t-1}$ | -0.006 | -0.026 | -0.027 | | |
| | (0.022) | (0.022) | (0.021) | | |
| $ubiquity_{k,t-1}$ | 0.011** | 0.012** | 0.008*** | 0.008 | 0.011 |
| | (0.004) | (0.004) | (0.002) | (0.004) | (0.007) |
| $\rho_{ik,t-1}^{M}$ | 0.233 | 0.278 | 0.368*** | 0.476*** | -0.805 |
| | (0.501) | (0.432) | (0.050) | (0.052) | (0.668) |
| $\rho_{ik,t-1}^{\omega}$ | 0.027 | 0.020 | 0.004 | 0.000 | 0.037 |
| | (0.028) | (0.036) | (0.022) | (0.058) | (0.044) |
| $\log(pop_{i,t})$ | 0.064 | 0.268* | 0.243* | | |
| | (0.047) | (0.116) | (0.114) | | |
| FE: period | Y | Y | Y | | |
| FE: region | | Y | Y | | |
| FE: category | | | Y | Y | |
| FE: period-region | | | | Y | |
| FE: region-period-broad category | | | | | Y |
| FE: category-period | | | | | Y |
| Observations | 6216 | 6216 | 6216 | 6193 | 3948 |
| Pseudo-R2 | 0.111 | 0.129 | 0.136 | 0.148 | 0.225 |
| BIC | 5670.3 | 6693.6 | 6861.2 | 7487.4 | 9574.5 |

Standard errors are clustered by period and region. *asinh()* denotes the inverted hyperbolic sine function. * p < 0.1, ** p < 0.05, *** p < 0.01

*Table S31.* Logistic regression models explaining exits, decomposing RCA values

|  | Dependent Variable: $Exit_{ik,t}$ | | | | |
|---|---|---|---|---|---|
|  | (1) | (2) | (3) | (4) | (5) |
| asinh $(N^{immi}_{ik,t-1})$ | -0.294*** | -0.323*** | -0.324*** | -0.414*** | -0.378*** |
|  | (0.040) | (0.036) | (0.045) | (0.011) | (0.092) |
| asinh $(\sum_k N^{immi}_{ik,t-1})$ | 0.221*** | 0.445** | 0.421** |  |  |
|  | (0.044) | (0.148) | (0.151) |  |  |
| asinh $(\sum_i N^{immi}_{ik,t-1})$ | 0.150 | 0.407 | 0.533 | 0.510 | -2.546 |
|  | (0.516) | (0.537) | (0.702) | (0.967) | (2.330) |
| asinh $(N^{emi}_{ik,t-1})$ | -0.070 | -0.056 | -0.029 | -0.015 | 0.046 |
|  | (0.047) | (0.048) | (0.049) | (0.045) | (0.206) |
| asinh $(\sum_k N^{emi}_{ik,t-1})$ | -0.290** | -0.700* | -0.790** |  |  |
|  | (0.111) | (0.315) | (0.296) |  |  |
| asinh $(\sum_i N^{emi}_{ik,t-1})$ | 0.192 | -0.040 | -0.101 | -0.114 | 3.110 |
|  | (0.530) | (0.565) | (0.847) | (1.136) | (2.377) |
| asinh $(\sum_k N^{births}_{ik,t})$ | -0.285** | -0.298 | -0.312 |  |  |
|  | (0.141) | (0.207) | (0.210) |  |  |
| asinh $(\sum_i N^{births}_{ik,t})$ | -0.499*** | -0.521*** | -0.422*** | -0.479*** | -0.721*** |
|  | (0.069) | (0.077) | (0.100) | (0.083) | (0.099) |
| $\omega^{immi}_{ik,t-1}$ | -0.019*** | -0.024** | -0.021** | -0.047*** | -0.072*** |
|  | (0.003) | (0.010) | (0.010) | (0.014) | (0.017) |
| $\omega^{emi}_{ik,t-1}$ | 0.014* | 0.015* | 0.014 | -0.005 | -0.042 |
|  | (0.008) | (0.009) | (0.009) | (0.017) | (0.054) |
| $\omega^{births}_{ik,t-1}$ | -0.003 | -0.004 | -0.003 | -0.037*** | -0.018 |
|  | (0.009) | (0.010) | (0.011) | (0.008) | (0.032) |
| $diversity_{i,t-1}$ | -0.010 | -0.006 | 0.000 |  |  |
|  | (0.015) | (0.032) | (0.038) |  |  |
| $ubiquity_{k,t-1}$ | -0.011 | -0.015** | -0.023*** | -0.021** | -0.052** |
|  | (0.007) | (0.007) | (0.007) | (0.010) | (0.020) |
| $\rho^M_{ik,t-1}$ | 1.815** | 2.308** | 1.788* | 2.195** | 7.230*** |
|  | (0.825) | (0.913) | (0.921) | (1.086) | (0.575) |
| $\rho^\omega_{ik,t-1}$ | -0.029 | -0.003 | 0.021 | 0.039 | 0.139 |
|  | (0.018) | (0.038) | (0.043) | (0.054) | (0.161) |
| $\log(pop_{i,t})$ | -0.030** | -0.058 | -0.062 |  |  |
|  | (0.011) | (0.167) | (0.173) |  |  |
| FE: period | Y | Y | Y |  |  |
| FE: region |  | Y | Y |  |  |
| FE: category |  |  | Y | Y |  |
| FE: period-region |  |  |  | Y |  |
| FE: region-period-broad category |  |  |  |  | Y |
| FE: category-period |  |  |  |  | Y |
| Observations | 2070 | 2048 | 2042 | 2009 | 1056 |
| Pseudo-R2 | 0.111 | 0.129 | 0.136 | 0.148 | 0.225 |
| BIC | 2619.9 | 3475.0 | 3615.5 | 4040.7 | 3665.7 |

Standard errors are clustered by period and region. *asinh()* denotes the inverted hyperbolic sine function. * $p < 0.1$, ** $p < 0.05$, *** $p < 0.01$

**Table S32**. *Logistic regression models explaining entries defined as first births in a specific activity, decomposing RCA values*

|  | Dependent Variable: $Entry2_{ik,t}$ | | | | |
|---|---|---|---|---|---|
|  | (1) | (2) | (3) | (4) | (5) |
| asinh $(N^{immi}_{ik,t-1})$ | 0.203* | 0.195 | 0.256** | 0.317** | 0.325*** |
|  | (0.117) | (0.136) | (0.118) | (0.143) | (0.114) |
| asinh $(\sum_k N^{immi}_{ik,t-1})$ | 0.051 | 0.338*** | 0.347** |  |  |
|  | (0.056) | (0.095) | (0.117) |  |  |
| asinh $(\sum_i N^{immi}_{ik,t-1})$ | 1.704** | 1.706* | 0.918** | 0.809** | 3.208*** |
|  | (0.842) | (0.912) | (0.316) | (0.374) | (1.103) |
| asinh $(\sum_k N^{emi}_{ik,t-1})$ | 0.024 | -0.481 | -0.582* |  |  |
|  | (0.061) | (0.299) | (0.341) |  |  |
| asinh $(\sum_i N^{emi}_{ik,t-1})$ | -1.366 | -1.345 | -0.780 | -0.729 | -2.791 |
|  | (0.866) | (0.990) | (0.482) | (0.522) | (1.640) |
| $\omega^{immi}_{ik,t-1}$ | 0.017*** | 0.015*** | 0.017*** | 0.023*** | 0.005 |
|  | (0.002) | (0.004) | (0.004) | (0.006) | (0.006) |
| $\omega^{emi}_{ik,t-1}$ | 0.017*** | 0.012** | 0.012* | -0.004 | 0.012 |
|  | (0.003) | (0.006) | (0.007) | (0.018) | (0.028) |
| $\omega^{births}_{ik,t-1}$ | -0.010 | -0.005 | -0.009 | 0.004 | -0.017 |
|  | (0.007) | (0.007) | (0.007) | (0.010) | (0.015) |
| $diversity_{i,t-1}$ | 0.017*** | 0.015*** | 0.017*** | 0.023*** | 0.005 |
|  | (0.018) | (0.048) | (0.049) |  |  |
| $ubiquity_{k,t-1}$ | -0.007 | -0.007 | -0.005 | -0.005 | -0.016 |
|  | (0.005) | (0.006) | (0.009) | (0.010) | (0.014) |
| $\rho^M_{ik,t-1}$ | 1.704*** | 1.872*** | 2.651*** | 3.018*** | 2.988*** |
|  | (0.326) | (0.355) | (0.527) | (0.698) | (0.520) |
| $\rho^\omega_{ik,t-1}$ | 0.053 | 0.047 | 0.085* | 0.098* | 0.187 |
|  | (0.043) | (0.072) | (0.038) | (0.050) | (0.124) |
| $\log(pop_{i,t})$ | 0.282*** | 0.383* | 0.419 |  |  |
|  | (0.068) | (0.197) | (0.248) |  |  |
| FE: period | Y | Y | Y |  |  |
| FE: region |  | Y | Y |  |  |
| FE: category |  |  | Y | Y |  |
| FE: period-region |  |  |  | Y |  |
| FE: region-period-broad category |  |  |  |  | Y |
| FE: category-period |  |  |  |  | Y |
| Observations | 5600 | 5570 | 5570 | 5517 | 3541 |
| Pseudo-R2 | 0.125 | 0.154 | 0.231 | 0.245 | 0.325 |
| BIC | 5367.5 | 6292.7 | 6042.0 | 6558.2 | 8411.8 |

Standard errors are clustered by period and region. *asinh()* denotes the inverted hyperbolic sine function. * $p < 0.1$, ** $p < 0.05$, *** $p < 0.01$

**Table S33.** Logistic regression models explaining exits defined as last births in a specific activity, decomposing RCA values

|  | Dependent Variable: $Exit2_{ik,t}$ | | | | |
|---|---|---|---|---|---|
|  | (1) | (2) | (3) | (4) | (5) |
| asinh ($N_{ik,t-1}^{immi}$) | -0.341*** | -0.360*** | -0.382*** | -0.410*** | -0.311** |
|  | (0.053) | (0.067) | (0.084) | (0.076) | (0.120) |
| asinh ($\sum_k N_{ik,t-1}^{immi}$) | -0.250*** | -0.422** | -0.401** |  |  |
|  | (0.047) | (0.168) | (0.193) |  |  |
| asinh ($\sum_i N_{ik,t-1}^{immi}$) | -1.640** | -1.498** | -1.259** | -1.574** | -14.026*** |
|  | (0.661) | (0.584) | (0.527) | (0.544) | (3.014) |
| asinh ($N_{ik,t-1}^{emi}$) | -0.082 | -0.089 | -0.055 | -0.046 | 0.115 |
|  | (0.064) | (0.063) | (0.091) | (0.105) | (0.177) |
| asinh ($\sum_k N_{ik,t-1}^{emi}$) | 0.053 | 0.151 | 0.109 |  |  |
|  | (0.103) | (0.321) | (0.320) |  |  |
| asinh ($\sum_i N_{ik,t-1}^{emi}$) | 0.699 | 0.390 | 0.542 | 0.651 | 12.561** |
|  | (0.797) | (0.686) | (0.585) | (0.707) | (3.796) |
| $\omega_{ik,t-1}^{immi}$ | 0.001 | -0.011** | -0.015*** | -0.016 | -0.064*** |
|  | (0.008) | (0.005) | (0.005) | (0.014) | (0.024) |
| $\omega_{ik,t-1}^{emi}$ | 0.006 | 0.010 | 0.008 | 0.008 | 0.014 |
|  | (0.010) | (0.009) | (0.008) | (0.016) | (0.012) |
| $\omega_{ik,t-1}^{births}$ | 0.002 | -0.014 | -0.013 | -0.023 | 0.005** |
|  | (0.012) | (0.012) | (0.013) | (0.019) | (0.002) |
| $diversity_{i,t-1}$ | -0.044 | 0.071 | 0.076 |  |  |
|  | (0.037) | (0.041) | (0.047) |  |  |
| $ubiquity_{k,t-1}$ | 0.012** | 0.012 | -0.008 | -0.002 | -0.005 |
|  | (0.005) | (0.007) | (0.013) | (0.016) | (0.027) |
| $\rho_{ik,t-1}^M$ | -0.616 | -0.749 | -0.270 | -0.709 | -2.410 |
|  | (1.058) | (0.939) | (0.929) | (0.870) | (1.892) |
| $\rho_{ik,t-1}^\omega$ | 0.053 | 0.128 | 0.094 | 0.112 | 0.127 |
|  | (0.042) | (0.076) | (0.052) | (0.065) | (0.096) |
| log ($pop_{i,t}$) | -0.341*** | -0.058 | -0.118 |  |  |
|  | (0.034) | (0.336) | (0.311) |  |  |
| FE: period | Y | Y | Y |  |  |
| FE: region |  | Y | Y |  |  |
| FE: category |  |  | Y | Y |  |
| FE: period-region |  |  |  | Y |  |
| FE: region-period-broad category |  |  |  |  | Y |
| FE: category-period |  |  |  |  | Y |
| Observations | 2687 | 2687 | 2659 | 2547 | 1498 |
| Pseudo-R2 | 0.209 | 0.263 | 0.327 | 0.332 | 0.405 |
| BIC | 3118.7 | 3947.3 | 3853.5 | 4278.8 | 4655.8 |

Standard errors are clustered by period and region. $asinh()$ denotes the inverted hyperbolic sine function. * p < 0.1, ** p < 0.05, *** p < 0.01